\documentclass[aps,prd,superscriptaddress,eqsecnum,nofootinbib,preprint,tightenlines,floatfix,titlepage]{revtex4-2}

\usepackage{float}
\usepackage{amsmath,amssymb}
\usepackage{multirow}
\usepackage[normalem]{ulem}
\usepackage{centernot}
\usepackage{booktabs}
\usepackage{youngtab}
\usepackage{tikz-cd}
\usepackage[compat=1.1.0]{tikz-feynman}
\usepackage[section]{placeins}
\usepackage[sort&compress]{natbib}
\usepackage[hyperfootnotes=false]{hyperref}
\hypersetup{
    colorlinks=true,     
    linkcolor=blue,      
    citecolor=blue,      
    filecolor=blue,      
    urlcolor=blue        
}
\usepackage[capitalise]{cleveref}
\usepackage{xcolor,cancel}

\crefname{section}{Sec.}{Secs.}
\Crefname{section}{Section}{Sections}
\crefrangeformat{equation}{Eqs.~(#3#1#4)--(#5#2#6)}
\Crefrangeformat{equation}{Equations~(#3#1#4)--(#5#2#6)}
\crefrangeformat{figure}{Figs.~#3#1#4--#5#2#6}
\Crefrangeformat{figure}{Figures~#3#1#4--#5#2#6}

\newcommand{\nn}{\nonumber}

\newcommand{\Ctconn}{C^{ll}(t)(\mathrm{conn.})}

\newcommand{\amuL}{a_\mu^{ll}(\mathrm{conn.})}
\newcommand{\amuHVP}{a_\mu^{\mathrm{HVP,LO}}}

\newcommand{\fnalaf}{Theory Division, Fermi National Accelerator Laboratory, Batavia, Illinois, 60510, USA}

\tikzfeynmanset{warn luatex=false}

\begin{document}
\count\footins = 1000

\title{The two-pion contribution to the hadronic vacuum polarization with staggered quarks.}

\author{Shaun Lahert}\email{shaun.lahert@gmail.com}
\affiliation{Department of Physics and Astronomy, University of Utah, Salt Lake City, UT, USA}
\affiliation{Department of Physics, University of Illinois, Urbana, Illinois, 61801, USA}

\author{Carleton DeTar}
\affiliation{Department of Physics and Astronomy, University of Utah, Salt Lake City, UT, USA}

\author{Aida X.~El-Khadra}
\affiliation{Department of Physics, University of Illinois, Urbana, Illinois, 61801, USA}
\affiliation{Illinois Center for the Advanced Studies of the Universe, University of Illinois, Urbana, Illinois, 61801, USA}

\author{Steven~Gottlieb}
\affiliation{Department of Physics, Indiana University, Bloomington, IN 47405, USA}

\author{Andreas S.~Kronfeld}
\affiliation{\fnalaf}

\author{Ruth~S.~\surname{Van~de~Water}}\affiliation{\fnalaf}

\date{\today}

\begin{abstract}
We present results from the first lattice QCD calculation of the two-pion contributions to the light-quark connected vector-current correlation function obtained from staggered-quark operators.  We employ the MILC collaboration's gauge-field ensemble with $2+1+1$ flavors of highly improved staggered sea quarks at a lattice spacing of $a\approx 0.15$~fm with a light sea-quark mass at its physical value.  
The two-pion contributions allow for a refined determination of the noisy long-distance tail of the vector-current correlation function, which we use to compute the light-quark connected contribution to HVP with improved statistical precision.  
We compare our results with traditional noise-reduction techniques used in lattice QCD calculations of the light-quark connected HVP, namely the so-called fit and bounding methods. We observe a factor of roughly three improvement in the statistical precision in the determination of the HVP contribution to the muon's anomalous magnetic moment over these approaches. We also lay the group theoretical groundwork for extending this calculation to finer lattice spacings with increased numbers of staggered two-pion taste states.

\end{abstract}

\maketitle

\allowdisplaybreaks

\newpage

\tableofcontents

\section{Introduction}\label{sec:intro}

The final result from the Fermilab Muon $g-2$ experiment has enabled a world-average determination of the muon anomalous magnetic moment $a_\mu \equiv (g_\mu-2)/2$ with the exquisite precision of $124$~ppb~\cite{Muong-2:2025xyk}. In contrast, the latest Standard Model (SM) prediction obtained by the Muon $g-2$ Theory Initiative in its 2025 White Paper (WP25) \cite{Aliberti:2025beg} has an uncertainty that is roughly four times larger than the experimental average, underscoring the need for improved theoretical control over the hadronic vacuum polarization (HVP) contribution to $a_\mu$, which is the dominant source of theoretical uncertainty.
Data-driven dispersive evaluations of the HVP rely on experimental measurements of the $e^+e^- \to \text{hadrons}$ cross sections (the $R$~ratio), which now have significant unresolved discrepancies in the crucial $\pi^+\pi^-$ channel \cite{Aliberti:2025beg}, making a meaningful average to obtain a new, data-driven estimate for $a_\mu^{\mathrm{HVP}}$ presently impossible.\footnote{This situation is expected to change in the coming years, with ongoing and new measurements as well as theoretical work to improve MC generators and radiative corrections \cite{Aliberti:2025beg}.}
In light of this situation, the Muon $g-2$ Theory Initiative instead relied on lattice Quantum Chromodynamics (lattice QCD) calculations of this quantity. Improving the overall precision of lattice HVP calculations, therefore, continues to be essential, both for interpreting the muon $g-2$ measurement and for clarifying the status of the data-driven dispersive approach.

Lattice QCD offers an {\em ab initio} approach to computing the hadronic corrections and hence allows for an entirely SM-based evaluation.\footnote{The only external inputs are those required to fix the quark masses and lattice spacing in the QCD Lagrangian, typically hadron masses.} Significant progress has been made in recent years toward a precise lattice-QCD determination of $\amuHVP$ \cite{RBC:2018dos,Giusti:2019xct,Borsanyi:2020mff,Lehner:2020crt,Wang:2022lkq,Aubin:2022hgm,Ce:2022kxy,ExtendedTwistedMass:2022jpw,RBC:2023pvn,Kuberski:2024bcj,Boccaletti:2024guq,Spiegel:2024dec,RBC:2024fic,Djukanovic:2024cmq,ExtendedTwistedMass:2024nyi,MILC:2024ryz,FermilabLatticeHPQCD:2024ppc}. The long-distance (LD) contribution to $\amuHVP$ has been identified as the dominant source of uncertainty in lattice calculations, and has recently been the focus of dedicated calculations by the RBC/UKQCD collaboration~\cite{RBC:2024fic}, the Mainz group~\cite{Djukanovic:2024cmq}, and the Fermilab Lattice, HPQCD, and MILC collaborations~\cite{FermilabLatticeHPQCD:2024ppc}, all of which were included in the WP25 lattice average~\cite{Aliberti:2025beg}. A hybrid result combining a lattice QCD calculation with a data-driven evaluation of the LD contributions was also presented in Ref.~\cite{Boccaletti:2024guq}, though it was published only recently, after Ref.~\cite{Aliberti:2025beg}. Despite this progress, the LD contribution remains the dominant bottleneck for reaching the precision needed to fully exploit the final Fermilab experimental result. The purpose of this paper is to develop an improved methodology to better control the systematic uncertainty of long-distance contributions to HVP, as part of a larger undertaking by the Fermilab Lattice, HPQCD and MILC collaborations~\cite{Davies:2019efs,Bazavov:2023has,MILC:2024ryz,FermilabLatticeHPQCD:2024ppc}.

The HVP is typically computed in lattice QCD as an integral over Euclidean time $t$ of two-point correlation functions with vector-current operators (representing the corresponding electromagnetic current) at the source and sink \cite{Blum:2002ii, Bernecker:2011gh}. As is well known, vector-current correlation functions of light-quark operators suffer from rapidly increasing statistical uncertainty at large Euclidean times, which in turn limits statistical precision of the integral. Noise-reduction methods, such as the truncated solver method, low-mode averaging or improvement \cite{TSM1,TSM2,Giusti:2004yp,DeGrand:2004wh,Shintani:2014vja} coupled with high-statistics computations have been used to improve statistical precision at large Euclidean times. In addition, analysis methods such as the bounding \cite{RBC:2018dos} and fit \cite{Davies:2019efs} methods can yield further improvements. However, to obtain lattice QCD results of $\amuHVP$ at the required few permille level, better control over the LD tail of the correlation function is needed. 
In the spectral decomposition of the vector-current correlation function, the dominant contributions at large Euclidean times come from two-pion states (where the pions have back-to-back momenta) with energies below the $\rho^0$ meson. Hence, a robust strategy is to compute additional correlation functions to obtain the energies and amplitudes of all contributing low-energy two-pion states. This approach, which requires the computation of two-, three-, and four-point functions, has already been implemented for order-$a$-improved Wilson \cite{Erben:2019nmx,Djukanovic:2024cmq} and domain wall fermions \cite{Bruno:2019nzm,RBC:2024fic}.

In this work, we perform the first such computation for the case of staggered quarks with the full set of staggered two-meson operators. First results from this study were presented in Ref.~\cite{Lahert:2021xxu}.\footnote{See also Ref.~\cite{Lahert:2023ore} for a detailed description of the group theoretic derivation, analysis steps, and additional background information. Preliminary results from a similar effort were reported in Ref.~\cite{Frech:2023bzy}.} 
The staggered formulation \cite{Banks:1975gq,Susskind:1976jm,Sharatchandra:1981si,Kawamoto:1981hw} of lattice QCD, 
which uses the so-called doubling symmetry of the naively discretized Dirac action to reduce the number of spin degrees of freedom from 4 to 1, results in a more complicated group structure with an additional quantum number which is called `taste.' Hence, careful treatment of the modified group structure is needed to correctly resolve the low-lying spectra. This includes obtaining the irreducible representations of the staggered group, computing the corresponding Clebsch-Gordan coefficients and constructing the multi-particle two-pion operators.  With the staggered operator basis in hand, the remaining steps are similar to those in Refs.~\cite{Bruno:2019nzm,Erben:2019nmx}. After computing the correlation functions on the $a\approx0.15$~fm HISQ ensemble with a light-sea quark mass at its physical value \cite{MILC:2012znn}, we obtain the spectrum of the two-pion energies and amplitudes from a generalized-eigenvalue-problem (GEVP) analysis and finally use it to reconstruct the two-point vector-current correlation function at large~$t$. We find a significant reduction in the statistical uncertainty of the resulting $\amuHVP$ over the traditional methods of large-$t$ noise reduction, in agreement with the studies using other discretizations. Hence, we plan to incorporate this approach into our ongoing effort within the Fermilab Lattice, HPQCD, and MILC Collaborations~\cite{Davies:2019efs,Bazavov:2023has} to compute $\amuHVP$ at the less than $0.5\%$ level.  

The rest of the paper is organized as follows.
\Cref{sec:LQC} introduces the vector-current correlation function and its relation to $\amuHVP$. \Cref{sec:method} details our calculation strategy, from constructing our operator basis (\cref{sec:opconstruction,sec:OpBasis}), computing the correlation functions (\cref{sec:corrFuncs,sec:numerical}) and determination of the two-pion energies and amplitudes from the GEVP (\cref{sec:gevp}). In \cref{sec:results}, we present our final results on the $a\approx 0.15$ ensemble, for the two-pion spectrum (\cref{sec:twoPiSpectrum}) and our subsequent reconstruction of the correlation function and computation of $\amuHVP$ (\cref{sec:corrRec}). \Cref{sec:outlook} provides a summary and outlook of the potential impact of this approach. In \cref{sec:stagGroupPrimer,sec:charTables,sec:stagIrrepsLists}, we cover the prerequisite details of the staggered-quark formalism, namely the notation employed for the irreducible representations, treatment of staggered states at non-zero momentum and connection to the continuum. We include the theoretical details to perform this calculation at any lattice spacing with any number of two-pion states. We note that our work builds on the results presented in Refs.~\cite{sharpe.1,Golterman:1984cy,Golterman:1986} and we restate the pertinent parts using our notation (and include minor corrections). \Cref{sec:stag2piStatesApp} contains tables of the Clebsch-Gordan coefficients and \Cref{sec:rooting} discusses the correct weighting of connected and disconnected diagrams with rooted staggered quarks.
\Cref{sec:timeSplit} details a slight modification made to the two-pion operators and how it impacts the analysis. Finally, \cref{sec:matrixFit} explores an alternative approach to the GEVP method using a direct multi-exponential matrix fit to the correlator matrix, which serves as a cross check on our analysis strategy.

\section{Preliminaries}\label{sec:LQC}

In lattice QCD, the hadronic vacuum polarization contribution to the muon's anomalous magnetic moment, $\amuHVP$, is, typically, obtained from weighted integrals of Euclidean vector-current correlation functions \cite{Blum:2002ii,Bernecker:2011gh},
\begin{align} 
    C(t)&= \frac{1}{3} \sum_{\vec{x}, k}\left\langle J^{k}(\vec{x}, t) J^{k}(0)\right\rangle ,
    \quad
    k=1,2,3,
    \label{eq:corrFunc2pt}\\
    &J^{\mu}(x)= \sum_{f} q_{f} \bar{\psi}_{f}(x) \gamma^{\mu} \psi_{f}(x), \label{eq:vecCurrent}
\end{align}
where the electromagnetic current $J^{\mu}(x)$ is summed over all quark flavors $f=\{u,d,s,c,b,t\}$ and $q_f$ are the corresponding electric charges in units of $e$. The RHS of \cref{eq:corrFunc2pt} contains both quark-line connected and disconnected Wick contractions. The leading-order HVP contribution to $a_\mu$ is obtained from the following formulae \cite{Bernecker:2011gh}:
\begin{align}
    a_{\mu}^{\mathrm{HVP}, \mathrm{LO}} &= 4\alpha^{2} \int_{0}^{\infty} \mathrm{d} t\, C(t)  \tilde K(t)
    \label{eq:amuTint}\\
    \tilde{K}(t) &= 2 \int_{0}^{\infty} \frac{\mathrm{d} Q}{Q}\, K_{E}(Q^{2})
        \left[Q^{2} t^{2}-4 \sin ^{2}\left(\frac{Q t}{2}\right)\right].
    \label{eq:Ktilde}
\end{align}
The integration kernel $K_{E}(Q^{2})$ \cite{Blum:2002ii}, which contains the muon mass dependence, is given as
\begin{align}
    K_E\left(Q^2\right) = \frac{m_\mu^2Q^2Z^3 (1-Q^2Z)}{1+m_\mu^2Q^2 Z^2},  \quad\quad
    \label{eq:KE}
\end{align}
where $Z = [(Q^4 + 4m_\mu^2 Q^2)^{1/2} -Q^2]/(2m_\mu^2Q^2)$. 
In lattice-QCD calculations of $\amuHVP$ the contributions from each quark flavor and from connected and disconnected Wick contractions are typically computed separately and then summed up. Here we focus on the dominant light-quark connected contribution in the isospin-symmetric limit, $\amuL$. Therefore, our electromagnetic vector current $J^{\mu}(x)$ includes only the up and down terms with both masses equal, $m_l = (m_u + m_d)/2$. Additionally, the correlation function $C(t)$ includes only the connected contractions. 
This can be straightforwardly related to the pure isospin 1 contribution. Splitting the flavor components of the vector current operator from \cref{eq:vecCurrent} into isospin 1 and isospin 0 components,
$J_{i}=   J_{i}^{I=1} +  J_{i}^{I=0}$, 
gives
\begin{align}
    J_{i}^{I=1} = \rho^0_i = \frac{1}{2} \left( \bar u \gamma_i u - \bar d \gamma_i d \right) , \quad \quad \label{eqn:I1VecOp} \\
    J_{i}^{I=0} = \frac{1}{6}\left(\bar{u} \gamma_{i} u+\bar{d} \gamma_{i} d-2 \bar{s} \gamma_{i} s+\ldots\right). 
\end{align}
We note that $J_{i}^{I=1}$ has $I_3=0$ and is equivalent to a $\rho^0$ meson bilinear. (In most of the rest of this work the $\rho^0$ notation is employed.) Hence, once charge factors are accounted for, the following linear relationship between the light-quark connected and $I=1$ correlation functions is obtained,
\begin{align}
    \left\langle J^{l}_{i}(x) J^{l}_{i}(0)\right\rangle_{\textrm{conn.}} &= \frac{10}{9}\left\langle \rho_{i}^{0}(x) \rho_{i}^{0}(0)\right\rangle,\\
    \Rightarrow \Ctconn &= \frac{10}{9}C_{\rho \to \rho}(t).
\end{align}
The light-quark connected correlation function, therefore, has the following spectral representation,
\begin{align}
    \Ctconn &=  \frac{10}{9}\sum_{n=0} \left|\langle 0 | \rho^{0} | n \rangle \right|^2e^{-E_n t} \, . \label{eq:corrFuncSpecRep}
\end{align}
 The average over the spatial direction in \cref{eq:corrFuncSpecRep} is implicit. The overlap amplitudes $\langle 0 | \rho^{0} | n \rangle$ select the states $| n \rangle$ of the Hamiltonian with the same quantum numbers as the $\rho^0$.

 The signal-to-noise issue discussed in the introduction can be traced to the fact that the variance of this correlation function falls off with an exponent of $ 2m_\pi$ \cite{Lepage:1989hd}
\begin{align}
     \lim_{t \to \infty} \sigma^2_{ \Ctconn} \sim \lim_{t \to \infty} \left[ \langle \left(\rho^0_i(t)\rho^0_i(0)\right)^2 \rangle -\langle \rho^0_i(t)\rho^0_i(0) \rangle^2 \right]&\sim \lim_{t \to \infty}  \langle \left(\rho^0_i(t)\rho^0_i(0)\right)^2 \rangle \sim e^{-2m_\pi t} \, . \label{eq:varcorrFuncSpecRep}
\end{align}
while the signal falls off with the lowest energy $I=1$, $I_3=0$ state, which is a two-pion state with the smallest non-zero back-to-back momentum possible in the finite volume of the lattice,
\begin{align}
    \lim_{t \to \infty} \Ctconn &\sim  e^{-E^0_{\pi \pi}(p \neq 0)t} \, . \label{eq:corrFuncSpecRepLargeT}
\end{align}
The noise, the square root of the variance in \cref{eq:varcorrFuncSpecRep}, falls off more slowly and overwhelms the two-pion signal in the large-time region.

At present, there are two commonly employed analysis-based approaches to address the signal-to-noise issue when computing $\amuL$, namely, the ``bounding'' and ``fit'' methods:
\begin{itemize}
\item \textbf{Bounding method} \cite{RBC:2018dos}: Two series of $\amuL$ values are obtained by replacing the correlation function, $\Ctconn$, with 
    \begin{align}
        C_{\textrm{bounded}}(t) = 
        \begin{cases}
            \Ctconn &\text{if  } t \leq t_\textrm{cut}, \\
            C^{ll}(t_\textrm{cut})(\textrm{conn.}) e^{-E_{\textrm{bound}}(t-t_\textrm{cut})} &\text{if  } t > t_\textrm{cut},
        \end{cases}
    \label{eqn:bounding}
\end{align}
for upper and lower bounding energies, resulting in lower and upper bounds on $\Ctconn$, respectively. Here, $t_\textrm{cut}$ is a free parameter that ranges over the temporal extent of the lattice. The final result for $\amuL$ is obtained at the value of $t_\textrm{cut}$ ( or range of values of $t_\textrm{cut}$) where the two bounds meet. The lower energy bound is taken to be the free, lattice two-pion energy \cite{RBC:2018dos,Borsanyi:2020mff}, $E_{\textrm{bound}}= 2\sqrt{(2\pi/L)^2+M_\pi^2}$.  The energy appearing in \cref{eq:corrFuncSpecRepLargeT} is the interacting energy, which is smaller than the free energy due to the binding energy of the $\pi \pi$ state.
However, this approximation is reasonable because the binding energy is small enough to shift $\amuL$ by only a small fraction of the total uncertainties currently achievable \cite{Budapest-Marseille-Wuppertal:2017okr}.\footnote{We find that this is true for the differences between interacting and free energies obtained in this work.} 
The upper energy bound, usually taken to be $E_{\textrm{bound}}=\infty$, can be improved by, instead, taking the ground state from a fit to $\Ctconn$. In the case of staggered fermions, the final choice of $t_\textrm{cut}$ is complicated by the presence of oscillations in the correlation function. Compared with direct integration, the bounding method improves the statistical precision of $\amuL$. However, the improvement is limited, because the bounds typically meet well into the noisy part of the tail, at roughly 2.5--3.5 fm for physical-mass correlators. 

\item \textbf{Fit method} \cite{Chakraborty:2016mwy,FermilabLatticeHPQCD:2024ppc}: For this approach, the correlation function is fit over a time range suitable for determining the spectrum.
The determined spectrum can be improved via combined fits to, for example, smeared correlated functions. The energies and $\Ctconn$ amplitudes are then used to reconstruct it after some time $t \ge t^\star$. Here, there is a systematic uncertainty associated with how well the fit correctly parameterizes the behavior of the lowest-energy states that determine $\Ctconn$ in the region where it is being replaced. This systematic effect has been studied extensively in Ref.~\cite{FermilabLatticeHPQCD:2024ppc}.
\end{itemize}

In this work, we treat the signal-to-noise problem by obtaining an accurate
spectral representation of the vector-current correlation function at large Euclidean times. For this purpose, we generate correlation functions using suitably constructed two-pion operators, from which the following matrix of correlation functions is formed,
\begin{align}
\mathbf{C}(t)=\left(\begin{array}{ll}
C_{\rho, \tilde{\rho} \rightarrow \rho, \tilde{\rho}}(t) & C_{\rho, \tilde{\rho} \rightarrow \pi \pi}(t) \\
C_{\pi \pi \rightarrow \rho, \tilde{\rho}}(t) & C_{\pi \pi \rightarrow \pi \pi}(t)
\end{array}\right). \label{eqn:ctmat}
\end{align}
The upper left $2\times2$ block, $C_{\rho, \tilde{\rho} \rightarrow \rho, \tilde{\rho}}(t)$, contains the correlation function constructed with the $\rho^0$ operator of \cref{eqn:I1VecOp} along with additional correlation functions obtained by including a smeared version of the operator $\tilde{\rho}^0$. This smearing improves the overlap with the ground state \cite{Davies:2019efs}. The bottom right block consists of the two-pion to two-pion four-point correlation functions, and the size of the block is given by the number of two-pion operators included. The off-diagonal blocks are correlation functions constructed from the ($\rho^0$,$\tilde{\rho}^0$) and two-pion operators.  
With this matrix, the lowest lying states for the $I=1$, $I_3=0$ channel can be precisely resolved and the tail of $\Ctconn$ can be reconstructed from this information. A similar approach was implemented in Ref.~\cite{Fu:2016itp} in a study of the $\rho$ resonance parameters with staggered valence quarks, where, however, only the simplest case of Goldstone-boson pion operators was considered. Ours is the first study of this system based on a complete description of the staggered two-pion operators.

\section{Methodology}\label{sec:method}

In this section, we describe all the steps of the calculation. 
\Cref{sec:opconstruction} describes the computation of the Clebsch-Gordan coefficients for the symmetry group of the staggered-fermion transfer matrix and, hence, the construction of the two-pion operators used here. In \cref{sec:corrFuncs}, we give the required Wick contractions corresponding to the correlation functions in the matrix of \cref{eqn:ctmat}. We tabulate the complete staggered operator bases on the physical-mass HISQ ensembles in \cref{sec:OpBasis}. \Cref{sec:numerical} describes the numerical strategy we employ to compute the Wick contractions of \cref{sec:corrFuncs}. In \cref{sec:finiteT}, we give our preferred approach for dealing with finite-time effects in the diagonal four-point correlation functions of \cref{eqn:ctmat}. Finally, in \cref{sec:gevp}, we discuss our GEVP based approach for extracting the desired energies and amplitudes from our matrix using a correlated fit.

\subsection{Operator construction}\label{sec:opconstruction}

For the case of staggered quarks, the two-pion states that couple to the $\rho^0$ need to transform correctly under isospin and the staggered symmetry group.
Under isospin, the two-pion operators need to transform as $I=1$, $I_3=0$, where
the single pion operators have the following $I=1$ quantum numbers,
\begin{align}
\pi^{+}  &= -\bar d u \quad \quad \quad \quad \quad \quad \ \  I_3=1,\\
\pi^{-}  &= \bar u  d \quad \quad \quad \quad \quad \quad \quad \ I_3=-1,\\
\pi^{0}  &= \frac{1}{\sqrt{2}}\left(\bar u u - \bar d d \right) \quad \quad \,  I_3=0.
\end{align}
and the two-pion operator then takes the form
\begin{align}
    \left(\pi \pi\right)^{I=1, I_3=0}=\frac{1}{\sqrt{2}}\pi^{+} \pi^{-} - \frac{1}{\sqrt{2}}\pi^{-} \pi^{+}.
    \label{eqn:TwoPiIsoState}
\end{align}
The $\pi^\pm$ states transform into each other under charge conjugation, so the minus sign on the right-hand side of \cref{eqn:TwoPiIsoState} ensures that these two-pion states have $C=-1$, just like~$\rho^0$.
The factors $\pm1/\sqrt{2}$ are $\text{SU}(2)$ Clebsch-Gordan coefficients---the rest of this subsection explains how to set up the analogous construction for the staggered symmetry group, to obtain two-pion operators with the same quantum numbers as staggered $\rho^0$ states.

Here, ``quantum numbers'' refer to irreducible representations (irreps) of the symmetry group of the transfer matrix of staggered quarks for $N_s^3$ spatial lattices.
This group is
\begin{equation}
    G_{T_0}(N_s) = Z_{N_s/2}^3\rtimes(\Gamma_{4,1}\rtimes \text{O}_\text{h}),
\end{equation}
where $T_0$ refers to the two timeslice transfer matrix~\cite{Sharatchandra:1981si}, and $\rtimes$ denotes semi-direct product.
The factors are, respectively, two-hop translations, the Clifford group of taste and charge conjugation, and the symmetry group of a cube.
Eigenstates of translations are labeled by momentum $\vec{p}$, where each component satisfies,
\begin{equation}
    p_i = \frac{2\pi}{aN_s}\ell_i, \quad \ell_i = -\frac{N_s}{4}+1, \ldots, -1, 0, 1, \ldots, \frac{N_s}{4},
    \label{eqn:allowedMomentum}
\end{equation}
for periodic boundary conditions; below it is more convenient to use $\vec{\ell}$ to label irreps.
For mesons, taste is denoted by a four-vector with entries $\pm1$ or, equivalently, $e^{i\xi_\mu}$, $\xi_\mu\in\{0,\pi\}$.
Similarly, charge conjugation is $e^{i\xi_C}=\pm1$.
The irreps of $\text{O}_\text{h}$ are $A_0^\pm$, $A_1^\pm$, $E^\pm$, $T_0^\pm$, $T_1^\pm$, with the superscript for parity.
We denote a general (bosonic) irrep
\begin{equation}
    (\ell_1,\ell_2,\ell_3) \rtimes [(\xi_0,\xi_1,\xi_2,\xi_3),\xi_C] \rtimes R, \label{eqn:stagIrrep}
\end{equation}
where the $\rtimes$ is reminder that the formalism of semi-direct groups is needed to construct the  irrep; the last factor $R$ denotes an irrep of $\text{O}_\text{h}$ or a so-called little group appropriate to $\vec{\ell}$ and $\xi_\mu$.
\Cref{sec:stagGroupPrimer,sec:charTables,sec:stagIrrepsLists,sec:stag2piStatesApp} contains a full discussion of the staggered group irreps; below we refer to them for details.

The staggered two-pion operators must transform under the same irreducible representation (irrep) as the vector current operator.
At zero momentum, the sixteen tastes are collected into five irreps; see \cref{sec:rhoIrreps}.
We choose the taste-singlet $\rho^0$ (see \cref{eqn:1lrho}) because it couples to a ground two-pion state of two pseudo-Goldstone boson pions, the lowest-energy two-pion state possible for any taste. Furthermore, all two-pion states which couple to the taste-singlet $\rho^0$ are taste singlets as well, meaning the two pions in the two-pion state must be in the same taste irrep.
Alongside the taste irreps, we must also consider the momentum and rotation irreps.
This is achieved by computing the Clebsch-Gordan coefficients (CGs) for $G_{T_0} \times G_{T_0} \to G_{T_0}$ as follows:
\begin{align}
    \pi^{(\alpha)}(\vec{p}) \otimes \pi^{(\alpha')}(-\vec{p}) \to (0, 0, 0) \rtimes [(0, 0, 0, 0), \pi]  \rtimes  T_0^{-} ,
    \label{eqn:SGDpSingle}
\end{align}
where $\pi^{(\alpha)}(\vec{p})$ denotes some non-zero momentum pion irrep, with the full list of such irreps given in \cref{sec:pionIrreps}.
The right-hand side of \cref{eqn:SGDpSingle} is the taste-singlet $\rho^0$ irrep from \cref{eqn:1lrho}.

In order to compute these CGs, the irreps from \cref{sec:stagIrreps} must be constructed.
For values of $N_s$ typically used in numerical simulations, $G_{T_0}(N_s)$ is enormous, but we are interested only in two-pion states below $\rho^0$ threshold (see \cref{sec:OpBasis}).
For the MILC HISQ ensembles with spatial size around 5--6~fm, that means we can restrict our attention to states with momenta $\ell_i \in \{0,\pm 1\}$.
According to \cref{eqn:allowedMomentum}, the smallest group needed to construct these irreps has $N_s=6$. The corresponding transfer-matrix symmetry group (\cref{eqn:stagGroupStruc}) has only $|G_{T_0}(6)|=82944$ elements. 

The non-zero momentum irreps correspond to matrices of dimensions between $6$ and $24$, depending on the taste- and momentum-dimension. Happily, one does not need to construct and store matrices for each of the 82944 elements of the group. A smaller subset can be used to form the tensor product representations in \cref{eqn:SGDpSingle} and decompose them into irreps. If this decomposition contains the taste singlet irrep of \cref{eqn:SGDpSingle}, we then compute the CGs.

For the first step, forming the tensor product representation and decomposing it, one needs only a representative element for each conjugacy class to perform the character decomposition \cite{Cornwell}. For $N_s/2=3$, this corresponds to 404 classes.
The second step, computing the CGs, is typically done by summing over all the group elements. However, for semi-direct product groups, the sum can be reduced by breaking up the group into subgroups and corresponding cosets \cite{Sakata:1974hd}.
The Clebsch-Gordan matrix $U$ relates a tensor product representation to the block-diagonal reduced matrix $D_R$,
\begin{align}
    D^{(\alpha)}(g) \otimes D^{(\beta)}(g)=U D_R(g) U^{\dagger}, \quad \quad \forall g \in G .
\end{align}
where the two representations on the left correspond to the two single-pion representations on the LHS of \cref{eqn:SGDpSingle}.
The approach to obtaining $U$, given in Ref.~\cite{Sakata:1974hd}, is summarized by the following equation,
\begin{align}
    U=\text{unitarity: } \sum_{g \in G}\left[D^{(\alpha)}(g) \otimes D^{(\beta)}(g)\right] A D_R(g)^{\dagger}, \label{eqn:cgCompute}
\end{align}
where $A$ is a matrix of entries to be determined from the unitarity constraint. As mentioned, the staggered group has the natural structure (nested semi-direct product) to reduce this sum to one over subgroups and cosets. First, the cosets of the full staggered group under the subgroup $\Gamma_{4,1} \rtimes \text{O}_\text{h}$ are obtained. This gives $(N_s/2)^3$ momentum cosets, with only one representative element from each coset needed. Then, the cosets of the group $\Gamma_{4,1} \rtimes \text{O}_\text{h}$ under $\text{O}_\text{h}$ are obtained, of which there are 64. So in total, for $N_s/2=3$, one needs only to store $27+64+48=139$ matrices for this step, where 48 is the order of $\text{O}_\text{h}$, the final subgroup.
The sum is thus reduced as
\begin{align}
    U=\text{unitarity: } \sum_{\tilde{z} \in \tilde{Z}_{N_s/2}^3}  D(\tilde{z}) \left[\sum_{ \tilde{\gamma} \in \tilde{\Gamma}_{4,1}} D(\tilde{\gamma})\left[\sum_{o \in \text{O}_\text{h}} D(o) A D^{\dagger}(o) \right] D^{\dagger}(\tilde{\gamma}) \right]D^{\dagger}(\tilde{z}),
\end{align}
where the tilde, for example $\tilde{\Gamma}_{4,1}$, denotes the coset representatives of the corresponding set. Hence, in combination with the 404 class representative elements, the total number of essential matrices needed per irrep, for $N_s/2=3$, is around $500$. This is a significant storage and computational cost reduction over the total order of the group.

To illustrate the steps outlined above, we perform the procedure for two specific cases of \cref{eqn:SGDpSingle}.
The first is the case of the staggered pion irreps that are one-dimensional at zero momentum, i.e., the irreps of \cref{eqn:pion1,eqn:pion50,eqn:pion0,eqn:pion5}.
As taste singlets, they have the same CGs as Wilson fermions.
The second case is for the staggered pion irreps that are three-dimensional at zero momentum, \emph{i.e.}, the irreps of \cref{eqn:pioni,eqn:pion5i,eqn:pioni0,eqn:pionij}.
As described in \cref{sec:nonzeromom}, these irreps can undergo ``taste-orbit splitting" at non-zero momentum, typically, into a one- and two-dimensional taste-orbit irrep.
The one-dimensional taste irrep is, again, akin to Wilson quarks, while the two-dimensional irrep is unique to staggered quarks.

To help illustrate these examples, we introduce the familiar notation for a general staggered operator with momentum and spin and taste quantum numbers $\Gamma_S$ and $\Gamma_T$, respectively,
\begin{align}
    \mathcal{O}^{\Gamma_S \otimes \Gamma_T}(\ell_1,\ell_2,\ell_3) ,
\end{align}
which are described in \cref{sec:stagOPS}, with the precise meaning of this notation defined through \cref{eqn:stBasisOP,eqn:phaseShiftOp,eqn:phiTrace,eqn:OpShift,eqn:opSym}.
The operators excite the states of the staggered irreps of \cref{eqn:stagIrrep}.
Hence, just as we label a staggered irrep by a single representative state of that irrep, as in \cref{eqn:stagIrrep}, we can correspondingly label the irrep by a representative operator which excites this specific state.
The correspondence between staggered irrep states and the operators which excite them is given by \cref{eqn:tasteOP,eqn:parityOp,eqn:rotOP,eqn:chargeOp}.
We denote this correspondence with the : symbol throughout the rest of this work, for example $\mathcal{O}^{\gamma_5\otimes\gamma_5}(0,0,1)~:~(0,0,1)\rtimes[(\pi,\pi,\pi,\pi), 0]\rtimes A_0$ for the irrep with pseudoscalar spin and taste and one unit of momentum.

\paragraph{Example 1}
For the first case, we take the above-mentioned pseudoscalar with one unit of momentum, \cref{eqn:100pion5}, as the representative example irrep. We have the following decomposition of the tensor product representation into irreps: 
\begin{align}
&\mathcal{O}^{ \gamma_5  \otimes  \gamma_5}(0, 0, 1) \otimes \mathcal{O}^{ \gamma_5  \otimes  \gamma_5}(0, 0, 1)  \;:\;  \nn \\
& (0, 0, 1)  \rtimes  [(\pi, \pi, \pi, \pi), 0]    \rtimes    A_0 \otimes (0, 0, 1)  \rtimes  [(\pi, \pi, \pi, \pi), 0]    \rtimes    A_0 \nn \\
& = \ (0, 0, 0)  \rtimes  [(0, 0, 0, 0), \pi]    \rtimes    A_0^{+}\quad \;:\; \quad \mathcal{O}^{ \gamma_0  \otimes  1}(0, 0, 0) \nn \\
& \oplus \ (0, 0, 0)  \rtimes  [(0, 0, 0, 0), \pi]    \rtimes    E_0^{+} \quad \;:\; \nn \\
& \oplus \ (0, 0, 0)  \rtimes  [(0, 0, 0, 0), \pi]    \rtimes    T_0^{-}\quad \;:\; \quad \mathcal{O}^{ \gamma_i  \otimes  1}(0, 0, 0) \label{eqn:decomposition} \\
& \oplus \ (0, 0, 1)  \rtimes  [(0, 0, 0, 0), \pi]    \rtimes    A_0\quad \;:\; \quad \mathcal{O}^{ \gamma_3  \otimes  1}(0, 0, 1) \nn \\
 &\oplus \ (1, 1, 0)  \rtimes  [(0, 0, 0, 0), \pi]    \rtimes    A_0\quad \;:\; \quad \mathcal{O}^{ \gamma_0  \otimes  1}(1, 1, 0) \nn \\
 &\oplus \ (1, 1, 0)  \rtimes  [(0, 0, 0, 0), \pi]    \rtimes    A_2\quad \;:\; \quad \mathcal{O}^{ \gamma_1\gamma_2  \otimes  1}(1, 1, 0) .\nn
\end{align}

Here we have implicitly incorporated the form of \cref{eqn:TwoPiIsoState} in the above direct product to ensure the desired staggered charge conjugation, $\xi_C=\pi$, is obtained in the decomposition. At zero momentum, there are $16\times 16=256$ staggered bilinears and $448$ irrep rows (states) in total (see \cref{table:tasteorbit}). Hence, some irreps, like the irrep including $E_0^+$ above, have no associated simple staggered bilinear which excite them. This irrep is instead excited by a more complicated staggered operator with a derivative insertion.

We are interested in the zero-momentum taste-singlet vector irrep which is the third irrep in the decomposition on the right-hand side of \cref{eqn:decomposition}, with the corresponding operator $\mathcal{O}^{\gamma_i\otimes1}(0,0,0)$.
The CGs are computed for the states of this irrep using \cref{eqn:cgCompute} and are given in \cref{tab:cgg5p001}. For clarity, we use the more familiar, operators instead of the states to label the rows and columns in the table.
\begin{table}
\centering
\caption{Clebsch-Gordan table for $(0, 0, 1) \, \rtimes \, [(\pi, \pi, \pi, \pi), 0] \,   \rtimes \,   A_0\ \otimes \ (0, 0, 1) \, \rtimes \, [(\pi, \pi, \pi, \pi), \pi] \,   \rtimes \,   A_0 = (0, 0, 0) \, \rtimes \, [(0, 0, 0, 0), \pi] \,   \rtimes \,   T_0^{-}  \oplus\cdots$. The irreps in the rows and columns are labeled by the corresponding operators.}\label{tab:cgg5p001}
\begin{tabular}{lccc}
\hline\hline
Tensor product row & $\mathcal{O}^{ \gamma_1 \, \otimes \, 1}\,(0, 0, 0)$ & $\mathcal{O}^{ \gamma_2 \, \otimes \, 1}\,(0, 0, 0)$ & $\mathcal{O}^{ \gamma_3 \, \otimes \, 1}\,(0, 0, 0)$ \\
\hline
$\mathcal{O}^{ \gamma_5 \, \otimes \, \gamma_5}\,(1, 0, 0) \ \otimes \ \mathcal{O}^{ \gamma_5 \, \otimes \, \gamma_5}\,(-1, 0, 0)$ &                                                               $\frac{1}{\sqrt{2}}$ &                                                                0 &                                                                0 \\
$\mathcal{O}^{ \gamma_5 \, \otimes \, \gamma_5}\,(-1, 0, 0) \ \otimes \ \mathcal{O}^{ \gamma_5 \, \otimes \, \gamma_5}\,(1, 0, 0)$ &                                                              $-\frac{1}{\sqrt{2}}$ &                                                                0 &                                                                0 \\
$\mathcal{O}^{ \gamma_5 \, \otimes \, \gamma_5}\,(0, 1, 0) \ \otimes \ \mathcal{O}^{\gamma_5 \, \otimes \, \gamma_5}\,(0, -1, 0)$ &                                                                0 &                                                               $\frac{1}{\sqrt{2}}$ &                                                                0 \\
$\mathcal{O}^{ \gamma_5 \, \otimes \, \gamma_5}\,(0, -1, 0) \ \otimes \ \mathcal{O}^{ \gamma_5 \, \otimes \, \gamma_5}\,(0, 1, 0)$ &                                                                0 &                                                              $-\frac{1}{\sqrt{2}}$ &                                                                0 \\
$\mathcal{O}^{ \gamma_5 \, \otimes \, \gamma_5}\,(0, 0, 1) \ \otimes \ \mathcal{O}^{ \gamma_5 \, \otimes \, \gamma_5}\,(0, 0, -1)$ &                                                                0 &                                                                0 &                                                               $\frac{1}{\sqrt{2}}$ \\
$\mathcal{O}^{ \gamma_5 \, \otimes \, \gamma_5}\,(0, 0, -1) \ \otimes \ \mathcal{O}^{ \gamma_5 \, \otimes \, \gamma_5}\,(0, 0, 1)$ &                                                                0 &                                                                0 &                                                              $-\frac{1}{\sqrt{2}}$ \\
\hline\hline
\end{tabular}
\end{table}

The two-pion operators are constructed from linear combinations of products of staggered single-pion operators with the CGs as coefficients. The staggered single-pion and two-pion operators are defined as
\begin{align}
\pi^{+}_{\otimes \gamma_\xi}\left(\vec{p},t\right)& \equiv-\frac{1}{N^{3/2}_S}\sum_{\vec{n}} e^{ia\vec{p} \cdot \vec{n}}\bar d(n) \gamma_5 \otimes \gamma_\xi u(n), \label{eqn:stagPionOP}\\
\pi^{-}_{\otimes \gamma_\xi}\left(\vec{p},t\right)& \equiv \frac{1}{N^{3/2}_S}\sum_{\vec{n}} e^{ia\vec{p} \cdot \vec{n}}\bar u(n) \gamma_5 \otimes \gamma_\xi d(n),\\
\mathcal{O}_{\pi \pi}(\vec{p}_1 + \vec{p}_2, t)\equiv\sum_{\xi_{1}, \xi_{2}, I_3^1, I_3^2,  \vec{p}_1, \vec{p}_2 }& \text{CG}_{G_{T_0} \text {, iso.}}(\xi_{1}, \xi_{2}, \vec{p}_1, \vec{p}_2, I_3^1, I_3^2) \pi_{\otimes \gamma_{\xi_{1}}}^{I_3^1}(\vec{p}_1, t) \pi^{I_3^2}_{\otimes \gamma_{\xi_{2}}}(\vec{p}_2, t).
\end{align}
Combining the results of \cref{tab:cgg5p001} with \cref{eqn:TwoPiIsoState} 
we obtained the normalized\footnote{We are interested only in the overall overlap amplitudes of the $\rho^0$ operator, so are free to normalize the two-pion operators as we choose.} staggered-isospin two-pion operator, built from $\otimes \gamma_5$ taste pions, that couples to the third component of the taste-singlet $\rho^0$:
\begin{align}
     \left[\mathcal{O}^{\otimes \, \gamma_5}_{\pi \pi}(\vec{0}, t )\right]&^{ \gamma_3 \, \otimes \, \gamma_1,\,I=1, I_3=0}= \nn \\
      &\frac{1}{\sqrt{2}}\left[\pi^{+}_{\otimes \gamma_5}\left((0,0,1),t\right) \pi^{-}_{\otimes \gamma_5}\left((0,0,-1),t\right)-\pi^{-}_{ \otimes \gamma_5}\left((0,0,-1),t\right) \pi^{+}_{ \otimes \gamma_5}\left(0,0,1),t\right)\right]. \label{eqn:Opg52pi001} 
\end{align} 

\paragraph{Example 2}
Differences from the Wilson case appear only when one considers spin-pseudoscalar irreps that have a larger dimension than one at zero momentum. For example, starting with the taste pseudo-vector \cref{eqn:pion5i}, which is three-dimensional, giving it one unit of momentum and taking the irrep where taste orbit is two-dimensional as our starting point (second line of \cref{eqn:pion001G5I}). Here the taste-vector is orthogonal to the momentum . The tensor product representation then has the following decomposition,
\begin{align}
 \mathcal{O}^{ \gamma_5  \otimes  \gamma_5\gamma_{i\neq 3}}(0, 0, 1)& \otimes \mathcal{O} ^{ \gamma_5  \otimes  \gamma_5\gamma_{i\neq 3}}(0, 0, 1) \;:\; \nn \\
&\ (0, 0, 1)  \rtimes  [(0, 0, \pi, 0), 0]    \rtimes    A_2 \ \otimes \ (0, 0, 1)  \rtimes  [(0, 0, \pi, 0), 0]    \rtimes    A_2 \nn \\
& = \ (0, 0, 0)  \rtimes  [(0, 0, 0, 0), \pi]    \rtimes    A_0^{+} \;:\; \mathcal{O}^{ \gamma_0  \otimes  1}(0, 0, 0) \nn \\
& \oplus \ (0, 0, 0)  \rtimes  [(0, 0, 0, 0), \pi]    \rtimes    E_0^{+} \nn \\
& \oplus \ (0, 0, 0)  \rtimes  [(0, 0, 0, 0), \pi]    \rtimes    E_0^{+} \nn \\
& \oplus \ (0, 0, 0)  \rtimes  [(0, 0, 0, 0), \pi]    \rtimes    E_0^{+} \nn \\
& \oplus \ (0, 0, 0)  \rtimes  [(0, 0, 0, 0), \pi]    \rtimes    T_0^{-} \;:\; \mathcal{O}^{ \gamma_i  \otimes  1}(0, 0, 0) \\
& \oplus\  \ldots, \nn 
\end{align}
where again we use the form of \cref{eqn:TwoPiIsoState} to obtain the desired charge conjugation in the decomposition. There are now multiple copies of the same irrep appearing in the tensor product representation, as it corresponds to a $12 \times 12 =144$ dimensional reducible matrix. The sixth irrep listed is the one we are after, and the CGs for this are given in \cref{tab:cgg5ip001}.
\begin{table}
\centering
\caption{Clebsch-Gordan table for $(0, 0, 1) \, \rtimes \, [(0, 0, \pi, 0), 0] \,   \rtimes \,   A_2\ \otimes \ (0, 0, 1) \, \rtimes \, [(0, 0, \pi, 0), \pi] \,   \rtimes \,   A_2 = (0, 0, 0) \, \rtimes \, [(0, 0, 0, 0), \pi] \,   \rtimes \,   T_0^{-}  \oplus\cdots$. The irreps in the rows and columns are labeled by the corresponding operators.} \label{tab:cgg5ip001}
\begin{tabular}{lccc}
\hline\hline
Tensor product row & $\mathcal{O}^{ \gamma_1 \, \otimes \, 1}\,(0, 0, 0)$ & $\mathcal{O}^{ \gamma_2 \, \otimes \, 1}\,(0, 0, 0)$ & $\mathcal{O}^{ \gamma_3 \, \otimes \, 1}\,(0, 0, 0)$ \\
\hline
$\mathcal{O}^{ \gamma_5 \, \otimes \, \gamma_5\gamma_2}\,(1, 0, 0) \ \otimes \ \mathcal{O}^{ \gamma_5 \, \otimes \, \gamma_5\gamma_2}\,(-1, 0, 0)$ &                                                               $\frac{1}{\sqrt{4}}$ &                                                                0 &                                                                0 \\
$\mathcal{O}^{ \gamma_5 \, \otimes \, \gamma_5\gamma_3}\,(1, 0, 0) \ \otimes \ \mathcal{O}^{ \gamma_5 \, \otimes \, \gamma_5\gamma_3}\,(-1, 0, 0)$ &                                                               $\frac{1}{\sqrt{4}}$ &                                                                0 &                                                                0 \\
$\mathcal{O}^{ \gamma_5 \, \otimes \, \gamma_5\gamma_2}\,(-1, 0, 0) \ \otimes \ \mathcal{O}^{ \gamma_5 \, \otimes \, \gamma_5\gamma_2}\,(1, 0, 0)$ &                                                              $-\frac{1}{\sqrt{4}}$ &                                                                0 &                                                                0 \\
$\mathcal{O}^{ \gamma_5 \, \otimes \, \gamma_5\gamma_3}\,(-1, 0, 0) \ \otimes \ \mathcal{O}^{ \gamma_5 \, \otimes \, \gamma_5\gamma_3}\,(1, 0, 0)$ &                                                              $-\frac{1}{\sqrt{4}}$ &                                                                0 &                                                                0 \\
$\mathcal{O}^{ \gamma_5 \, \otimes \, \gamma_5\gamma_3}\,(0, 1, 0) \ \otimes \ \mathcal{O}^{ \gamma_5 \, \otimes \, \gamma_5\gamma_3}\,(0, -1, 0)$ &                                                                0 &                                                               $\frac{1}{\sqrt{4}}$ &                                                                0 \\
$\mathcal{O}^{ \gamma_5 \, \otimes \, \gamma_5\gamma_1}\,(0, 1, 0) \ \otimes \ \mathcal{O}^{ \gamma_5 \, \otimes \, \gamma_5\gamma_1}\,(0, -1, 0)$ &                                                                0 &                                                               $\frac{1}{\sqrt{4}}$ &                                                                0 \\
$\mathcal{O}^{ \gamma_5 \, \otimes \, \gamma_5\gamma_3}\,(0, -1, 0) \ \otimes \ \mathcal{O}^{ \gamma_5 \, \otimes \, \gamma_5\gamma_3}\,(0, 1, 0)$ &                                                                0 &                                                              $-\frac{1}{\sqrt{4}}$ &                                                                0 \\
$\mathcal{O}^{ \gamma_5 \, \otimes \, \gamma_5\gamma_1}\,(0, -1, 0) \ \otimes \ \mathcal{O}^{ \gamma_5 \, \otimes \, \gamma_5\gamma_1}\,(0, 1, 0)$ &                                                                0 &                                                              $-\frac{1}{\sqrt{4}}$ &                                                                0 \\

$\mathcal{O}^{ \gamma_5 \, \otimes \, \gamma_5\gamma_1}\,(0, 0, 1) \ \otimes \ \mathcal{O}^{ \gamma_5 \, \otimes \, \gamma_5\gamma_1}\,(0, 0, -1)$ &                                                                0 &                                                                0 &                                                               $\frac{1}{\sqrt{4}}$ \\
$\mathcal{O}^{ \gamma_5 \, \otimes \, \gamma_5\gamma_2}\,(0, 0, 1) \ \otimes \ \mathcal{O}^{ \gamma_5 \, \otimes \, \gamma_5\gamma_2}\,(0, 0, -1)$ &                                                                0 &                                                                0 &                                                               $\frac{1}{\sqrt{4}}$ \\
$\mathcal{O}^{ \gamma_5 \, \otimes \, \gamma_5\gamma_1}\,(0, 0, -1) \ \otimes \ \mathcal{O}^{ \gamma_5 \, \otimes \, \gamma_5\gamma_1}\,(0, 0, 1)$ &                                                                0 &                                                                0 &                                                              $-\frac{1}{\sqrt{4}}$ \\
$\mathcal{O}^{\gamma_5 \, \otimes \, \gamma_5\gamma_2}\,(0, 0, -1) \ \otimes \ \mathcal{O}^{ \gamma_5 \, \otimes \, \gamma_5\gamma_2}\,(0, 0, 1)$ &                                                                0 &                                                                0 &                                                              $-\frac{1}{\sqrt{4}}$ \\
\hline\hline
\end{tabular}
\end{table}
When combining these results with \cref{eqn:TwoPiIsoState}, we obtain the following normalized two-pion operator which couples to the third component of the taste-singlet $\rho^0$,
{\interdisplaylinepenalty=10000
\begin{align}
     &\left[\mathcal{O}^{\otimes \, \gamma_5 \gamma_{i \neq 3}}_{\pi \pi}(\vec{0}, t )\right]^{ \gamma_3 \, \otimes \, \gamma_1,\,I=1, I_3=0}= \nn \\
      &\frac{1}{\sqrt{4}}\left[\pi^{+}_{\otimes \gamma_5 \gamma_1}\left((0,0,1),t\right) \pi^{-}_{\otimes \gamma_5 \gamma_1}\left((0,0,-1),t\right) + \pi^{+}_{\otimes \gamma_5 \gamma_2}\left((0,0,1),t\right) \pi^{-}_{\otimes \gamma_5 \gamma_2}\left((0,0,-1),t\right) \right.\nn\\
      &\left.-\pi^{-}_{ \otimes \gamma_5 \gamma_1}\left((0,0,-1),t\right) \pi^{+}_{ \otimes \gamma_5 \gamma_1}\left(0,0,1),t\right) -\pi^{-}_{ \otimes \gamma_5 \gamma_2}\left((0,0,-1),t\right) \pi^{+}_{ \otimes \gamma_5 \gamma_2}\left(0,0,1),t\right)\right], \label{eqn:Opg5i2pi001}
\end{align}
}
When computing the three-point correlation function of this operator and the $\rho^0$ operator that appear in the matrix of \cref{eqn:ctmat}, all four terms in \cref{eqn:Opg5i2pi001} give identical contributions to the correlation functions, which follows from the taste and rotation symmetry.
However, for the $C(t)_{\pi \pi \to \pi \pi}$ four-point correlation functions that also appear in \cref{eqn:ctmat}, there are cross terms which are not equivalent.
The Clebsch-Gordan coefficients for all the other cases (momentum and taste) are given in \cref{sec:stag2piStatesApp}.

\subsection{Correlation functions}\label{sec:corrFuncs}

\subsubsection{Two-points}\label{subsubsec:2pts}

With the operators in hand, the correlation functions in \cref{eqn:ctmat} can be constructed and the Wick contractions computed. The $\rho$ operator, \cref{eqn:I1VecOp}, at zero momentum is given by 
\begin{align}
 \rho^0_k(\vec{0}, t) = \frac{1}{2N^{3/2}_S}\sum_{\vec{n}} \bar u(n) \gamma_k \otimes 1\, u(n) - \bar d(n) \gamma_k \otimes 1\, d(n),   
\end{align}
The taste-singlet $\rho^0$ two-point correlation function, in the isospin-symmetric limit, is
\begin{align}
&C(t)_{\rho \to \rho} = \frac{1}{3}\sum_{k}\left\langle \rho^0_{k}(\vec{0}, t) \rho_{k}^{0 \dagger}(\vec{0}, 0)\right\rangle_{\textrm{conn.}} = \sum_n \left|\langle 0 | \rho^0 | n \rangle \right|^2  e^{-E_n t} \label{eqn:twoPointTimedep}\\
&=  2\cdot \frac{1}{4}\cdot  \frac{1}{4} \cdot  \frac{1}{N^3_S} \cdot \frac{1}{3}  \sum_{k}\  \vcenter{\hbox{\begin{tikzpicture}
  \begin{feynman}
    \vertex[empty dot, label=below:{($\vec{0},0)$ }] (l) at (0.0,0) {$\gamma_i\otimes 1$};
    \vertex[empty dot, label=below:{($\vec{0},t)$ }] (r) at (4,0) {$\gamma_i\otimes 1$};
    \diagram* {
      (l) -- [fermion, bend left=20, edge label=\(D^{-1}_{l}\)] (r) -- [fermion, bend left=20, edge label=\(D^{-1}_{l}\)] (l)};
  \end{feynman}
\end{tikzpicture}}} \\
&=\frac{1}{24 N^3_S} \sum_{k, \vec{n}_0, \vec{n}_1, \{\pm \delta_{j}\}} \varphi^{\gamma_i \otimes 1}(n) \textrm{tr}\left[ D_l^{-1}(\vec{n}_0 + \delta^{\gamma_i \otimes 1}, 0| \vec{n}_1 + \delta^{\gamma_i \otimes 1},t) D_l^{-1}( \vec{n}_1 + \delta^{\gamma_i \otimes 1},t| \vec{n}_0, 0) \right], \label{eqn:twoPoint}  
 \end{align}
where $D_l^{-1}$ are staggered light-quark propagators.
The formulas for obtaining $\varphi(n)$ and $\delta$ from the spin and taste structure are given in \cref{eqn:phiTrace,eqn:OpShift} with $\varphi^{\gamma_i \otimes 1}(n)$, $\delta^{\gamma_i \otimes 1}$ given explicitly in \cref{eqn:onelinkOperator}. The $\{\pm \delta_{j}\}$ in the sum is a symmetrization over all components of each $\delta$ that appear. We leave the gauge fields implicit, with the trace just over the color index. The individual multiplicative factors are left explicit in the second line to illustrate the different sources of normalization. The factor of two arises from taking the isospin symmetric limit. The first factor of $\frac{1}{4}$ is from the operator normalization in \cref{eqn:I1VecOp}, and the $1 /{N^3_S}$ is from the Fourier transformation of these operators to momentum space. The second factor of $\frac{1}{4}$ comes from the staggered rooting procedure (\cref{sec:rooting}). 

\subsubsection{Three-points}\label{subsubsec:3pts}

The two-pion operators, \cref{eqn:Opg52pi001,eqn:Opg5i2pi001} and all others considered here, are built out of hermitian sub-operators of the form,
\begin{align}
\mathcal{O}^{\otimes \gamma_\xi}_{\pi \pi}(\vec{0},t) &=  \pi_{\otimes \gamma_\xi}^{+}(\vec{p}, t)\pi_{\otimes \gamma_\xi}^{-}(-\vec{p}, t)- \pi_{\otimes \gamma_\xi}^{-}(\vec{p}, t)\pi_{\otimes \gamma_\xi}^{+}(-\vec{p}, t). \label{eqn:subOp}
\end{align}
Hence, all correlation functions containing two-pion operators can be broken up into a linear combination of sub-correlation functions each containing an operator of this form. In following discussions, for simplicity, we just use \cref{eqn:subOp} when computing the Wick contractions. 

The $C(t)_{\pi \pi \to \rho}$ three-point function, in the isospin symmetric limit, is then
\begin{align}
C(t)_{\pi \pi \to \rho} &=  \frac{1}{3}\sum_{k}\left\langle \rho^0_k(\vec{0}, t)  \mathcal{O}^{\otimes \gamma_\xi}_{\pi \pi}(\vec{0},0)\right\rangle = \sum_n \langle 0 | \rho^0 | n \rangle  \langle n |  \mathcal{O}^{\otimes \gamma_\xi}_{\pi \pi} | 0 \rangle  e^{-E_n t} \label{eqn:threePointTimedep}\\
&=   4\cdot  \frac{1}{2} \cdot \frac{1}{4}  \cdot \frac{1}{N_S^{9/2}}  \cdot  \frac{1}{3}\sum_{k}\  \vcenter{\hbox{\begin{tikzpicture}
  \begin{feynman}
    \vertex[empty dot, label=below:{($\vec{p},0$)}]  (l) at (0.0,0) {$\gamma_5\otimes \gamma_\xi$};
    \vertex[empty dot, label=below:{($-\vec{p},0$)}] (r) at (3,0) {$\gamma_5\otimes \gamma_\xi$};
    \vertex[empty dot, label=above:{($\vec{0},t$)}]  (t) at (1.5,2.25) {$\gamma_k\otimes 1$};
    \diagram* {
      (l) -- [fermion, edge label'=\(D^{-1}_{l}\)] (r) -- [fermion, edge label'=\(D^{-1}_{l}\)] (t) -- [fermion, edge label'=\(D^{-1}_{l}\)] (l)};
  \end{feynman}
\end{tikzpicture}}}\\
&=\frac{1}{6N_S^{9/2}}\sum_{k,\vec{n}_0,\vec{n}_1,\vec{n}_2, \{\pm \delta_j \}} \varphi^{\gamma_5 \otimes \gamma_\xi}(n_0) \varphi^{\gamma_5 \otimes \gamma_\xi}(n_1) \varphi^{\gamma_i \otimes 1}(n_2) e^{i a\vec{p} \cdot \left(\vec{n}_0-\vec{n}_1\right)} \nn \\
&\times \textrm{tr} \left[ D^{-1}_l(\vec{n}_0 + \delta^{\gamma_5 \otimes \gamma_\xi}, 0| \vec{n}_1  + \delta^{\gamma_5 \otimes \gamma_\xi}, 0) D^{-1}_l(\vec{n}_1 + \delta^{\gamma_5 \otimes \gamma_\xi}, 0| \vec{n}_2, t) D^{-1}_l(\vec{n}_2 + \delta^{\gamma_i \otimes 1}, t| \vec{n}_0  + \delta^{\gamma_5 \otimes \gamma_\xi}, 0) \right].\label{eqn:threePoint}
\end{align}
Disconnected Wick contributions cancel in the isospin symmetric limit. The factor of four in the second line comes from four connected Wick contractions, of which we only show one, which are all equivalent under isospin and parity. The $\frac{1}{2}$ is the normalization from the $\rho^0$ operator, and the $\frac{1}{4}$ is from the rooting procedure. The factor of $1/{N_S^{9/2}}$ arises from the Fourier transform of the three operators. We do not generate the $C(t)_{\rho \to \pi \pi} $ correlation functions, because they are significantly noisier with the random-wall source approach used here (see \cref{sec:numerical}), and it is equivalent to $C(t)_{\pi \pi \to \rho}$ under time-reversal symmetry.

\subsubsection{Four-points}\label{subsubsec:4pts}

The $\pi \pi \to \pi \pi$ four point function, in the isospin-symmetric limit, is 
\begin{align}
&C(t)_{\pi \pi \to \pi \pi} =\left\langle \mathcal{O}^{\otimes \gamma_{\xi_2}}_{\pi \pi}(\vec{0},t)  \mathcal{O}^{\otimes \gamma_{\xi_1}, \dagger}_{\pi \pi}(\vec{0},0)\right\rangle =  \sum_n  \langle 0 | \mathcal{O}^{\otimes \gamma_{\xi_2}}_{\pi \pi} | n\rangle  \langle n | \mathcal{O}^{\otimes \gamma_{\xi_1}}_{\pi \pi} | 0\rangle  e^{-E_n t} \label{eqn:fourPointTimedep} \\
&=  -4\cdot \frac{1}{4} \cdot \frac{1}{N_S^6} \times \vcenter{\hbox{\begin{tikzpicture}
  \begin{feynman}
    \vertex[empty dot, label=below:{($\vec{p},0)$ }] (bl) at (0.0,0) {$\gamma_5 \otimes \gamma_{\xi_1}$};
    \vertex[empty dot, label=below:{($-\vec{p},0)$ }] (br) at (2.5,0) {$\gamma_5 \otimes \gamma_{\xi_1}$};
    \vertex[empty dot, label=above:{($-\vec{p},t)$ }] (tl) at (0,2.5) {$\gamma_5 \otimes \gamma_{\xi_2}$};
    \vertex[empty dot, label=above:{($\vec{p},t)$ }] (tr) at (2.5,2.5) {$\gamma_5 \otimes \gamma_{\xi_2}$};
    \diagram* {
      (bl) -- [fermion] (br) -- [fermion] (tr) -- [fermion] (tl) -- [fermion] (bl)};
  \end{feynman}
\end{tikzpicture}}}
\quad \quad +4\cdot \frac{1}{4}  \cdot \frac{1}{N_S^6} \times \vcenter{\hbox{\begin{tikzpicture}
  \begin{feynman}
    \vertex[empty dot, label=below:{($\vec{p},0)$ }] (bl) at (0.0,0) {$\gamma_5 \otimes \gamma_{\xi_1}$};
    \vertex[empty dot, label=below:{($-\vec{p},0)$ }] (br) at (2.5,0) {$\gamma_5 \otimes \gamma_{\xi_1}$};
    \vertex[empty dot, label=above:{($-\vec{p},t)$ }] (tl) at (0,2.5) {$\gamma_5 \otimes \gamma_{\xi_2}$};
    \vertex[empty dot, label=above:{($\vec{p},t)$ }] (tr) at (2.5,2.5) {$\gamma_5 \otimes \gamma_{\xi_2}$};
    \diagram* {
      (bl) -- [fermion] (br) -- [fermion] (tl) -- [fermion] (tr) -- [fermion] (bl)};
  \end{feynman}
\end{tikzpicture}}} \nn \\
 & \ \ +2 \cdot \frac{1}{16} \cdot \frac{1}{N_S^6} \times \vcenter{\hbox{\begin{tikzpicture}
  \begin{feynman}
    \vertex[empty dot, label=below:{($\vec{p},0)$ }] (bl) at (0.0,0) {$\gamma_5 \otimes \gamma_{\xi_1}$};
    \vertex[empty dot, label=below:{($-\vec{p},0)$ }] (br) at (2.5,0) {$\gamma_5 \otimes \gamma_{\xi_1}$};
    \vertex[empty dot, label=above:{($-\vec{p},t)$ }] (tl) at (0,2.5) {$\gamma_5 \otimes \gamma_{\xi_2}$};
    \vertex[empty dot, label=above:{($\vec{p},t)$ }] (tr) at (2.5,2.5) {$\gamma_5 \otimes \gamma_{\xi_2}$};
    \diagram* {
      (bl) -- [fermion, bend left=20] (tl) -- [fermion, bend left=20] (bl),
      (br) -- [fermion, bend left=20] (tr) -- [fermion, bend left=20] (br)};
  \end{feynman}
\end{tikzpicture}}}
\quad \quad  -2 \cdot \frac{1}{16} \cdot \frac{1}{N_S^6} \times \vcenter{\hbox{\begin{tikzpicture}
  \begin{feynman}
    \vertex[empty dot, label=below:{($\vec{p},0)$ }] (bl) at (0.0,0) {$\gamma_5 \otimes \gamma_{\xi_1}$};
    \vertex[empty dot, label=below:{($-\vec{p},0)$ }] (br) at (2.5,0) {$\gamma_5 \otimes \gamma_{\xi_1}$};
    \vertex[empty dot, label=above:{($-\vec{p},t)$ }] (tl) at (0,2.5) {$\gamma_5 \otimes \gamma_{\xi_2}$};
    \vertex[empty dot, label=above:{($\vec{p},t)$ }] (tr) at (2.5,2.5) {$\gamma_5 \otimes \gamma_{\xi_2}$};
    \diagram* {
      (bl) -- [fermion, bend left=20] (tr) -- [fermion, bend left=20] (bl),
      (br) -- [fermion, bend left=20] (tl) -- [fermion, bend left=20] (br)};
  \end{feynman}
\end{tikzpicture}}}, \nn \\
&= \frac{1}{N_S^6}\sum_{\vec{n}_0,\vec{n}_1,\vec{n}_2, \vec{n}_3, \{\pm \delta_j \}} \varphi^{\gamma_5 \otimes \gamma_{\xi_1}}(n_0) \varphi^{\gamma_5 \otimes \gamma_{\xi_1}}(n_1) \varphi^{\gamma_5 \otimes \gamma_{\xi_2}}(n_2) \varphi^{\gamma_5 \otimes \gamma_{\xi_2}}(n_3) e^{i a\vec{p} \cdot \left(\vec{n}_0-\vec{n}_1+\vec{n}_2-\vec{n}_3\right)} \Big[ \nn\\
&\quad \quad \quad \quad \quad \quad \quad \quad -\textrm{tr} \left[D^{-1}_l(\vec{n}_0 + \delta^{\gamma_5 \otimes \gamma_{\xi_1}}, 0| \vec{n}_1  + \delta^{\gamma_5 \otimes \gamma_{\xi_1}}, 0) D^{-1}_l(\vec{n}_1 + \delta^{\gamma_5 \otimes \gamma_{\xi_1}}, 0| \vec{n}_2, t) \right.\nn \\
& \quad \quad \quad \quad \quad \quad \quad \quad \quad \quad \left. \times D^{-1}_l(\vec{n}_2 + \delta^{\gamma_5 \otimes \gamma_{\xi_2}}, t| \vec{n}_3, t) D^{-1}_l(\vec{n}_3 + \delta^{\gamma_5 \otimes \gamma_{\xi_2}}, t| \vec{n}_0  + \delta^{\gamma_5 \otimes \gamma_{\xi_1}}, 0)\right] \nn\\
&\quad \quad \quad \quad \quad \quad \quad \quad +\textrm{tr} \left[D^{-1}_l(\vec{n}_0 + \delta^{\gamma_5 \otimes \gamma_{\xi_1}}, 0| \vec{n}_1  + \delta^{\gamma_5 \otimes \gamma_{\xi_1}}, 0) D^{-1}_l(\vec{n}_1 + \delta^{\gamma_5 \otimes \gamma_{\xi_1}}, 0| \vec{n}_3, t) \right.\nn \\
& \quad \quad \quad \quad \quad \quad \quad \quad \quad \quad \left. \times D^{-1}_l(\vec{n}_3 + \delta^{\gamma_5 \otimes \gamma_{\xi_2}}, t| \vec{n}_2, t) D^{-1}_l(\vec{n}_2 + \delta^{\gamma_5 \otimes \gamma_{\xi_2}}, t| \vec{n}_0 + \delta^{\gamma_5 \otimes \gamma_{\xi_1}}, 0)\right] \nn\\
&\quad \quad \quad \quad \quad \quad \quad \quad +\frac{1}{8}\textrm{tr} \left[D^{-1}_l(\vec{n}_0 + \delta^{\gamma_5 \otimes \gamma_{\xi_1}}, 0| \vec{n}_2, t) D^{-1}_l(\vec{n}_2 + \delta^{\gamma_5 \otimes \gamma_{\xi_2}}, t| \vec{n}_0 + \delta^{\gamma_5 \otimes \gamma_{\xi_1}}, 0) \right]\nn \\
& \quad \quad \quad \quad \quad \quad \quad \quad \quad \quad \times \textrm{tr} \left[  D^{-1}_l(\vec{n}_1 + \delta^{\gamma_5 \otimes \gamma_{\xi_1}}, 0| \vec{n}_3, t) D^{-1}_l(\vec{n}_3 + \delta^{\gamma_5 \otimes \gamma_{\xi_2}}, t| \vec{n}_1 + \delta^{\gamma_5 \otimes \gamma_{\xi_1}}, 0)\right] \nn\\
&\quad \quad \quad \quad \quad \quad \quad \quad -\frac{1}{8}\textrm{tr} \left[D^{-1}_l(\vec{n}_0 + \delta^{\gamma_5 \otimes \gamma_{\xi_1}}, 0| \vec{n}_3, t) D^{-1}_l(\vec{n}_3 + \delta^{\gamma_5 \otimes \gamma_{\xi_2}}, t| \vec{n}_0 + \delta^{\gamma_5 \otimes \gamma_{\xi_1}}, 0) \right]\nn \\
& \quad \quad \quad \quad \quad \quad \quad \quad \quad \quad \times \textrm{tr} \left[  D^{-1}_l(\vec{n}_1 + \delta^{\gamma_5 \otimes \gamma_{\xi_1}}, 0| \vec{n}_2, t) D^{-1}_l(\vec{n}_2 + \delta^{\gamma_5 \otimes \gamma_{\xi_2}}, t| \vec{n}_1 + \delta^{\gamma_5 \otimes \gamma_{\xi_1}}, 0)\right] \Big]. \label{eqn:fourPoints}
\end{align}
The individual factors of $4,4,2,2$ in front of the respective diagrams are independent-diagram multiplicities. The $\frac{1}{4}, \frac{1}{4}, \frac{1}{16}, \frac{1}{16}$ are rooting factors ($\frac{1}{4}$ for each trace). 

The numerical simulation presented here actually employs time-split two-pion operators instead of \cref{eqn:fourPointTimedep} to address a potential Fierz-rearrangement issue discussed in Ref.~\cite{Fukugita:1994ve}. This modification and the additional considerations it introduces are described in \cref{sec:timeSplit}. It turns out that the Fierz-rearrangement problem does not arise with the random-wall sources used in this work, hence our ongoing studies employ \cref{eqn:fourPointTimedep}.

\subsubsection{Effective energies and amplitudes}

We make use of the following formula for extracting the effective energy and amplitudes from the correlation functions used in this work. The effective energy is obtained from
\begin{align}
    aE_{0, \textrm{eff}}(t) &= \frac{1}{2}\operatorname{arccosh} \left[\frac{C(t+2) + C(t-2)}{2C(t)}\right], \label{eqn:effectiveMass}
\end{align}
where the averaging is performed over time separations of $(t\pm 2)$ to remove staggered oscillatory effects. The effective amplitude is then given by the following,
\begin{align}
    Z^2_{0, \textrm{eff}}(t) = e^{N_t E_{0, \textrm{eff}} /2}\frac{C(t)}{\cosh(E_{0, \textrm{eff}}(N_t/2-t))}, \label{eqn:effectiveAmp}
\end{align}
where the parameter $E_{0, \textrm{eff}}$ is obtained from \cref{eqn:effectiveMass}, once the function has plateaued.

\subsection{Choosing the operator basis}\label{sec:OpBasis}

\begin{figure}
  \centering
    \includegraphics[width=0.75\textwidth]{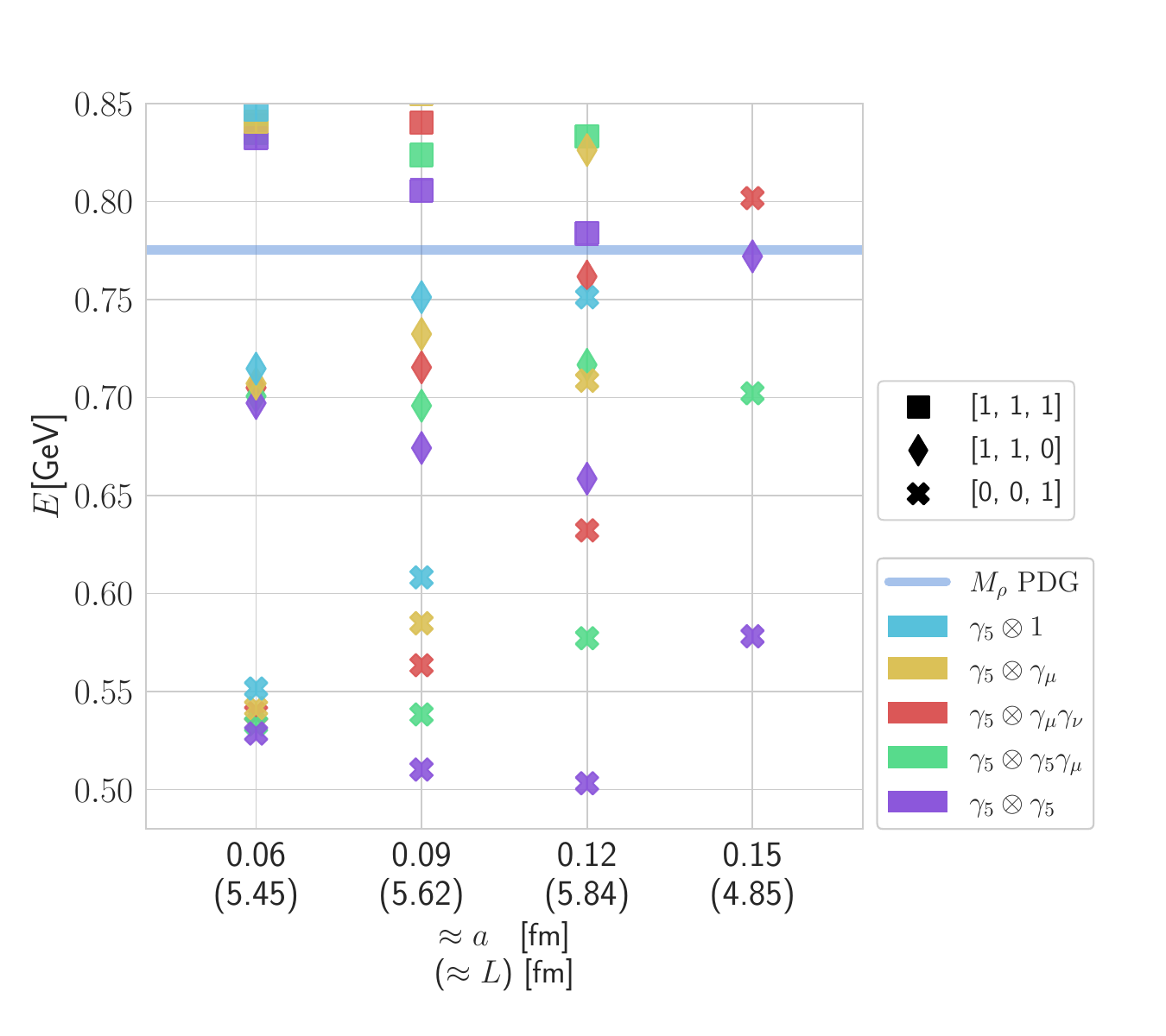}
    \caption{Continuum, non-interacting energy spectrum for $\pi_\xi\pi_\xi\to\rho$ case on four physical-mass HISQ ensembles for the relevant tastes, $\xi$. The ensembles' parameters are given in Ref.~\cite{Bazavov:2023has}. The tastes of the single pions in the two-pion state are indicated by the color. The back-to-back momentum is indicated by the symbol shape. Also given is the $\rho^0$ Particle Data Group mass (blue band) \cite{ParticleDataGroup:2022pth} which is used as the cut-off energy to select the operator basis.}
    \label{fig:naiveSpectrum}
\end{figure}

For the two-pion operators in the matrix, \cref{eqn:ctmat}, we choose a range of pion momenta and tastes corresponding to two-pion energies up to the 
mass of the $\rho^0$ meson \cite{ParticleDataGroup:2024cfk}. 
For this purpose, we construct the non-interacting two-pion energies using 
\begin{align}
    E_{\textrm{free}}=2 \sqrt{\vec{p}^2+M^2_{\xi}},
\end{align} 
where $M_\xi$ are the measured ground-state masses of pion correlation functions obtained from   \cref{eqn:pion1,eqn:pion50,eqn:pion0,eqn:pion5,eqn:pioni,eqn:pion5i,eqn:pioni0,eqn:pionij} and $p_i=2\pi \ell_i/L$, $\ell_i=0,1 \ldots$. This spectrum is shown in \cref{fig:naiveSpectrum} for the four physical mass HISQ ensembles currently used in related $g-2$ work from the Fermilab Lattice, HPQCD and MILC Collaborations~\cite{Bazavov:2023has}. \Cref{fig:naiveSpectrum} does not account for interactions or the taste-orbit splittings described in \cref{sec:nonzeromom}, but suffices for deciding which $\pi\pi$ operators to use.

\begin{table}
\caption{Operator basis on the $a\approx 0.15$~fm ensemble. The single pion operators in the two-pion states have equal taste and equal-but-opposite momentum. We indicate the irrep splitting by listing the operators separately.}
\label{table:opBasis}
\centering
\begin{tabular}{ll}
\hline\hline
Operator & Momenta (back-to-back) \\
\hline
$\rho^0, \, \tilde \rho^0$ & \\
$\mathcal{O}_{\pi \pi}^{\otimes \gamma_{5}}$ & ${(0,0,1),(1,1,0)}$ \\
$\mathcal{O}_{\pi \pi}^{\otimes \gamma_{5} \gamma_{ 1 / 2}}$, $\mathcal{O}_{\pi \pi}^{\otimes \gamma_{5} \gamma_{ 3}}$ & ${(0,0,1)}$ \\
$\mathcal{O}_{\pi \pi}^{\otimes \gamma_{1/2} \gamma_{ 0}}$, $\mathcal{O}_{\pi \pi}^{\otimes \gamma_{3} \gamma_{ 0}}$ & ${(0,0,1)}$ \\
\hline\hline
\end{tabular}
\end{table}

At $0.15$ fm, which is the focus of this work, we see there are four states below or near the threshold. We include the taste-tensor (red cross) even though it is above the $\rho^0$ mass. We select an eight-operator basis, shown in \cref{table:opBasis}. From the irreps listed in \cref{eqn:pion1,eqn:pion50,eqn:pion0,eqn:pion5,eqn:pioni,eqn:pion5i,eqn:pioni0,eqn:pionij}, we leave out operators for the taste pseudo-vector with a temporal taste component, \cref{eqn:pion50}, and the taste tensors without a temporal taste component \cref{eqn:pionij}. This choice is based on the fact these operators, in the form of $\mathcal{O}^{\gamma_5 \otimes \, \xi}$, have links in the time-direction. Averaging over the forward and backward time-links removes the oscillating contribution, but results in non-local time dependence in the corresponding correlation functions. The time-link can be removed by modifying the operators as $\mathcal{O}^{\gamma_5  \otimes \xi} \to \mathcal{O}^{\gamma_5 \gamma_0 \otimes \xi}$, which preserves the original quantum numbers (see \cref{eqn:pionTimeLinkTreat}). We generate additional correlation functions with these modified operators to check that the variational basis in \cref{table:opBasis} is complete, \textit{i.e.}, to check that including them does not resolve any additional states beneath the threshold. This is expected based on the degeneracy's at zero-momentum, and confirmed in our analysis. These $\mathcal{O}^{\gamma_5 \gamma_0 \otimes \xi}$ operators, however, are found to have significant overlap with excited states, resulting in very noisy correlation functions; hence, they are not included in the main analysis presented here.

\subsection{Numerical setup}\label{sec:numerical}

\begin{table*}
\centering
\caption{Ensemble parameters used in this work; from Ref.~\cite{Davies:2019efs}. Shown are the approximate lattice spacing in~fm, the spatial length $L$ of the lattice in~fm, the size of the lattice, the sea-quark masses in lattice-spacing units, the gradient-flow scale $w_0/a$~\cite{Borsanyi:2012zs,Davies:2019efs}, the taste-Goldstone pion mass \cite{MILC:2012znn}, the number of configurations analyzed, the number of loose-residual solves per configuration used in the truncated solver method \cite{TSM1, TSM2}, and the time-slice range computed for the four point functions. We take $w_0=0.1715(9)$~fm~\cite{Dowdall:2013rya}.}
\label{table:ens2pi}
\begin{tabular}{lllcccccc}
\hline \hline
$\approx a$ (fm) & $L$ (fm) & $N_s^3 \times N_t$ & $a m_{l}^\text{sea} / a m_{s}^\text{sea} / a m_{c}^\text{sea}$ & $w_{0} / a$ & $M_{\pi_{5}}$ (MeV) & $N_{\text{conf}}$ & $N_{\text{loose}}$ & $t_{\text{sep}}$ \\
\hline
$0.15$ & $4.85$ & $ 32^3 \times 48$ & 0.002426/0.0673/0.8447 & $1.13215(35)$ & 134.73(71) & 3473 & 16 & [3,17] \\ 
\hline \hline
\end{tabular}
\end{table*}

The 0.15 fm HISQ ensemble parameters are given in \cref{table:ens2pi}. We use the same numerical strategy as described in Ref.~\cite{Bazavov:2023has} for the two-point correlation functions in \cref{eqn:twoPoint,eqn:fourPoints}. For the three-point, \cref{eqn:threePoint}, and single-trace four-point contractions of \cref{eqn:fourPoints}, we employ sequential sources \cite{FermilabLattice:2016ipl}. For these four-point contractions, this approach requires an additional solve for each time separation in the correlation function. To reduce computational expense, we generate a subset of the total possible time separations, which are shown in the last column of \cref{table:ens2pi}. The total number of configurations for which we compute the correlation matrix \cref{eqn:ctmat} is given in the third-to-last column of \cref{table:ens2pi}. We calculate two-point correlators \cref{eqn:twoPoint} on $\approx 6000$ additional configurations, as the reconstructed tail accounts for only about $\approx20\%$ of the total value of the integrand, \cref{eq:amuTint}, with the rest coming from the two-point data. We renormalize the vector operator using the results from Ref.~\cite{Hatton:2019gha}. Uncertainties are propagated through the analysis using the \textsc{gvar} package \cite{gvarGitHub}. We find that the \textsc{gvar} uncertainties are in excellent agreement with jackknife resampling, while being computationally faster.

\subsection{Finite time effects}\label{sec:finiteT}

A complication which must be addressed with the matrix \cref{eqn:ctmat} is the non-negligible wrap-around contribution that arises in the diagonal $C(t)_{\pi \pi \to \pi \pi}$ correlation functions due to the finite temporal size of the lattice employed. In general, the spectral decomposition of $C(t)=\langle\mathcal{O}(t)\mathcal{O}^\dagger(0)\rangle$ is
\begin{align}
    C(t) &= \frac{1}{Z} \sum_{mn} \langle m|\hat{\mathcal{O}}|n\rangle \langle n|\hat{\mathcal{O}}^\dagger|m\rangle  e^{-E_nt-E_m(T-t)} , \\
    Z    &= \sum_n e^{-E_nT} ,
\end{align}
with the states ordered by increasing energy $E_0<E_1\le E_2\le\cdots$.

Inserting a general $I=1, I_Z=0$ hermitian two-pion operator, $\mathcal{O}_{\pi \pi} = \pi^+(\vec{p}) \pi^-(-\vec{p}) -  \pi^+(-\vec{p}) \pi^-(\vec{p})$, gives
\begin{align}
    =\frac{\sum_{n, m} \left|\langle n | \pi^+(\vec{p}) \pi^-(-\vec{p}) -  \pi^+(-\vec{p}) \pi^-(\vec{p}) | m \rangle\right|^2  e^{- E_n(T-t)} e^{- E_m t} }{\sum_{n} e^{-T E_n}}. \label{eq:spectraldecomp}
\end{align}
In order to understand the spectrum excited by this operator we need the following,
\begin{align}
   \pi^+(\vec{p}) &=  b_{-\vec{p}} + a^\dagger_{\vec{p}} \\
   \pi^-(\vec{p}) &=  a_{-\vec{p}} + b^\dagger_{\vec{p}} 
\end{align}
where $a^\dagger_{\vec{p}}$ and $a_{\vec{p}}$ are creation and annihilation operators of the $\pi^+$ states:
\begin{align}
    a_{\vec{p}}^\dagger | 0 \rangle &= | \pi^+(\vec{p}) \rangle \\
    a_{\vec{p}} |\pi^+(\vec{p}) \rangle   &= | 0 \rangle   
\end{align}
and likewise with $b^\dagger_{\vec{p}}$ and $b_{\vec{p}}$ for $\pi^-(\vec{p})$. The amplitude in \cref{eq:spectraldecomp}, after normal-ordering, becomes
\begin{align}
     &= \langle n | b_{-\vec{p}} a_{\vec{p}} | m \rangle + \langle n |  b^\dagger_{-\vec{p}} b_{-\vec{p}}| m \rangle + \langle n |  a^\dagger_{\vec{p}}  a_{\vec{p}}| m \rangle + \langle n |a^\dagger_{\vec{p}} b^\dagger_{-\vec{p}} | m \rangle \nn\\
     &- \langle n | b_{\vec{p}} a_{-\vec{p}} | m \rangle - \langle n | b^\dagger_{\vec{p}}  b_{\vec{p}}| m \rangle - \langle n | a^\dagger_{-\vec{p}}a_{-\vec{p}}   | m \rangle  - \langle n |a^\dagger_{-\vec{p}} b^\dagger_{\vec{p}} | m \rangle \label{eqn:ampOpBreakdown}
\end{align}
where the commutator term that appears in the first line cancels with the commutator from the second line. For vacuum initial and final states, $|n\rangle , | m\rangle =|0\rangle$, all the terms are zero. There cannot be an odd combined-total number of pions in $|n\rangle$ and $|m\rangle$ as the operator has two pions, so the first non-zero contribution to the correlation function appears for $|n\rangle , |m\rangle = |\pi^\pm(\pm\vec{p}) \rangle$. There are 16 possibilities but only the four cases where $|n\rangle=|m\rangle$ are non-zero. In particular, the second, third, sixth and seventh terms of \cref{eqn:ampOpBreakdown} are only each non-zero once for each of the four choices, giving
\begin{align}
    \Big[ &\langle \pi^+(\vec{p}) |  \mathcal{O}_{\pi \pi}(\vec{0}) | \pi^+(\vec{p}) \rangle^2 + \langle \pi^+(-\vec{p}) |  \mathcal{O}_{\pi \pi}(\vec{0}) | \pi^+(-\vec{p}) \rangle^2\\
    &+ \langle \pi^-(\vec{p}) |  \mathcal{O}_{\pi \pi}(\vec{0}) | \pi^-(\vec{p}) \rangle^2 + \langle \pi^-(-\vec{p}) |  \mathcal{O}_{\pi \pi}(\vec{0}) | \pi^-(-\vec{p}) \rangle^2\Big]  e^{- E_{\pi_{\vec{p}}} T}  \\
    &=4\langle \pi^+(\vec{p}) |  \mathcal{O}_{\pi \pi}(\vec{0}) | \pi^+(\vec{p}) \rangle^2 e^{- E_{\pi_{\vec{p}}} T} \label{eq:FTn1m1}
\end{align}
where the symmetry of the operator (squared) under $+ \to -$ and $\vec{p} \to -\vec{p}$ was used. With the weakly-interacting approximation~\cite{Dudek:2012gj},
\begin{align}
\langle\pi^\pm|\hat{\mathcal{O}}_{\pi^+\pi^-}|\pi^\pm\rangle \approx \left|\langle 0 |\pi^{\pm}\big| \pi^{\pm} \rangle\right|^2,
\end{align}
which is valid for diagonal correlation functions where the first diagram in \cref{eqn:fourPoints} dominates, the constant term $4 \left|\big\langle 0 \big|\pi^{\pm}\big| \pi^{\pm} \big\rangle\right|^4 e^{-E_{\pi^{\pm}} T}$ is then the leading wrap-around contribution. 

The next non-zero contributions are from from $|n\rangle=|\pi^+(\pm\vec{p})\pi^-(\mp\vec{p}) \rangle$, $|m\rangle=|0 \rangle$
and $|m\rangle=|\pi^+(\pm\vec{p})\pi^-(\mp\vec{p}) \rangle$, $|n\rangle=|0 \rangle$.
Given the forms of the 1st and 5th (4th and 8th) terms in \cref{eqn:ampOpBreakdown}, there are two incoming (outgoing) states,
\begin{align}
&2\langle 0 |  \mathcal{O}_{\pi \pi}(\vec{0}) | \pi^+(\vec{p})\pi^-(-\vec{p}) \rangle^2 e^{- E_{\pi_{\vec{p}} \pi_{\vec{p}}} t}  + 2\langle \pi^+(\vec{p})\pi^-(-\vec{p}) |  \mathcal{O}_{\pi \pi}(\vec{0}) | 0 \rangle^2 e^{- E_{\pi_{\vec{p}} \pi_{\vec{p}}}(T-t)} \nn\\
&=2\langle 0 |  \mathcal{O}_{\pi \pi}(\vec{0}) | \pi^+(\vec{p})\pi^-(-\vec{p}) \rangle^2 \left(e^{-E_{\pi_{\vec{p}} \pi_{\vec{p}}} t} + e^{- E_{\pi_{\vec{p}} \pi_{\vec{p}}}(T-t)} \right),
\end{align}
where the symmetries of the operator were again employed to obtain the first line. This is the standard cosh form of the spectral representation of a correlation function on a periodic lattice. This is taken into account through \cref{eqn:effectiveMass,eqn:effectiveAmp} and in the fit ansatz of \cref{sec:fits}. 

Going to higher energies, we find repetition of these two types of finite-time contributions but with both additional non-interacting states, which are removed by the denominator of \cref{eq:spectraldecomp}, and interacting states which are suppressed by factors of $e^{-E_{\pi_{\vec{p}}} T} \approx \frac{1}{30}$.

While the leading $t$-independent term of \cref{eq:FTn1m1} is formally small, it is not small in practice in the region of interest, where $t$ is a bit shorter than $T/2$. This contribution is especially relevant for this calculation as $T$ on the $0.15$ fm physical mass HISQ ensemble is $\approx 0.8$ fm smaller than on the other HISQ ensembles in \cref{fig:naiveSpectrum}. 

\begin{figure}
  \centering
    \includegraphics[width=0.6\textwidth]{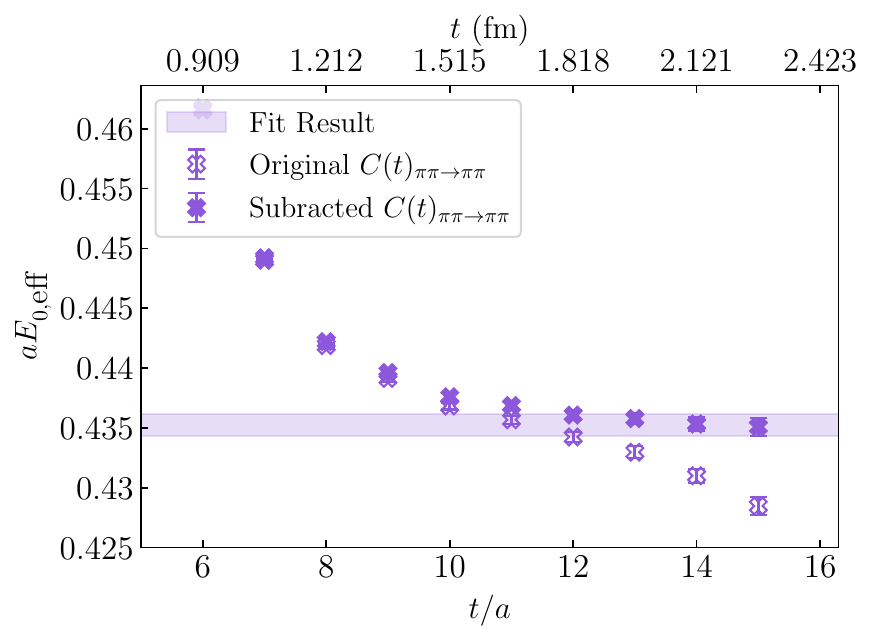}
    \caption{Effective mass of the correlation function $C(t)_{\pi \pi \to \pi \pi}:\langle \mathcal{O}^{\otimes \, \gamma_5}_{\pi \pi}(\vec{0}, t ) \mathcal{O}^{\otimes \, \gamma_5}_{\pi \pi}(\vec{0}, 0) \rangle $ uncorrected (empty) and corrected (filled) purple crosses for the wrap around contribution using the method described in this section. Also shown is the result of a fit to corrected correlation function.}
    \label{fig:wrapAround}
\end{figure}

We explicitly subtract this term from the diagonal correlators in \cref{eqn:ctmat} after obtaining $\langle 0 \big|\pi^{\pm}\big| \pi^{\pm} \big\rangle$ and $E_{\pi^{\pm}}$ from a fit to the single-pion two-point correlation functions (third diagram in \cref{eqn:fourPoints}). Shown in \cref{fig:wrapAround} is the result of applying this procedure to the ground state correlation function $\langle \mathcal{O}^{\otimes \, \gamma_5}_{\pi \pi}(\vec{0}, t ) \mathcal{O}^{\otimes \, \gamma_5}_{\pi \pi}(\vec{0}, 0) \rangle$,
where we plot the effective energy, \cref{eqn:effectiveMass}, for the original and subtracted correlation functions.

In the limit $t \to \infty \sim T/2$, the effective energy, $aE_{0, \textrm{eff}}(t)$, should plateau to the ground state energy if there is no constant term. From the plot, one sees this is indeed the case for the subtracted version.  Moreover, the effective energy of the subtracted correlation function now agrees with the fit result (purple band), while the unsubtracted effective energy shows clear contamination from the wrap-around contribution. All following results use the subtracted version of \cref{eqn:ctmat} in which the diagonal two-pion correlators are replaced with the  versions that have the leading wrap-around contribution subtracted.

In \cref{sec:matrixFit} we determine the sub-leading finite-time contributions not accounted for here. Namely, the finite-time effects in the diagonal $C_{\rho \to \rho}(t)$ two-points and the off-diagonal correlators of \cref{eqn:ctmat}, as well as the sub-leading constant-in-time contributions not included \cref{eq:FTn1m1}. We find that, at this level of precision, our results are insensitive to these effects.

\subsection{The GEVP and optimized operators}\label{sec:gevp}

\begin{figure}
  \centering
    \includegraphics[width=0.8\textwidth]{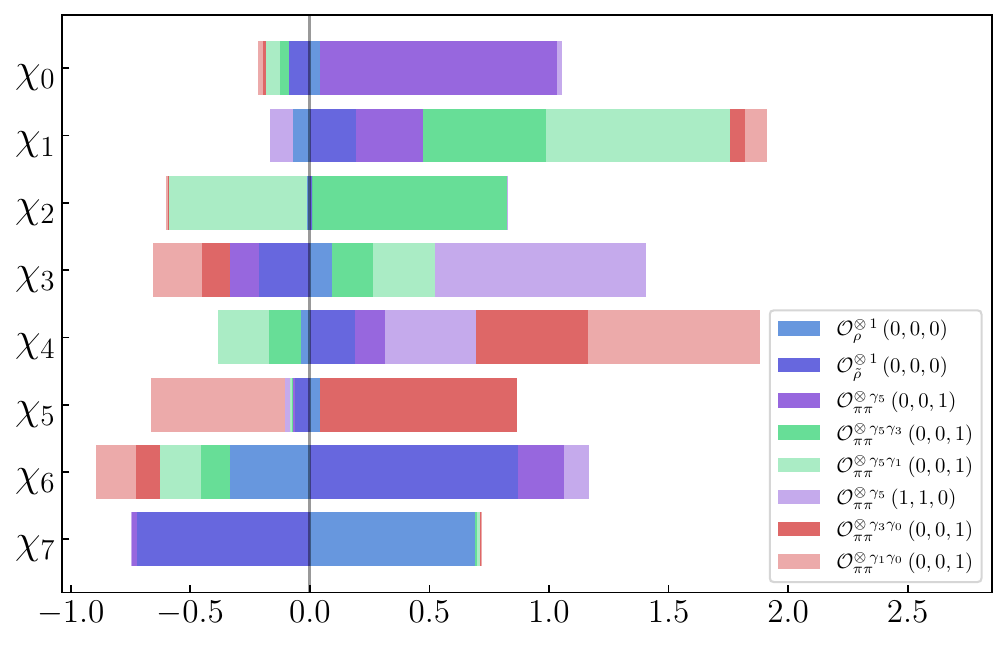}
    \caption{Qualitative picture of the composition of the optimized operators $\chi_n$ as constructed from the original operator basis in \cref{table:opBasis} for a value of $t_0/a=5$. The picture is qualitative, as the operators all must be normalized to some common value. This is achieved by setting their corresponding diagonal correlation functions equal at a reference time $t$. Here $t/a=10$ is chosen, the picture varies slightly depending on the choice of reference time. The horizontal line divides the negative and positive contributions to the optimized operators\footnote{The matrix $C(t)$ is symmetric and real, hence the eigenvectors are real.}.}
    \label{fig:optimOps}
\end{figure}
 
To obtain the energies and amplitudes from the matrix, \cref{eqn:ctmat}, eigenvectors $v_n(t,t_0)$ are first extracted through a generalized eigenvalue problem (GEVP) \cite{Blossier:2009kd},
\begin{align}
\mathbf{C}(t) v=\lambda \mathbf{C}(t_{0}) v . \label{eqn:gevp}
\end{align}
Here, the reference time $t_0$ is a free parameter, which we vary later in the analysis to check for stability. A smaller value of $t_0$ yields eigenvectors and eigenvalues with better statistical precision, albeit with potentially larger excited state contamination. The resulting $v_n(t)$ are functions of Euclidean time. Their asymptotic values, $v_n$, at large enough $t=t'$ are the coefficients of the `optimized operators' \cite{bruno2019exclusive}. We find that at $t^\prime/a = 10$ all the $v_n(t)$ appear to have plateaued to constants. The optimized operators with maximal overlap with the states $|n \rangle$ are, then: 
\begin{align}
\chi_{n}(t)=\sum_i\left(v_{n}\right)_{i} \mathcal{O}_{i}(t) .
\end{align}
\Cref{fig:optimOps} provides a visual display of the components of $v_n$ for the full operator basis in \cref{table:opBasis}. In the plot, the relative contributions of the original operators to the $\chi_n$ are shown. For this purpose, the original operators are first normalized, so their diagonal correlators are equal at time $t/a = 10$. One observes that the ground state optimized operator, $\chi_0$, is predominantly made up of $\mathcal{O}_{\pi \pi}^{\otimes \gamma_{5}}$ with $(0,0,1)$ back-to-back momenta as expected. The first and second excited states are primarily built out of the taste-pseudo vector, one-link operators, where the first excited state is an additive combination while the second is subtractive. The third operator is primarily the $\mathcal{O}_{\pi \pi}^{\otimes \gamma_{5}}$ operator with $[1,1,0]$ momentum, but with significant mixing from the other taste operators. The fourth and fifth are analogous to the first and second but for the taste-tensor, two-link operators. The sixth is primarily the smeared $\rho^0$ operator and the last is essentially a ``junk" operator with the normal and smeared  $\rho^0$ operators almost cancelling out.

The lowest energies $E_{n}$ and overlap amplitudes $\langle 0 | \rho^0 | n\rangle$, that appear in \cref{eq:corrFuncSpecRep}, are obtained from the following correlation functions constructed from the optimized operators
\begin{align}
v_n^\dagger \mathbf{C}(t) v_n =  \left\langle\chi_{n}(t) \chi_{n}^{\dagger}(0)\right\rangle&=\sum_{n}\left[Z_{n}^{2} e^{-E_{n} t} + (-1)^t Z_{n, \textrm{osc}}^{2} e^{-E_{n, \textrm{osc}} t}\right],\label{eqn:gevpCorrsDiag}\\
\left(\mathbf{C}(t) v_n\right)_0 =  \left\langle\chi_{n}(t)\, \rho^{0 \dagger}(0)\right\rangle&=\sum_{n}\left[ Z_{n}\left\langle 0\left|\rho^{0}\right| n\right\rangle e^{-E_{n} t} + (-1)^t  Z_{n, \textrm{osc}}\left\langle 0\left|\rho^{0}\right| n, \textrm{osc}\right\rangle e^{-E_{n, \textrm{osc}} t}\right]  . \label{eqn:gevpCorrsOff}
\end{align}
The $t \to T-t$ terms, from periodic boundary conditions, in the spectral representation are implicit. In the following sections, we will also consider variations of the original operator basis which do not contain the $\rho^{0}$, in this case we simply pad the $v_n$ with a zero as the first element so that these formulas still hold.

\subsubsection{Extracting the energies and amplitudes}\label{sec:fits}

\begin{figure}
  \centering
    \includegraphics[width=1\textwidth]{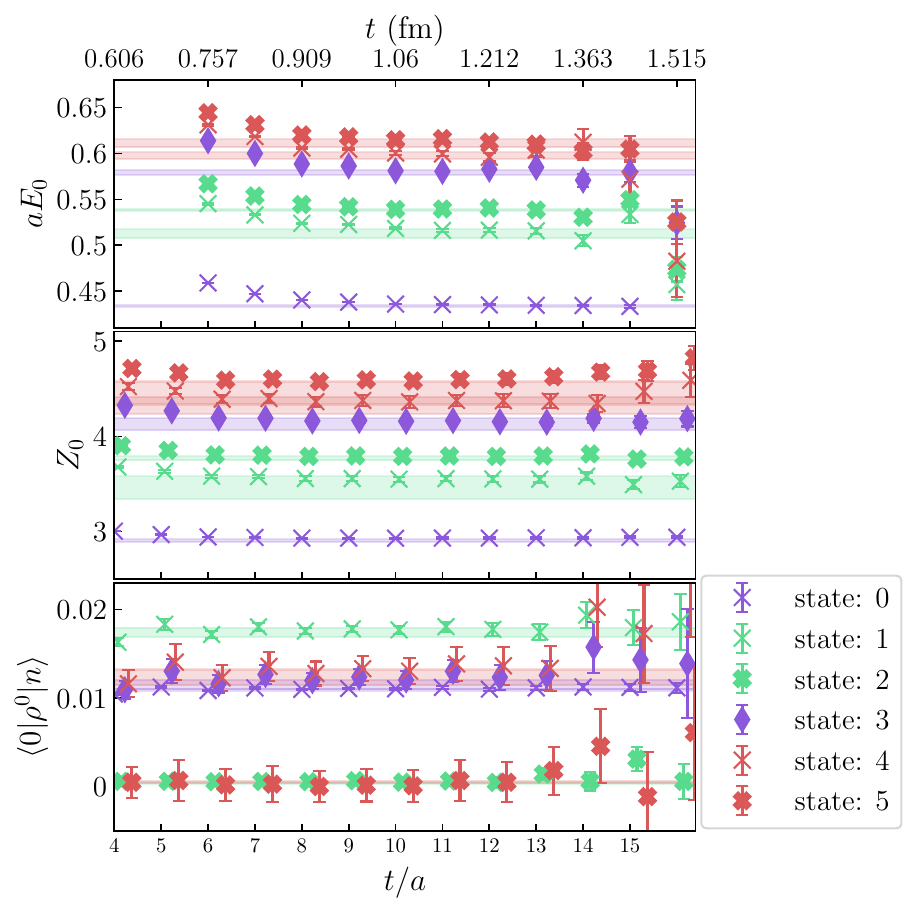}
    \caption[Spectrum and amplitudes from the generalized eigenvalue problem]{Energies (top), optimized operator overlap amplitudes (middle) and $\rho^0$ operator overlap amplitudes (bottom) for the two-pion states, extracted from a generalized eigenvalue analysis on the $\approx 0.15$~fm ensemble with $t_0/a=5$. Bands are results from fits, data points are effective masses and amplitudes. The sixth state is left out due to the significant oscillations and overlapping error bands, rendering the plot unclear.}
    \label{fig:gevp}
\end{figure}

\begin{table}
\centering
\caption{Fit parameters used to extract energies and amplitudes from  \cref{eqn:gevpCorrsDiag,eqn:gevpCorrsOff}. The D (diagonal) and OD (off-diagonal) labels correspond to the first and second equations, respectively. We display the fit quality through the $\chi^2/\textrm{DoF}$ value, which does not include the contribution from priors. We also display the BAIC weight \cite{Neil:2023pgt}, which we use to select these preferred fit parameters over other variations.}
\label{table:fitParams}
\begin{tabular}{lccccccc}
\hline\hline
state & $t_{\text{min, D}}/a$ \ & $t_{\text{max, D}}/a$ \ & $t_{\text{min, OD}}/a$  \ & $t_{\text{max, OD}}/a$ \ & ($N_{\text{states}}$, $N_{\text{osc.\ states}}$) & $\chi^2/\text{DoF}$ & BAIC \\ \hline
0 & 6 & 17 &  6 &  20 &  (2, 1) & 0.51  & 67.2 \\
1 & 6 & 14 &  6 &  20 &  (2, 1) & 0.58 & 72.7 \\
2 & 6 & 14 &  6 &  20 &  (2, 1) & 1.14  & 81.4 \\
3 & 6 & 17 &  6 &  20 &  (2, 1) & 0.55  & 81.4 \\
4 & 6 & 17 &  6 &  20 &  (2, 1) & 0.71  & 70.9 \\
5 & 6 & 17 &  6 &  19 &  (2, 1) & 0.98  & 76.6 \\
6 & 6 & 17 &  7 &  20 &  (2, 1) & 0.62  & 70.6 \\
\hline\hline
\end{tabular}
\end{table}

In order to extract the energies and amplitudes from \cref{eqn:gevpCorrsDiag,eqn:gevpCorrsOff}, we perform a combined fit to the functional forms on the right-hand side of these equations, including the $t \to T-t$ contributions. The sum is truncated with independent limits for the regular and oscillating states, $N_{\text{states}}$ and $N_{\text{osc.\ states}}$. With a Bayesian fit approach, we use prior information for the ground state energies and overlap amplitudes extracted from the plateaus of the effective energy, \cref{eqn:effectiveMass}, and amplitude, \cref{eqn:effectiveAmp}. These effective energies and amplitudes are shown in \cref{fig:gevp}. We take the results of these as estimates for the prior central values and assign a 20\% width.  The effective amplitude for $\left\langle 0\left|\rho^{0}\right| n\right\rangle$ is obtained by taking a ratio of the respective effective amplitudes of \cref{eqn:gevpCorrsDiag,eqn:gevpCorrsOff}. We use a prior of $\Delta E = 0.5(0.5)$ GeV for the energy splitting to higher states. The higher-state amplitudes are given the same prior as the ground state but with 100\% widths. These higher state priors have little effect on the fits beyond helping with stability in some cases. Fits are performed up to $N_{\text{states}} \leq 3$ and $N_{\text{osc.\ states}} \leq N_{\textrm{states}}$ but we find that states beyond the first excited state and the first oscillating state are not well determined, even when including the earliest time-slices. Additionally, we vary $t_{\textrm{min}}$ and $t_{\textrm{max}}$ independently on the two datasets for \cref{eqn:gevpCorrsDiag,eqn:gevpCorrsOff}. This is beneficial as the two correlation functions have differing excited state contamination and noise-to-signal profiles. The stability of the fit results with respect to these parameter variations (as well as $t_0$ and operator basis variation) is discussed in \cref{sec:stability}. In order to select our preferred set of fit parameters, we simply choose the fit for each $n$ with the highest weight according to the Bayesian Akaike information criterion (BAIC) \cite{Neil:2023pgt}. In general, they correspond to what one would obtain from a more traditional `stability analysis,' \text{i.e.}, the lowest $t_{\textrm{min}}$ and highest $t_{\textrm{max}}$ in the region of fit stability. Applying a full model-averaging procedure, discussed in Ref.~\cite{Neil:2023pgt}, yields consistent results. \Cref{table:fitParams} lists the fit parameters for our preferred reference time $t_0/a = 5$ (reasoning discussed in \cref{sec:stability}) as well as corresponding fit-quality measures, namely the $\chi^2/\text{d.o.f.}$ (where d.o.f.\ stands for degrees of freedom) and the BAIC value. As an additional cross-check of the extracted spectrum and the GEVP method in general, we perform an alternative MEM fit to the full matrix \cref{eqn:ctmat} in \cref{sec:matrixFit}. Our results for both approaches are completely consistent.

\begin{figure}
  \centering
    \includegraphics[width=0.6\textwidth]{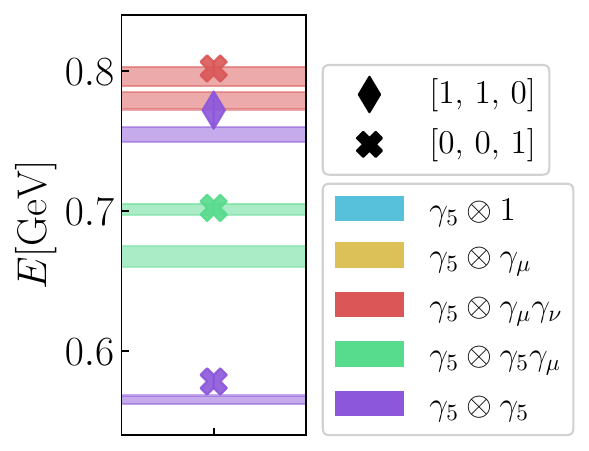} 
    \caption[]{Comparison of the free continuum spectrum of \cref{fig:naiveSpectrum} (symbols) with the interacting spectrum of \cref{fig:gevp} (bands).}
    \label{fig:spectrumCompare}
\end{figure}

\section{Results}\label{sec:results}

\subsection{Staggered two-pion spectrum}\label{sec:twoPiSpectrum}

In \cref{fig:gevp} we show the resultant GEVP fit energies and amplitudes for the first six states as color-coded bands. We find, as expected, that they agree very well with the effective mass and effective amplitude plateaus shown in this plot. We also find similar consistency for the highest well-determined state, $n=6$, which is not shown here \footnote{The $n=6$ state is included in the displayed spectra of \cref{fig:opBasisVariation,fig:t0variation1}.} as it renders the plot unclear. In \cref{fig:spectrumCompare} we compare the free, continuum energy spectrum (symbols) with these extracted energies (bands). We find that the ground state interacting energy (purple band) is roughly 2\% smaller than that of the free case. The expected taste-orbit splitting can be seen in the two-pion states built from the zero-momentum, three-dimensional single-pion irreps, \cref{eqn:pion001G5I,eqn:pion001GIT}. We see that these states, namely, $n$ = 1 and 2, and $n$ = 4 and 5 are non-degenerate. Of these, the two-pion states containing pions that are two-dimensional in the taste dimension are strongly interacting, while the opposite is true for the states that are one-dimensional (see \cref{fig:optimOps}). This enhanced (suppressed) interaction results in a larger (smaller) binding energy, and larger (smaller) overlap amplitudes, as seen in \cref{fig:gevp} (bottom panel).

\subsubsection{Stability}\label{sec:stability}

\begin{figure}
  \centering
    \includegraphics[width=0.8\textwidth]{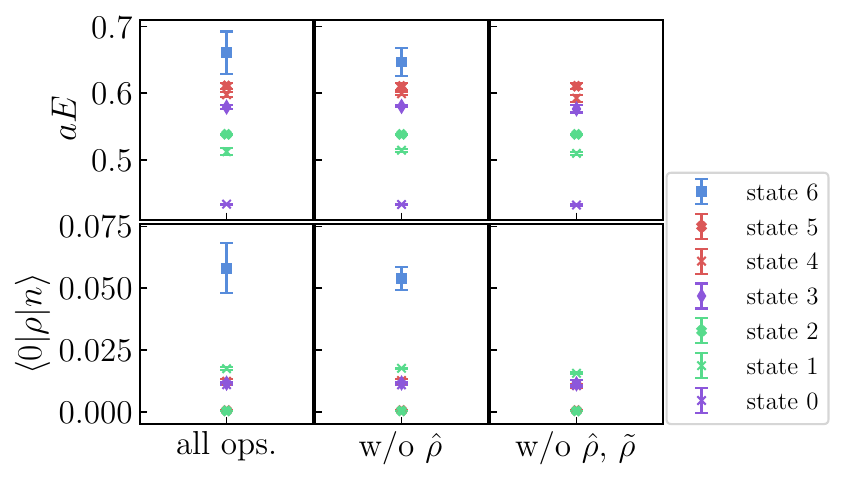}
    \caption[]{GEVP spectrum comparison for three different operator bases. The first is the full eight-operator basis, the second contains seven operators with the $\rho^0$ operator dropped, and the third has six operators with both the $\rho^0$ and $\tilde \rho^0$ omitted.}
    \label{fig:opBasisVariation}
\end{figure}

\begin{figure}
  \centering
    \includegraphics[width=0.6\textwidth]{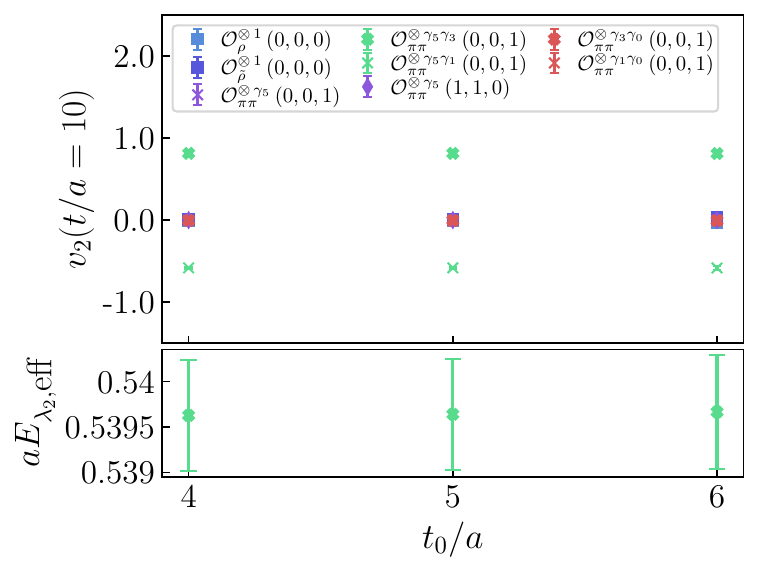}\hfill
    \includegraphics[width=0.32\textwidth]{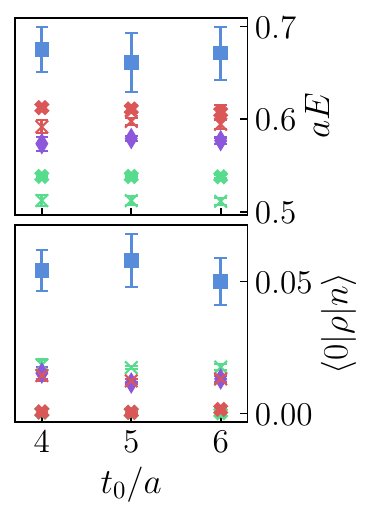}
    \caption[]{Left: Normalized eigenvector for the $n=2$ state as function of $t_0$ (top). The eigenvector is selected at $t^\prime/a=10$. The corresponding energy (bottom) is obtained by applying the effective mass formula, \cref{eqn:effectiveMass}, to the eigenvalue and performing a correlated average over the data points $t/a>10$. Right: Full GEVP spectrum comparison, energies (top) and amplitudes (bottom) for the same three choices of $t_0$. See legend of \cref{fig:opBasisVariation} for state labels. The ground state energy is omitted to improve the visibility of the rest of the spectrum, as the gap to the first excited state is large (see \cref{fig:opBasisVariation}).}
    \label{fig:t0variation1}
\end{figure}

\begin{figure}
\centering
    \includegraphics[width=0.9\textwidth]{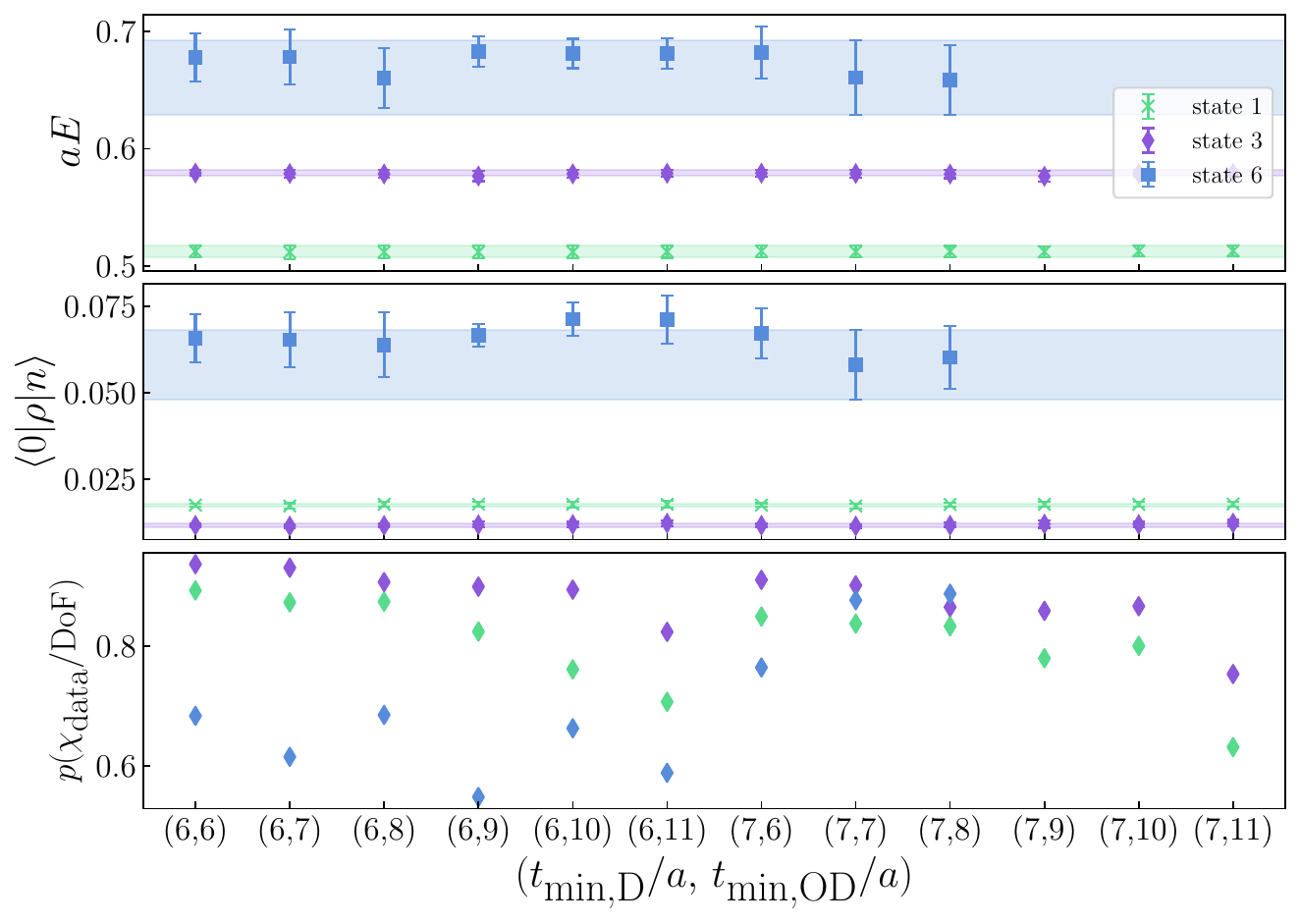}
    \caption[]{Stability of fit results with respect to $t_\text{min, D}/a$ and $t_\text{min, OD}/a$ for the $n=[1,3,6]$ states. All fits contain $2+1$ exponentials and $t_\text{max, D}/a$ and $t_\text{max, OD}/a$ are fixed to $18$ and $20$ which corresponds to the maximum available time-slice for $t_\text{max, D}/a$. Bands represent the fits from the preferred values listed in \cref{table:fitParams}. Top: Ground state energy. Middle: Overlap amplitude Bottom: Quality of fit as measured by the p value, excluding the prior contribution.}
    \label{fig:tminVariation}
\end{figure}

There are many choices that need to be made in order to arrive at a finalized energy spectrum and, hence, reconstruction of the vector-current correlation function and value for $\amuL$, namely, the choice the operator basis, the reference time $t_0$, asymptotic time $t^\prime$,
and the fit parameters for the optimized correlation functions, \cref{eqn:gevpCorrsDiag,eqn:gevpCorrsOff}, of which there are two sets of $t_{\text{min}}$ and $t_{\text{max}}$ for each $n$. Our selections are made as objectively as possible, using the BAIC weight, and after checking for stability under reasonable variations, among other considerations. 

We first consider variations in the operator basis by dropping the $\rho^0$ operators. In \cref{fig:opBasisVariation}, from left to right, we observe stability in the energies and amplitudes of the first six states as we drop first the $\rho^0$ operator (middle panel) and then the $\tilde \rho^0$ operator (right panel) from the basis. The energy and amplitude of the seventh state, $n=6$, changes slightly, albeit well within the uncertainty when the $\rho^0$ is omitted, which is not surprising given that the operator used to resolve it contains a significant contribution from the $\rho^0$ (see \cref{fig:optimOps}). As is well known, to obtain a reliable spectral decomposition of the first $n$ states, at least $n+1$ independent operators (correlators) are needed. Hence, in our following reconstructions of the vector current, which include the $n=7$ state, we use the full eight operator basis.

In the left-hand side of \cref{fig:t0variation1}, we show the eigenvector (top) and corresponding energy, extracted from the eigenvalue for the $n=2$ state as function of $t_0$. The full resultant spectrum for the same values of $t_0$ is shown in the right-hand side of \cref{fig:t0variation1}. The spectra are broadly consistent with each other as $t_0$ is varied with the $n=6$ state showing some fluctuation, although still being comfortably within uncertainties. We choose $t_0/a=5$ as our preferred choice for this parameter, as it is consistent with other choices and results in the best agreement with the raw correlator in the intermediate time range (see \cref{fig:recon}).

Finally, we examine the fit stability, in \cref{fig:tminVariation}, for a select number of states as a function of $t_\text{min, D}$ and $t_\text{min, OD}$, the lowest time included in the fit range for \cref{eqn:gevpCorrsDiag,eqn:gevpCorrsOff}, respectively. We find that the fit results are consistent across the values we consider, including the preferred choices from \cref{table:fitParams}, which are shown as bands. We do find for higher values of $t_\text{min, OD}$, namely, $9a$ that the fit to the $n=6$ state fails to converge.  Overall, all our stability checks indicate our final analysis choices result in well-determined energies and overlap amplitudes that are consistent with respect to reasonable parameter and operator basis variations.

\subsection{Correlator reconstruction and noise reduction}\label{sec:corrRec}

\begin{figure}
  \centering
    \includegraphics[width=0.7\textwidth]{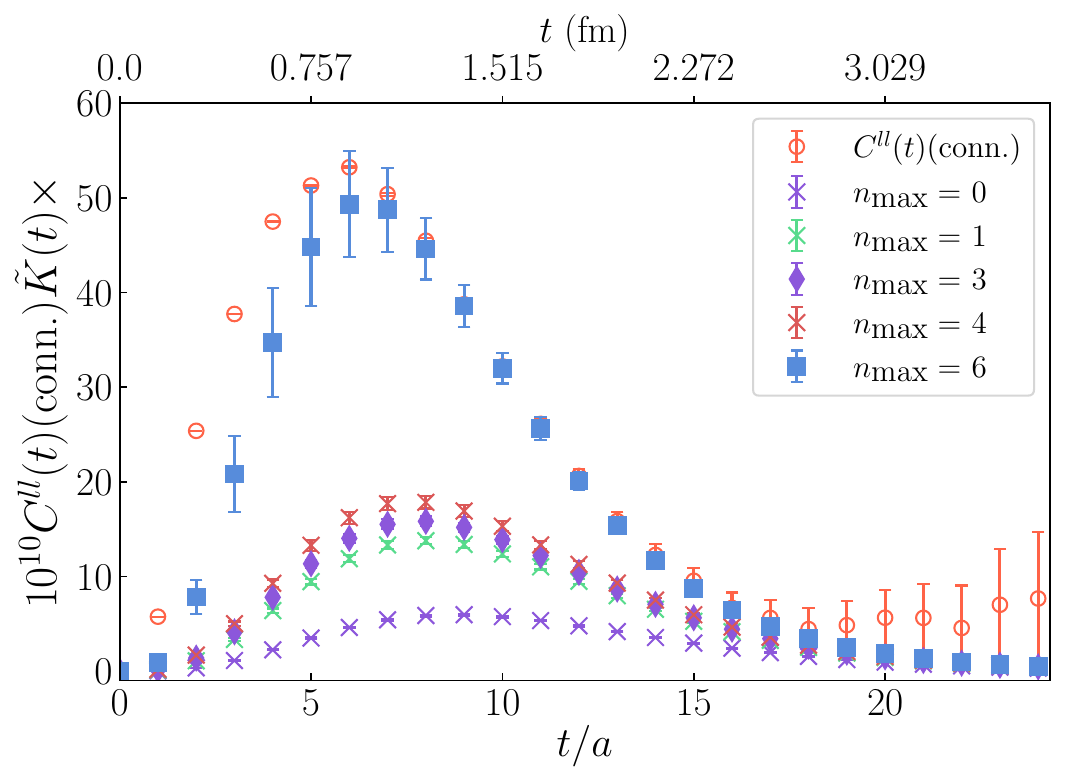}
    \caption{Results for the taste-singlet vector-current two-point, correlation function (orange). Reconstruction of the correlation function from determined parameters for states up to $n_\textrm{max}=[0,6]$.  In the region $t/a \in [7,13]$, the raw correlation function results are obscured by the $n_\textrm{max}=6$ reconstruction.}
    \label{fig:recon}
    \includegraphics[width=0.745\textwidth]{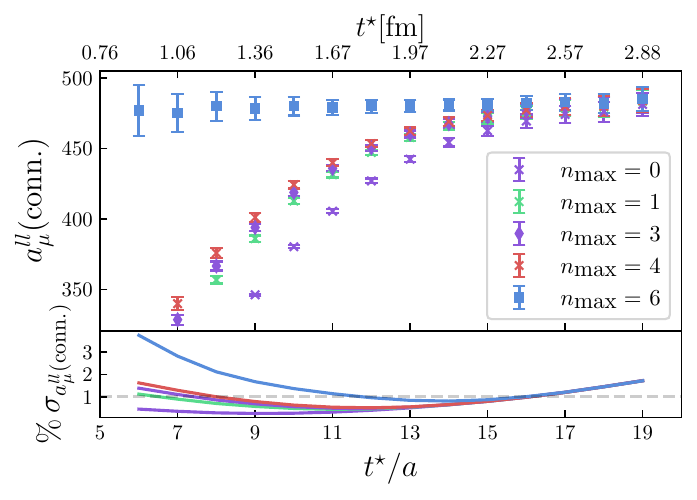}
    \caption[]{Top: Results for $\amuL$ as $t^\star$, the point at which the correlator is replaced by the reconstruction, is varied. Bottom: relative error in the determinations of $\amuL$ from the top figure.}
    \label{fig:reconStab}
\end{figure}
With the determined energies and overlap amplitudes, the correlation function is reconstructed using the sum in \cref{eq:corrFuncSpecRep} truncated to $n_\textrm{max}$. The corresponding reconstructed integrand of \cref{eq:amuTint} is given in \cref{fig:recon}. Reconstructions as more states are included up to the maximum at $n_\textrm{max}=6$ are shown. For visibility, we do not show the $n_\textrm{max}=2$ and $5$ reconstructions as they lie on top of the preceding reconstructions, due to the reduced overlap amplitudes (see bottom panel of \cref{fig:gevp}). Additionally, we do not include any oscillating contributions determined from the fits. The reconstructions are compared to the raw vector-current two-point data (orange open circles), after applying improved parity averaging \cite{Lehner:2020crt} to suppress the oscillatory behavior for better visualization.

For $n_\textrm{max}=4$, there is already good agreement between the raw data and the reconstruction at $t/a>16$. Once the highest state is included, we have agreement as early as $t/a=7$, but the reconstruction is actually noisier than the raw data here. In order to select a $t^\star$ at which to replace the vector-current correlator data with the reconstruction, we examine both the stability of $\amuL$ with respect to $t^\star$ and also the relative error. In \cref{fig:reconStab}, we show the value $\amuL$ as $t^\star$ is varied for a range of $n_\textrm{max}$. We see for $n_\textrm{max}=6$ we have stability starting around $t^\star/a>7$ in agreement with visual indication from \cref{fig:recon}. In the bottom panel, the relative error of these determinations is given. As mentioned, although the result stabilizes at $t^\star/a\approx 9$, precision is lost if the raw correlator data is replaced this early, as the reconstruction is noisier; hence, we select $t^\star/a=13$.

\begin{figure}
  \centering
    \includegraphics[width=0.8\textwidth]{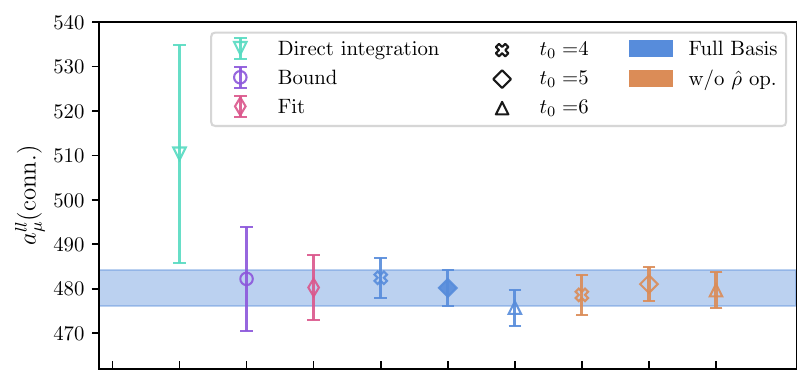}
    \caption[]{Results for $\amuL$ from the reconstruction strategy described in this work, using the 9600 configuration vector current dataset. We show results for the variations discussed in \cref{sec:stability}, choosing different values of $t_0$ and also dropping the $\rho^0$ operator before performing the GEVP. Our final choice of $\amuL$ is given by the full operator basis at $t_0/a=5$, indicated by the blue band and filled symbol. Also included are comparisons to $\amuL$ from the different noise-reduction strategies discussed in \cref{sec:LQC}, $a_\mu(\textrm{fit})$ and  $a_\mu(\textrm{bound})$, as well as $a_\mu(\textrm{direct})$, the result obtained by direct integration of the data with no noise-reduction applied.}
    \label{fig:amuCompare}
\end{figure}

\begin{table}
\centering
\caption{Numerical results for $\amuL$ from the different noise-reduction strategies discussed in this work. The values for $\amuL$ from the two-pion spectrum reconstruction are given in the column labelled by $a_\mu( \pi \pi \textrm{ recon.})$. The results from the fit and bounding methods discussed in \cref{sec:LQC} are given in the columns $a_\mu(\textrm{fit})$ and  $a_\mu(\textrm{bound})$, respectively. Also included is $a_\mu(\textrm{direct})$, the results obtained by direct integration of the data with no noise-reduction applied.}
\label{table:amuRes}
\begin{tabular}{lcccc}
\hline\hline
$N_{\textrm{conf}}$: $\left\langle J^{l}_{i}(x) J^{l}_{i}(0)\right\rangle_{\textrm{conn.}}$  &  $a_\mu( \pi \pi \textrm{ recon.})$ &  $a_\mu(\textrm{fit})$ &  $a_\mu(\textrm{bound})$ &  $a_\mu(\textrm{direct})$ \\ \hline
3473 & 476.6(5.3) & 479(11) & 470(17) & 550(41)\\
9800 & 480.2(4.0) & 480.3(7.2) & 482(12) & 510(25)\\
\hline\hline
\end{tabular}
\end{table}

Our results for the light-quark connected contribution to $\amuHVP$ are given in \cref{fig:amuCompare} for the analysis variations discussed in \cref{sec:stability}. Our preferred final result is the value obtained at the reference time $t_0/a=5$ using the full basis (blue band). We make this choice over $t_0/a=4$ to avoid possible excited state contamination from the $\rho^0$ operator at early times. However, we find all variations give consistent determinations of $\amuL$. The numerical value for our final result is given in \cref{table:amuRes}, second column, for the case of using the 3473 configs of the two-pion data (first row) and also for the case of using the additional $\approx 6300$ additional configurations for the vector current two-point function (second row). For comparison, shown in the third and fourth columns respectively and in \cref{fig:amuCompare} are results from the bounding and fit methods, discussed at the end of \cref{sec:LQC}. Also given is the result from direct integration of the raw data, which is in mild tension with the other results, albeit with a much larger uncertainty, due to the badly behaved tail of the correlation function, visible in \cref{fig:recon}. We find all noise-reduction strategies address this issue and indeed are all consistent; however, we obtain an improvement, from the two-pion reconstruction, in statistical precision over the bounding approach of roughly a factor of 2.5.

\section{Summary and outlook}\label{sec:outlook}

The last few years have seen great progress in lattice QCD calculations of $\amuHVP$, culminating in dedicated long-distance calculations~\cite{Djukanovic:2024cmq,RBC:2024fic,FermilabLatticeHPQCD:2024ppc}, which fed into the lattice average of the Muon $g-2$ Theory Initiative's second White Paper~\cite{Aliberti:2025beg}. Despite this progress, the long-distance contribution remains the dominant source of uncertainty in lattice HVP calculations, and improved methodology will be essential for reaching the precision needed to fully exploit the final Fermilab E989 $g-2$ result~\cite{Muong-2:2025xyk}.

In this paper, we address this issue by explicitly computing the contributions from exclusive channel two-pion states to the vector-current two-point function at large Euclidean times.  Ours is the first study of a staggered multi-hadron system which includes the full set of staggered operators. To construct the two-pion operators, we follow Refs.~\cite{sharpe.1,Golterman:1984cy,Golterman:1986} to obtain the irreducible representations of the staggered group and compute the Clebsch-Gordan coefficients. The detailed information needed to construct two-pion operators, transforming under any staggered vector-current irrep, is given in the Appendices. The $I=1$ three- and four-point correlation functions for $\rho \to \pi\pi$, $\pi\pi \to \rho$, and $\pi\pi \to \pi\pi$ are generated on the MILC collaboration's physical mass ensemble at $a\approx 0.15$~fm \cite{MILC:2012znn}. A GEVP analysis is used to extract the finite-volume amplitudes and energies of the interacting two-pion system. As shown in \cref{fig:reconStab}, the resulting spectral reconstructions of the vector-current correlation functions are obtained with greatly reduced statistical errors at large Euclidean times, while correctly reproducing the original vector-current correlation function over a range of Euclidean times down to $t\gtrsim 0.8$~fm. We find that results, for $\amuL$, obtained with the reconstructed correlation function are consistent with estimates using the bounding and fit methods, while improving the statistical precision by roughly a factor of $2.5$ (see \cref{table:amuRes} and \cref{fig:amuCompare}). In addition, we explore an alternative multi-exponential matrix fit approach as a cross-check on the GEVP analysis, finding consistent results. In summary, we show that the two-pion reconstruction offers a viable path towards lattice HVP calculations at the few permille level, also for simulations based on staggered fermions. 

The next step is to extend this study to finer lattice spacings so that the statistical gains survive the continuum limit. This poses new challenges, because the smaller taste splittings at finer lattice spacings result in an increasing number of two-pion operators (see \cref{fig:naiveSpectrum}). In particular, for the MILC collaboration's physical mass ensemble at the next-finest lattice spacing, $a\approx 0.12$~fm, a total of eighteen two-pion operators are needed to resolve the spectrum below the $\rho$-meson mass, including two-pion operators made of three-link (taste vector) and four-link (taste scalar) pions, which are expected to yield noisier correlation functions. These challenges will be investigated in a follow-up study on this ensemble that is already underway. 

Finally, the finite-volume amplitudes and energies of an interacting two-pion system can be related, in the L\"uscher formalism \cite{Luscher:1985dn}, to the corresponding $\pi\pi$ scattering parameters in infinite-volume. Utilizing this connection for the case at hand, is, however, not straightforward, because the staggered formulation employed in this work violates unitarity, a result of the taking the fourth root of the staggered-fermion determinant to represent one quark flavor in the generation of the gauge-field ensembles.  Since the unitarity violations enter as ${\cal O}(a^2)$ discretization errors \cite{Bernard:2007ma}, it may be possible to extend the L\"uscher formalism to incorporate them. This question was investigated in Ref.~\cite{Draper:2021wga} using partially-quenched ChPT for a non-unitary set-up involving twisted-mass fermions, while in Ref.~\cite{Hansen:2024cai} an extension of the  L\"uscher formalism to incorporate discretization effects was recently presented. Further investigations into this possibility are underway \cite{Valois:2026oqg}; if successful, they could enable ab-initio studies of scattering processes and resonance physics on the large library of HISQ ensembles generated by the MILC collaboration. 

\vfill

\begin{acknowledgments}
We thank Christine Davies, Peter Lepage, and all our collaborators in the Fermilab Lattice and MILC collaborations for useful discussions throughout the development of this project. 
Computations for this work were primarily carried out using resources provided by the Blue Waters sustained-petascale computing project, which is supported by NSF awards OCI-0725070 and ACI-1238993, the State of Illinois, and as of December 2019, the National Geospatial-Intelligence Agency. Blue Waters is a joint effort of the University of Illinois, Urbana-Champaign and its National Center for Supercomputing Applications.
Some additional computations were also performed using Delta advanced computing and data resource which is supported by the National Science Foundation (award OAC 2005572) and the State of Illinois through allocation MCA93S002: Lattice Gauge Theory on Parallel Computers from the Advanced Cyberinfrastructure Coordination Ecosystem: Services \& Support (ACCESS) program, which is supported by National Science Foundation grants \#2138259, \#2138286, \#2138307, \#2137603, and \#2138296.
This work was supported in part by the U.S.~Department of Energy under Award No. DE-SC0015655 (A.X.K. and S.L.) and No.~DE-SC0010120 (S.G.); by the Universities Research Association Visiting Scholarship awards 20-S-12 and 21-S-05 (S.L.); by the National Science Foundation under Grants PHY20-13064 and PHY23-10571. (C.D and S.L); by the Simons Foundation under their Simons Fellows in Theoretical Physics program (A.X.K.). A.~El-Khadra is grateful to the Pauli Center for Theoretical Studies and the ETH Z\"urich for support and hospitality. 
This document was prepared using the resources of the Fermi National Accelerator Laboratory (Fermilab), a U.S. Department of Energy, Office of Science, HEP User Facility. Fermilab is managed by Fermi Research Alliance, LLC (FRA), acting under Contract No. DE- AC02-07CH11359.

\end{acknowledgments}

\appendix

\section{Staggered quark theory primer}\label{sec:stagGroupPrimer}

As this work involves multiparticle states constructed from staggered mesons, a topic not often studied in detail, this Appendix serves as a primer on the group theoretical details and notation used here. We rely primarily on the methodology introduced in Ref.~\cite{sharpe.1}, as it includes a natural extension for studying states at non-zero momentum. The construction of the irreducible representations of the staggered group in that work is repeated here, including the aforementioned decomposition to states at non-zero momentum. Construction of the associated operators and the connection to continuum states is also repeated, correcting some examples discussed in that work and expanding on some pertinent results relevant here.

\subsection{Staggered lattice QCD}

The staggered action has one fermion component (per color) at each site~\cite{Susskind:1976jm,Kawamoto:1981hw,Sharatchandra:1981si}.
It can be obtained from the four-component naive action,
\begin{align}
S_{F}[q_f, \bar{q}_f, U]=a^{4}\sum_f \sum_{n \in \Lambda} \bar{q}_f(n)\left(\sum_{\mu=0}^{3} \gamma_{\mu} \frac{U_{\mu}(n) q_f(n+\hat{\mu})-U_{-\mu}(n) q_f(n-\hat{\mu})}{2 a}+m q_f(n)\right). \label{eqn:naiveFermionAction}
\end{align}
through the Kawamoto-Smit transformation \cite{Kawamoto:1981hw},
\begin{align}
    q(n) = \Omega(n) q^{\prime}(n)&, \quad
    \bar{q}(n) = \bar{q}^{\prime}(n) \Omega^{\dagger}(n), \label{eqn:KStransform}\\
    \Omega(n) \equiv& \left(\gamma_{0}\right)^{n_{0}} \left(\gamma_{1}\right)^{n_{1}} \left(\gamma_{2}\right)^{n_{2}} \left(\gamma_{3}\right)^{n_{3}}. \label{eqn:motoMatrix}
\end{align}
which diagonalizes the action as
\begin{align}
S_{F}\left[q_f^{\prime}, \bar{q}_f^{\prime}, U\right]=a^{4} \sum_f \sum_{n \in \Lambda} \bar{q}_f^{\prime}(n) \left(\sum_{\mu=0}^{3} \eta_{\mu}(n) \frac{U _ { \mu }(n)q_f^{\prime}(n+\hat{\mu})-U _ { \mu } ^ { \dagger } ( n - \hat { \mu } )q_f^{\prime}(n-\hat{\mu})}{2 a}+m q_f^{\prime}(n)\right), \label{eqn:stagFermionAction}
\end{align}
with
\begin{align}
\eta _ { \mu } ( n ) \equiv \Omega ^ { \dagger } ( n ) \gamma _ { \mu } \Omega ( n \pm \hat { \mu }  ) = ( - 1 ) ^ { \sum _ { \rho < \mu } n _ { \rho } }.   \label{eqn:etaDef}
\end{align}
The spacetime directions are ordered $(t,x,y,x)$ as in the MILC code, instead of the $(x,y,z,t)$ order in Ref.~\cite{sharpe.1}. Three of the four identical spin degrees of freedom are dropped to obtain the staggered quark action~\cite{Kawamoto:1981hw,Sharatchandra:1981si}
\begin{align}
S_{F}\left[\chi_f, \bar{\chi}_f, U\right]=a^{4} \sum_f \sum_{n \in \Lambda} \bar\chi_f(n) \left(\sum_{\mu=0}^{3} \eta_{\mu}(n) \frac{U _ { \mu }(n)\chi_f(n+\hat{\mu})-U _ { \mu } ^ { \dagger } ( n - \hat { \mu } )\chi_f(n-\hat{\mu})}{2 a}+m \chi_f(n)\right), \label{eqn:stagFermionActionReduced}
\end{align}
where the $\chi$ field has one fermion degree  of freedom per site.
The reason for the reduction is that the naive action leads to 16 Dirac fermions in the continuum limit. Now only four `tastes' survive.

 \subsection{Staggered symmetries and group structure}\label{sec:stagSyms}

This work employs a lattice with $N_t=48$ sites in the temporal direction and $N_s=32$ sites in the spatial directions, and $N_t>N_s$ holds on the other $2+1+1$-flavor HISQ ensembles [{\sl ref}] that will be used in the future.
Thus, symmetry between (Euclidean) time and space is absent \footnote{The effect of this symmetry breaking is not detectable in the analysis described here. We find, for example, that the 3-fold and 1-fold multiplet of spatial and temporal `one-link' pions are degenerate.}, which is fine, as the objective here is the transformation properties of eigenstates of the transfer matrix and the operators that create them.
Here, we show how the symmetries of staggered fermions combine to form the symmetry group of the transfer matrix.

\subsubsection{Symmetries}

The Kawamoto-Smit transformation in \cref{eqn:KStransform} depends on $n_\mu$, and hence modifies spin structure, differently at different spacetime points. Because of this, the original symmetries from the naive action, now have mixed spacetime-spin dependence when applied to $q^\prime(n)$. Translations acting on the fields in the diagonalized action, for example, become 
\begin{equation}
    q'(n) \to \zeta_\mu(n) \gamma_\mu q^\prime(n-\hat \mu),
    \label{eqn:transEx}
\end{equation}
where
\begin{align}
\zeta_{\mu}(n) \equiv \Omega^{-1}(n) \Omega(n \pm \hat{\mu} ) \gamma_{\mu}=(-1)^{\Sigma_{\sigma>\mu} n_{\sigma}}.
\label{eqn:zeta}
\end{align}
It is preferable to have symmetry transformations which are also diagonal in the spinor index, as these can be associated with the one-component staggered action in \cref{eqn:stagFermionActionReduced} and hence, can be used to classify the irreps (states) of the theory. The spin-diagonal set of transformations are obtained by combining the original symmetry transformations of the QCD action, after discretization, with the doubling transformations of the naive action,
\begin{align}
    q(n) \to e^{\omega^A B^A(x)} q(n),  \bar{q}(n) \to \bar{q}(n) e^{-\omega^A B^A(x)} .\label{eqn:doubleTrans}
\end{align}
The generating set $B^A(x)$ are anti-Hermitian and given by
\begin{align}
    B_\mu(n)=\gamma_\mu \gamma_5(-1)^{n_\mu},  B_5(n)=i \gamma_5 \varepsilon(n),  B_\mu(n) B_5(n),  B_\mu(n) B_\nu(n) (\mu<\nu), \label{eqn:doublegens}
\end{align}
where 
\begin{align}
    \varepsilon(n)=(-1)^{n_0+n_1+n_2+n_3}.
\end{align}

The resultant spin-diagonal symmetry operations are then
\begin{enumerate}
\item Translations $x\to x-a\hat{\mu}$ or $n\to n-\hat{\mu}$: \\
Choosing $B_\mu(n) B_5(n)$ results in a spin-diagonal operator, leading to the staggered \textit{shift}
\begin{align}
S _ { \mu } : 
\left\{ \begin{array} {l}
{\chi ( n ) \to \zeta _ { \mu } ( n ) \chi ( n - \hat \mu )}, \\
{ \overline { \chi } ( n )  \to \zeta _ { \mu } ( n ) \overline { \chi } \left( n  - \hat \mu \right)}.
\end{array} \right. \label{eqn:stagShifts}
\end{align}

\item Rotations by $\pi/2$ in the $\mu\nu$ plane, $R_{\mu\nu}$: \\
Choosing $B_{\mu}(n) B_\nu(n)$ leads to the transformation rule for the staggered field
\begin{align}
R _ { \mu \nu } : 
\left\{ \begin{array} {l}
{\chi ( n ) \to S _ {R_{\mu \nu} } \left( R ^ { - 1 } n \right) \chi \left( R ^ { - 1 } n \right)}\\
{ \overline { \chi } ( n )  \to S _ {R_{\mu \nu} } \left( R ^ { - 1 } n \right) \overline { \chi } \left( R ^ { - 1 } n \right)}
\end{array} \right. ,\label{eqn:stagRot}
\end{align}
where
\begin{align}
S _ { R_{\mu \nu} } \left( R ^ { - 1 } n \right) =
\frac{1}{2}&\left[1+\eta_{\mu}(R^{-1}_{\mu \nu} n) \eta_{\nu}(R^{-1}_{\mu \nu}n) \zeta_{\mu}(R^{-1}_{\mu \nu}n) \zeta_{\nu}(R^{-1}_{\mu \nu}n)
\right. \nonumber \\ & \left. \hspace{2em}
-\zeta_{\mu}(R^{-1}_{\mu \nu}n) \zeta_{\nu}(R^{-1}_{\mu \nu}n)+\eta_{\mu}(R^{-1}_{\mu \nu}n) \eta_{\nu}(R^{-1}_{\mu \nu}n)\right],
\end{align}
where
\begin{align}
\eta_{\mu}(n) \equiv (-1)^{\Sigma_{\sigma<\mu} n_{\sigma}}.
\label{eqn:eta}
\end{align}
Upon applying $R_{\mu\nu}$ four times, the product of the $S_{R_{\mu\nu}}$ factors yields $-1$, as it should for a fermion.

\item Spatial inversion $I_S: n_0\to n_0, n_i\to-n_i$: \\
Choosing $B_0 B_5$ leads to,
\begin{align}
I _ { S } :  
\left\{ \begin{array} {l}
{ \chi ( n ) \to (-1)^{n_1+n_2+n_3} \chi \left( I _ { S }n \right)}\\
{ \overline \chi ( n ) \to  (-1)^{n_1+n_2+n_3} \overline \chi \left( I _ { S } n \right)}
\end{array} \right. , \label{eqn:stagParity}
\end{align}
so staggered fermions at odd and even spatial sites have opposite intrinsic parity.
For inversion of a single axis, $I_\mu: \chi(n)\to(-1)^{n_\mu}\chi(I_\mu n)$ and similarly for $\overline{\chi}$.
As discussed below, $I_S=I_1I_2I_3$ is not quite the parity operator of the continuum limit.

\item Charge conjugation: \\
Choosing $ B_{2}(n) B_{5}(n) = i\gamma_2(-1)^{n_2} \varepsilon(n)$ gives the transformation rule for staggered charged conjugation  
\begin{align}
C _ { 0 } : 
\left\{ \begin{array} { l } 
{ \chi ( n ) \to \varepsilon ( n ) \overline { \chi } ( n ) } \\
{ \overline { \chi } ( n ) \to - \varepsilon ( n ) \chi ( n ) }
\end{array} \right. .
\end{align}
As discussed below, $C_0$ is not quite continuum-limit charge conjugation, hence the subscript.

\item Chiral symmetry: \\
The global chiral flavor symmetries also have spinor structure. In going to the reduced action, a remnant of this symmetry still exists as
\begin{align}
 \chi^{\prime} &=\mathrm{e}^{\mathrm{i} \alpha \varepsilon(n) T_{i}} \chi, & \bar{\chi}^{\prime} &=\bar{\chi} \mathrm{e}^{\mathrm{i} \alpha\varepsilon(n) T_{i}}, \label{eon:chiral-nonsinglet} \\
 \chi^{\prime} &=\mathrm{e}^{\mathrm{i} \alpha \varepsilon(n) } \chi, & \bar{\chi}^{\prime} &=\bar{\chi} \mathrm{e}^{\mathrm{i} \alpha \varepsilon(n) } , \label{eon:chiral-singlet}
\end{align}
where $T_i$ is a flavor-symmetry generator, and \cref{eon:chiral-singlet} show the flavor singlet case, which is \emph{not}, however, a taste singlet.
\end{enumerate}

\subsubsection{Group structure}\label{sec:groupStruct}
The symmetry group of the transfer matrix is generated by $\{R_{ij}, S_\mu, I_S, C_0\}$.%
\footnote{The temporal-spatial rotations $R_{0j}$ are not symmetries.}
It is necessary to know their commutation relations. As always, the rotations $R_{\mu \nu}$, \cref{eqn:stagRot}, and axis inversions, $I_\mu$, satisfy
\begin{align}
R_{\mu\nu} I_\mu R_{\mu\nu}^{-1} &= I_\nu = R_{\nu\mu} I_\mu R_{\nu\mu}^{-1} , \label{eqn:parityRot} \\
R _ {  \mu \nu  } I _ { \rho } R  _ {  \mu \nu  }^{-1} &= I _ { \rho },  \rho \neq \mu, \nu . \label{eqn:parityNonRot}
\end{align}
The shifts anti-commute, 
\begin{align}
    S _ { \mu } S _ { \nu } S _ { \mu }^{-1} = - S _ { \nu } ,  \nu\neq\mu .
\end{align}
With $N_s$ sites ($N_s$ must be even) in the spatial directions, repeating a spatial shift $N_s$ times yields $S_i^{N_s}=\pm1$, with the upper (lower) sign for (anti)periodic boundary conditions. Similarly, $S_0^{N_t}=\pm1$. The shifts and rotation-reflections satisfy
\begin{align}
R _ { \mu \nu }^ { - 1 } S _ { \mu } R _ { \mu \nu }  &= S _ { \nu }, \label{eqn:stagRotateShift} \\
R _ { \mu \nu }^ { - 1 } S _ { \nu } R _ { \mu \nu }  &= - S _ { \mu }^{-1},  \label{eqn:stagRotateShift2} \\
R _ { \mu \nu }^ { - 1 } S _ { \rho } R _ { \mu \nu }  &= S _ { \rho },  \rho\neq\mu,\nu  \label{eqn:stagRotateShift3} \\
I_S S_i I_S^{-1} &=  -S_i^{-1},  i = 1,2,3, \label{eqn:stagInvertShifti} \\
I_S S_0 I_S^{-1} &= S_0. \label{eqn:stagInvertShift0} 
\end{align}
Charge conjugation commutes with all reflections and rotations
and anti-commutes with the shifts,
\begin{align}
C _ { 0 } S _ { \mu } =  - S _ { \mu } C _ { 0 }.
\label{eqn:shift-C0}
\end{align}
The flavor and color symmetries commute with all geometric symmetries and charge conjugation.

The transfer matrix for staggered fermions, \cref{eqn:stagFermionActionReduced}, is a Hilbert-space operator acting on physical states, evolving them two temporal spacings forward~\cite{Sharatchandra:1981si}.
It is thus the Hilbert-space operator corresponding to
\begin{equation}
    T_0 = S_0^2.
\end{equation}
$T_0$, of course, commutes with $S_0$.
It is convenient to take the formal square root
\begin{equation}
    \Xi_0 = T_0^{-1/2} S_0,
\end{equation}
i.e., if the eigenvalue of $\hat{T}_0$ is $\mathrm{e}^{-2E}$, then $T^{-1/2}=\mathrm{e}^E$.
In the same vein, it is convenient to introduce the same construction for the spatial directions, $T_i=S_i^2$ and
\begin{equation}
    \Xi_i = T_i^{-1/2} S_i,
\end{equation}
where $T_i^{-1/2}$ is again defined via the eigenvalue of $T_i$.
It is customary to call the $\Xi_\mu$ taste operators, to distinguish them from the shifts.
They satisfy the same commutation rules as the $S_\mu$, \cref{eqn:stagRotateShift,eqn:stagRotateShift2,eqn:stagRotateShift3,eqn:stagInvertShifti,eqn:stagInvertShift0}.
In particular, the $\Xi_\mu$ generate the Clifford group $\Gamma_4$, or incorporating charge conjugation, \cref{eqn:shift-C0}, $\Gamma_{4,1}$~\cite{sharpe.1}.

Thus, ignoring flavor and color, the symmetry group of the staggered transfer matrix is 
\begin{align}
G_{T_0}  &= \left\{ T_i \right\} \rtimes \left[ \left\{ \Xi _ { \mu }, C_0 \right\} \rtimes \left\{ R _ { i j } , I _ { S } \right\} \right] \nn\\
&= Z_{N_s/2}^3 \rtimes \left(\Gamma_{4,1} \rtimes \text{O}_\text{h}\right),\label{eqn:stagGroupStruc}
\end{align}
with the octahedral group $\text{O}_\text{h}$ consisting of the rotation-reflection symmetries of the cube.

\subsection{Irreducible representations of the staggered group}\label{sec:stagIrreps}

Classifying the irreps of \cref{eqn:stagGroupStruc} involves applying Wigner's method \cite{Wigner:1939cj} for semi-direct products, $G=N\rtimes H$.
Wigner's method needs to applied twice, first with the normal subgroup given by $N=Z_{N_s/2}^3$ and $H=\Gamma_{4,1}\rtimes \text{O}_\text{h}$, and then with $N=\Gamma_{4,1}$ and $H=\text{O}_\text{h}$.
Only the bosonic representations are relevant for this work, as meson states appear exclusively. Considering just the bosonic representations of $\Gamma_{4,1}$ simplifies the construction, as the group homomorphism $\Gamma_{4,1} \to \text{Z}_2^5$ can be exploited in this case.

When $N$ is Abelian, Wigner's method proceeds as follows~\cite{Cornwell}:
\begin{enumerate}
\item Determine all (one-dimensional) irreps `$\sigma$' of the normal Abelian subgroup $N$.
\item For each irrep $\sigma$, determine the subgroup $H(\sigma)\subseteq H$ of elements $h$ satisfying the character equation
\begin{align}
\chi^{(\sigma)}_{N}(hnh^{-1})= \chi^{(\sigma)}_{N}(n) ,  \forall n \in N ,
\end{align}
where $\chi^{(\sigma)}_N$ denotes the character of $\sigma$ in the normal subgroup $N$.
The $H(\sigma)$ are the so-called little groups.
\item Classify the irreps, $\sigma$, into orbits(also called `stars'~\cite{Golterman:1986}), which
is achieved by breaking $H$ into right cosets under the little group $H(\sigma)$,
\begin{align}
H &= H(\sigma) h_1 + H(\sigma) h_2 + \ldots + H(\sigma) h_{|H|/|H(\sigma)|},
\label{eqn:disjoint-sum}
\end{align}
where $h_1=E$ (identity element) and $h_i \centernot \in H(\sigma) h_j$ for all $i \neq j$.
From \cref{eqn:disjoint-sum} we can then choose a set of coset representatives,
\begin{align}
\{ h_1, h_2, \ldots,  h_{|H|/|H(\sigma)|} \}.
\end{align}
The orbit is then the list of irreps, each with the same little group,
\begin{align}
\{  h_1(\sigma) = \sigma,  h_2(\sigma), \ldots,  h_{|H|/|H(\sigma)|} (\sigma) \},
\end{align}
determined from
\begin{align}
\chi^{( h_i(\sigma))}_N(n) &= \chi^{(\sigma)}_N( h_i n h_i^{-1}) ,  \forall n \in N .
\end{align}
\item Determine the irreps  $\rho$ of the little groups $H(\sigma)$.
\item Form irreps of the semi-direct groups $G(\sigma) = N \rtimes H(\sigma)$ for a single representative $\sigma$ in each orbit as
\begin{align}
D^{(\sigma, \rho)}_{G(\sigma)}(n h) &= \chi^{(\sigma)}_N(n) D^{(\rho)}_{H(\sigma)}(h).
\end{align}
\item Induce an irrep for the full group $G$ through the formula
\begin{align}
D^{(\gamma)}_{G} ( g ) _ { i t , j r } = \left\{ \begin{array} { l l } {  \chi^{( h_i(\sigma))}_N(n) D^{(\rho)}_{H(\sigma)}( h_i h  h_j ^ { - 1 })_ { t r } , } & { \text { if } h_i h  h_j ^ { - 1 } \in H(\sigma) } \\ { 0 , } & { \text { if }  h_i h  h_j ^ { - 1 } \notin H(\sigma)} \end{array} \right. .\label{12.83832}
\end{align}
\end{enumerate}
A complete set of irreps is obtained by preforming the above step for each orbit.

As mentioned, the approach described above needs to applied twice for the nested semi-direct products appearing in \cref{eqn:stagGroupStruc}. In the first case, where the normal subgroup is given by $N = Z_{N_s/2}^3$, and $H = \Gamma _ { 4,1 } \rtimes  \text{O}_\text{h}$, the one dimensional irreps of $Z_{N_s/2}^3$ are given by
\begin{align}
T_i|\vec{p}\rangle=\exp \left(i 2p_i\right)|\vec{p}\rangle, \quad  p_i = \frac{2\pi}{aN_s}\ell_i , \label{eqn:bosonIrrepTrans}
\end{align}
with $\ell_i$ as specified in \cref{eqn:allowedMomentum}. In the bosonic case, as mentioned above, the irreps of $\Gamma _ { 4,1 }$ can be obtained by the homomorphism to the Abelian group $\text{Z}_2^5$ and are
\begin{align}
D(\Xi_\mu) = e^{i (\pi_\xi)_\mu},&  \quad \pi_\xi \equiv (\xi_0, \xi_1, \xi_2, \xi_3), \quad  \xi_\mu = 0,\pi , \label{eqn:bosonTasteIrreps}\\ 
D(C_0) = e^{i \xi_{C}},&  \quad \xi_{C} = 0, \pi.
\label{eqn:bosonIrrepChageConj}
\end{align}
The elements of $\Gamma _ { 4,1 }$ leave the momentum invariant, hence orbits consist of the list of momenta obtained through application of elements of $\text{O}_\text{h}$ to the vector $p_i$.
The corresponding little groups are the subgroups of $\Gamma _ { 4,1 } \rtimes \text{O}_\text{h}$ which leave the orbit representatives invariant.
\begin{table}
\newcommand{\h}{\hphantom{2}}
\centering
\caption[Momentum orbits under the staggered rotation group]{Momentum orbits under the staggered rotation group. First column: The list of the momentum orbits (with $2\pi/ aN_s$ factored out) under the staggered rotation group, given by a representative element and the dimension. Second column: The corresponding little group given by its group generators and the corresponding group structure and order.}\label{table:orbits}
\begin{tabular}{cc@{}c@{}c}
\hline \hline
Orbit representative & Size & Little group generators and structure & Order  \\
\hline  
$(0, 0, 0)$ & $\h1$ & $\left\{\Xi_\mu, R_{ij}, I_S \right\}                             \cong \Gamma_{4,1} \rtimes \text{O}_\text{h}$                &  $48$ \\ 
$(0, 0, \ell)$ & $\h6$ & $\left\{\Xi_\mu, R_{12}, R_{13}^2 I_S \right\}                    \cong \Gamma_{4,1} \rtimes \text{D}_4$         & $\h8$ \\
$(\ell, \ell, 0)$ &  $12$ & $\left\{\Xi_\mu, R_{13}^2 R_{21}, R_{12}^2 I_S \right\}           \cong \Gamma_{4,1} \rtimes \text{C}_\text{2v}$ & $\h4$ \\
$(\ell, \ell, \ell)$ & $\h8$ & $\left\{\Xi_\mu, R_{13}^2 R_{12}I_S, R_{12}^2 R_{13}I_S \right\}  \cong \Gamma_{4,1} \rtimes \text{D}_3$         & $\h6$ \\
$(\ell, \ell, m)$ &  $24$ & $\left\{\Xi_\mu, R_{12}^2 I_S \right\}                            \cong \Gamma_{4,1} \rtimes \text{Z}_2$                & $\h2$ \\
$(0, \ell, m)$ &  $24$ & $\left\{\Xi_\mu, R_{23}^2 I_S \right\}                            \cong \Gamma_{4,1} \rtimes \text{Z}_2$                & $\h2$ \\
$(\ell, m, n)$ &  $48$ & $\left\{\Xi_\mu \right\}                                          \cong \Gamma_{4,1} \rtimes \{E\}$              & $\h1$ \\
\hline \hline 
\end{tabular}
\end{table}
The complete set of orbits and little groups are listed in \cref{table:orbits}.

To classify irreps of these little groups, Wigner's method must be employed again where the normal subgroup is now $N = \Gamma_{4,1}$ and $H$ are the rotation subgroups $ \text{O}_\text{h}$, $\text{D}_4$, $\text{C}_{2\rm v}$, $\text{D}_3$, $\text{Z}_2$.
From \cref{eqn:bosonTasteIrreps,eqn:bosonIrrepChageConj} there are $2^5=32$ one-dimensional bosonic irreps of $\Gamma_{4,1}$. By comparison of \cref{eqn:bosonIrrepTrans} to \cref{eqn:bosonTasteIrreps}, the spatial part of the taste irrep vector will behave similarly to the momentum orbits under the momentum little groups in \cref{table:orbits}. More specifically, labelling the irreps of $\Gamma_{4,1}$ by
\begin{align}
[\pi_\xi, \xi_{C}] = [(\xi_0, \xi_1, \xi_2, \xi_3), \xi_{C}],
\end{align}
the little group $H([\pi_\xi, \xi_{C}])$ is the group of all elements $h$ of $H$ such that $h: [(\xi_0, \xi_1, \xi_2, \xi_3),\xi_{C}]\linebreak \to [(\xi_0, \xi_1, \xi_2, \xi_3),\xi_{C}]$. And the orbits are then the unique set $\{[\pi_\xi, \xi_{C}]\}$ obtained from $h: [(\xi_0, \xi_1, \xi_2, \xi_3),\xi_{C}] \ \forall \ h \in H$. As $\xi_0$ commutes with everything (there are no $R_{0 i}$), $\xi_0 = 0$ and $\xi_0 =\pi$ always correspond to different orbits. Similarly, for $\xi_{C}=0$ and $\xi_{C}=\pi$. Because of the mod$2\pi$ associated with $\pi_\xi$ any element $h I_S$ in $H$ will be in the little group $H([\pi_\xi, \xi_{C}])$ if $h$ is. The complete list of bosonic taste orbits and taste little groups are given in the \cref{table:tasteorbit}. The character tables defining these bosonic  irreps\footnote{These (non-projective) irreps are labelled bosonic as they result in staggered bosonic irreps once combined with the Abelian irreps of $\Gamma_{4,1}$. Similarly, the non-Abelian irreps of $\Gamma_{4,1}$ combine with the projective irreps of the rotation little groups to give fermionic irreps.} are given in \cref{sec:charTablesTaste}.

\begin{table}
\newcommand{\h}{\hphantom{2}}
\centering
\caption[Taste orbits, little groups and irreps for each momentum orbit under the staggered rotation group.]{Taste orbits, little groups and irreps for each momentum orbit under the staggered group. The first column is the taste orbit, indicated by a representative element. The second column lists the orbit size.  The third and fourth columns show the taste little groups (giving their generating elements and conventional name) and order. The final column shows the irreps of these groups. Irreps labelled $A$ are 1D, $E$ are 2D and $T$ are 3D. Zero momentum irreps are also labelled by the sign under spatial inversion $\pm$. The character tables defining the irreps of the little groups are given in \cref{sec:charTablesTaste}. Also given is the total number of staggered group irreps resulting from these orbits and little group irreps and their dimensions.}\label{table:tasteorbit}
\begin{tabular}{c@{}cccc}
\hline\hline
Orbit representative & Orbit size &  Little group & Order & Little-group irreps \\
(momentum) [(taste), $C_0$] & & \multicolumn{3}{l}{total \# irreps (dimensions)} \\
\hline
\multicolumn{2}{l}{$(0,0,0)$} & \multicolumn{3}{l}{160 irreps ($32$ $1$D, $16$ $2$D, $96$ $3$D and $16$ $6$D)} \\
\begin{tabular}{l}
    $[(\xi_0,  0,  0,  0),   \xi_C]$ \\
    $[(\xi_0,\pi,\pi,\pi), \xi_C]$
\end{tabular} & 1
    & $ \left\{ R _ { i j } , I_S \right\} \cong  \text{O}_\text{h}$ & $48$ & $A_0^{\pm}$, $A_1^{\pm}$, $E_0^{\pm}$, $T_0^{\pm}$, $T_1^{\pm}$ \\[4pt]
\begin{tabular}{l}
    $[(\xi_0, \pi,0,  0),   \xi_C]$ \\
    $[(\xi_0, 0,\pi,\pi), \xi_C]$
\end{tabular} & 3
    & $\left\{ R _ { 12 } , R _ { 13 } ^ { 2 } , I_S \right\} \cong \text{D}_\text{4h}$ & $16$ & $A_0^{\pm}$, $A_1^{\pm}$, $A_2^{\pm}$, $A_3^{\pm}$, $E_0^{\pm}$ \\[6pt]
\multicolumn{2}{l}{$(0,0,\ell)$} & \multicolumn{3}{l}{112 irreps ($64$ $1$D and $48$ $2$D)} \\
\begin{tabular}{l}
    $[(\xi_0,  0,  0,\xi_3), \xi_C]$ \\
    $[(\xi_0,\pi,\pi,\xi_3), \xi_C]$
\end{tabular} & 1
    & $\left\{  R _ { 12 } , R _ { 13 } ^ { 2 } I_S \right\} \cong \text{D}_{4}$ & $\h8$ &  $A_0$, $A_1$, $A_2$, $A_3$, $E_0$ \\
\begin{tabular}{l}
    $[(\xi_0,  0,\pi,\xi_3),  \xi_C]$
\end{tabular} & 2
    & $\left\{ R _ { 12 } ^ { 2 } , R _ { 13 } ^ { 2 } I_S \right\} \cong \text{C}_\text{2v}$ & $\h4$ & $A_0$, $A_1$, $A_2$, $A_3$ \\[6pt]
\multicolumn{2}{l}{$(\ell,\ell,0)$} & \multicolumn{3}{l}{80 irreps ($64$ $1$D and $16$ $2$D)} \\
\begin{tabular}{l}
    $[(\xi_0,  0,  0,\xi_3), \xi_C]$ \\
    $[(\xi_0,\pi,\pi,\xi_3), \xi_C]$
\end{tabular} & 1
    & $\left\{  R _ { 13 } ^ { 2 } R _ { 21 } , R _ { 12 } ^ { 2 } I_S \right\} \cong \text{C}_\text{2v}$ & $\h4$ & $A_0$, $A_1$, $A_2$, $A_3$\\[4pt] 
\begin{tabular}{l}
    $[(\xi_0,  0,\pi,\xi_3),  \xi_C]$
\end{tabular} & 2
    & $\left\{  R _ { 12 } I_S \right\} \cong \text{Z}_2$ & $\h2$ & $A_0$, $A_1$ \\[6pt]
\multicolumn{2}{l}{$(\ell,\ell,\ell)$} & \multicolumn{3}{l}{40 irreps ($16$ $1$D,  $8$ $2$D and $16$ $3$D)} \\
\begin{tabular}{l}
    $[(\xi_0,  0,  0,  0), \xi_C]$ \\
    $[(\xi_0,\pi,\pi,\pi), \xi_C]$
\end{tabular} & 1
    &  $\left\{ R _ { 13 } ^ { 2 } R _ { 12 } I_S , R _ { 12 } ^ { 2 } R _ { 13 } I_S \right\} \cong \mathrm { D } _ { 3  }$ & $\h6$ &  $A_0$, $A_1$, $E_0$ \\ 
\begin{tabular}{l}
    $[(\xi_0,\pi,  0,  0), \xi_C]$ \\
    $[(\xi_0,  0,\pi,\pi), \xi_C]$
\end{tabular} & 3
    &  $ \left\{ R _ { 13 } ^ { 2 } R _ { 12 } I_S\right\} \cong \text{Z}_2$ & $\h2$ &  $A_0$, $A_1$ \\[6pt]
\multicolumn{2}{l}{$(\ell,\ell,m)$} & \multicolumn{3}{l}{40 irreps ($32$ $1$D and $8$ $2$D)} \\
\begin{tabular}{l}
    $[(\xi_0,  0,  0, \xi_3), \xi_C]$ \\
    $[(\xi_0,\pi,\pi, \xi_3), \xi_C]$
\end{tabular} & 1
    &   $ \left\{  R _ { 12 } ^ { 2 } I_S \right\} \cong  \text{Z}_2$ & $\h2$ &  $ A_0$, $ A_1$ \\
\begin{tabular}{l}
    $[(\xi_0,  0,\pi, \xi_3), \xi_C]$
\end{tabular} & 2
    &  $ \left\{ E \right\} $ & $\h1$ &  $A_0$ \\[6pt]
\multicolumn{2}{l}{$(0,\ell,m)$} & \multicolumn{3}{l}{64 irreps ($64$ $1$D)} \\
\begin{tabular}{l}
    $[(\xi_0, \xi_1, \xi_2, \xi_3), \xi_C]$
\end{tabular} & 1
    &   $ \left\{  R _ { 23 } ^ { 2 } I_S \right\} \cong Z _ { 2 }$ & $\h2$ &  $A _0$, $A_1$ \\[6pt]

\multicolumn{2}{l}{$(\ell,m,n)$} & \multicolumn{3}{l}{32 irreps ($32$ $1$D)} \\
\begin{tabular}{l}
    $[(\xi_0, \xi_1, \xi_2, \xi_3), \xi_C]$
\end{tabular} & 1
    &   $\left\{ E \right\}$ & $\h1$ & $ A_0 $ \\
\hline\hline
\end{tabular}
\end{table}

Staggered irreps are uniquely labelled by a momentum orbit representative, a taste-charge conjugation orbit representative and a rotation little group irrep. As an example, a zero momentum, taste-singlet, rotation-vector irrep, with negative staggered charge conjugation and negative staggered parity is denoted by
\begin{align}
    (0,0,0) \rtimes[(0, 0,0,0),\pi] \rtimes T_0^{-}. \label{eqn:tasteSingletVec}
\end{align}
In \cref{sec:stagOPS}, one sees this is excited by a one-link staggered spatial vector current operator. This is three-dimensional, which can be seen from the product of the dimensions of the two orbits and the rotation irrep dimension.
{\interdisplaylinepenalty=10000
\begin{align}
    &\text{Momentum orbit: } \{(0,0,0)\} -1 \text{D}, \nn\\
    &\text{Taste orbit: } \{(0, 0,0,0)\} -  1\text{D}, \nn\\
    &\text{O}_\text{h} \text{ irrep: } T_0^{-}        \ - 3\text{D}, \nn
\end{align}
}
giving a total irrep dimension of $1\times 1 \times 3$. As another example, a second irrep which also has the quantum numbers of the vector current operator is the zero momentum, taste-vector rotation-singlet irrep with positive charge conjugation and parity
\begin{align}
    (0,0,0) \rtimes [(\pi, 0, \pi, \pi), 0] \rtimes A_0^{+}, \label{eqn:tasteVecVec}
\end{align}
with the dimension breakdown
\begin{align}
    &\text{Momentum orbit: } \{(0,0,0)\}           - 1 \text{D}, \nn\\
    &\text {Taste orbit: } \{(\pi, 0, \pi, \pi),(\pi, \pi, 0, \pi),(\pi, \pi, \pi, 0)\}-3\text{D}, \nn\\
    &\text{D}_\text{4h} \text{ irrep: }  A_0^{+}                   - 1\text{D}. \nn
\end{align}
This also has the total dimension of 3, but it is now coming from the taste orbit rather than the rotation irrep. As a final example, a taste-vector, rotation-singlet irrep with one component of momentum, and negative charge conjugation,\footnote{Parity is not a good quantum number for states in flight.}
\begin{align}
    (0,0,\ell) \rtimes [( \pi, 0, \pi, \pi),\pi] \rtimes A_2, \label{eqn:tasteVecPion}
\end{align}
with the following breakdown,
\begin{align}
    &\text {Momentum orbit: } \{(0,0,\ell)+5 \text { perms.}\}-6\text{D}, \nn\\
    &\text {Taste orbit: }\{(\pi, 0, \pi, \pi),(\pi, \pi, 0, \pi)\} -2\text{D}, \nn\\
    &\text{C}_\text{2v}  \text { irrep: }  A_2     \        
\ -1\text{D}, 
\end{align}
giving a total dimension of 12. This irrep corresponds to a pseudo-scalar meson in flight in the continuum, i.e., a pion if one considers light-quark flavors.

In this work, we employ operators which excite the taste-singlet vector meson \cref{eqn:tasteSingletVec}, for the reasons discussed in \cref{sec:opconstruction}.  For each continuum bosonic state, there are $4 \times 4 = 16$ staggered states which have all the same quantum numbers except for the taste. The pseudoscalar irrep, \cref{eqn:tasteVecPion}, is one of the multiple tastes of pion we study here, the full set is given in \cref{sec:pionIrreps}. Depending on the taste and the momentum direction, these states can have degenerate or non-degenerate energies. In this work, the multi-particle states are built from single-particle states with momentum. Hence, understanding the relationship between staggered states at rest and states in flight is vital.

\subsubsection{Non-zero momentum decomposition}\label{sec:nonzeromom}

In order to decompose a staggered irrep at zero momentum to irrep(s) at non-zero momentum, one needs to
\begin{itemize}
\item Decompose the zero momentum taste orbit into the non-zero momentum taste orbit(s).
\item Restrict the zero momentum little group irrep to the corresponding non-zero momentum taste little group(s) and determine what irrep(s) are contained in the (now) reducible representation.
\end{itemize}
To illustrate this, consider giving momentum in the $z$-direction to the following zero momentum state
\begin{align}
    (0,0,0) \rtimes [(\pi, \pi,0, 0),\pi] \rtimes A_1^{+}-3 \text{D} ; \quad \text{O}_\text{h}\left(\vec{p}, [\pi_\xi, \xi_C]\right)= \text{D}_{4h} .\label{eqn:zeroMomIrrepEx}
\end{align}
The $\{(0,0,\ell)\}$ momentum little group only mixes  $\xi_1, \xi_2$ in orbits, so $\xi_3$ becomes independent. Hence, the $(0,0,0)$ momentum taste-orbit $\{(\pi, 0,0, \pi),(\pi, 0, \pi, 0),(\pi, \pi, 0,0)\}$ splits into two parts. A two-dimensional orbit $\{(\pi,0, \pi, 0),(\pi,\pi, 0, 0)\}$ with little group $\text{O}_\text{h}\left(\vec{p}, [\pi_\xi, \xi_C]\right)\cong\text{C}_\text{2v}$ and a one-dimensional orbit $\{(\pi,0,0, \pi)\}$ with little group $\text{O}_\text{h}\left(\vec{p}, [\pi_\xi, \xi_C]\right)\cong\text{D}_4$. One then restricts the original little group, $\text{D}_\text{4h}$, to the two new little groups giving the following irrep decomposition
\begin{align}
    \left.A_2^{+}\right|_{\text{D}_{4 \text{h}} \rightarrow \text{C}_\text{2v}} &= A_3,\\
    \left.A_2^{+}\right|_{\text{D}_{4 \text{h}} \rightarrow \text{D}_{4}}       &= A_1 .
\end{align}
This restriction is performed by considering the characters of the conjugacy classes which remain after removing the elements not contained in the respective subgroups. The standard character decomposition \cite{Cornwell} is employed to obtain the irreps of the subgroups. In both cases here, there is only one irrep contained, $A_3$ and $A_1$ respectively. Hence, there are now two $(0,0,\ell)$ momentum irreps from the original single zero-momentum irrep
\begin{align}
    &(0,0,\ell) \rtimes[(\pi, \pi, 0,0),\pi] \rtimes A_3-6\times 2\text{D}, \\
    &(0,0,\ell) \rtimes[(\pi, 0,0, \pi),\pi] \rtimes A_1-6\times 1\text{D}. \label{eqn:nonzeroMomIrrepEx}
\end{align}
This splitting of the taste-orbit into separate irreps is observed in the pion spectrum computed in \cref{sec:gevp}.

\subsection{Staggered operators}\label{sec:stagOPS}

Following the form of the staggered action in the hypercubic representation \cite{Gliozzi:1982ib,Kluberg-Stern:1983lmr}, one formally writes a staggered quark operator in the hypercubic representation (spin-taste basis) as
\begin{align}
    \mathcal{O}^{\Gamma_S \otimes \Gamma_T}(h)= \bar{q}(h) \Gamma_S \otimes \Gamma_T q(h). \label{eqn:stBasisOP}
\end{align}
This operator has a spin quantum number from $\Gamma_S$ and a taste quantum number from $\Gamma_T$.\footnote{The gauge links are left out for simplicity.} Numerical simulations, however, are typically performed in the representation of \cref{eqn:stagFermionActionReduced}. Recasting the operator in this form results in `phase-shift' operators, 
\begin{align}
    \mathcal{O}_{(\delta)}^{\varphi}(\vec{p}, t)=\sum_{\vec{n}} e^{-i a\vec{p} \cdot \vec{n}} \bar{\chi}(n) \chi(n+\delta) \varphi(n). \label{eqn:phaseShiftOp}
\end{align}
where the operator has now also been given a momentum $p$. The unbarred field is shifted by a spatial offset $\delta$ and there is an associated spacetime dependent staggered phase $\varphi(n)$. The relationship between these two representations is given by
\begin{align}
    \varphi(n)&=\frac{1}{4} \textrm{tr}\left(\Gamma_T^{\dagger} \Omega(n)^{\dagger} \Gamma_S \Omega(n+t+s)\right),
    \label{eqn:phiTrace} \\
    &\delta=t+s \bmod 2, \label{eqn:OpShift}
\end{align}
where $\Omega(n)$ is defined in \cref{eqn:motoMatrix} and $s$ and $t$ are four vectors which specify the spin and taste gamma structure. In this work, we use the $\Gamma_S \otimes \Gamma_T$ labelling to denote the operators, but the phase-shift form is used in the computation. Only symmetric-non-time-shift operators are considered,
\begin{align}
    \mathcal{O}_{( \pm \delta)}^{\varphi}(\vec{p},t)= \frac{1}{N_{\textrm{sym}}}\sum_{\pm \delta_i}  \mathcal{O}_{(\delta)}^{\varphi}(\vec{p}, t). \label{eqn:opSym}
\end{align}
With an average over forward and backward directions for each component of $\delta$ performed. Symmetrizing in this way guarantees the operators have well-defined parity and, for flavor singlets, charge conjugation.

The phase-shift operators can be straight-forwardly related to the rows of the irreps from above by acting on them with the staggered symmetry transformations and reading off the quantum numbers:
\begin{align}
    \Xi_i: \mathcal{O}_{( \pm \delta)}^{\varphi}(\vec{p}) &\to \zeta_i(\delta) \varphi(\hat\imath) \mathcal{O}_{( \pm \delta)}^{\varphi}(\vec{p}), \label{eqn:tasteOP}\\
    I_S: \mathcal{O}_{( \pm \delta)}^{\varphi}(\vec{p}) &\to (-1)^{\sum_i \delta_i} \mathcal{O}_{(\mp \delta)}^{\varphi}(-\vec{p}), \label{eqn:parityOp}\\
    {R}_{i j}: \mathcal{O}_{( \pm \delta)}^{\varphi}(\vec{p}) &\to  \mathcal{O}_{\left( \pm \delta^{\prime}\right)}^{\varphi^{\prime}}\left(\vec{p}'\right) \label{eqn:rotOP},\\
    C_0: \mathcal{O}_{( \pm \delta)}^{\varphi}(\vec{p}) &\to e^{i \vec{p} \cdot \vec{\delta}}(-1)^{\sum_i \delta_i} \varphi(\delta) \mathcal{O}_{(\mp \delta)}^{\varphi}(\vec{p}), \label{eqn:chargeOp}
\end{align}
where $\varphi'$ ($\vec{p}'$) is obtained from $\varphi$ ($\vec{p}$) via the given rotation. The momentum is specified through \cref{eqn:phaseShiftOp}. As the sum in \cref{eqn:phaseShiftOp} does not include~$t$, the operator is local in time and hence can excite irreps with any energy. Similarly, this results in $\xi_0$ not being fixed, hence the operators in \cref{eqn:phaseShiftOp} excite states with $\xi_0=0$ and $\xi_0=\pi$---without a full construction of the transfer matrix, operators of definite $\xi_0$ cannot be constructed. This is the source of the well known issue with local-time staggered operators, whereby states with both positive and negative continuum parity are excited, resulting in temporal oscillations in staggered correlation functions. Under the action of rotation group, \cref{eqn:rotOP}, the transformed operators $\{(\varphi^{\prime}_i,\delta^{\prime}_i)\}$  form a basis of the taste little group. Constructing the representation from this basis then allows one to determine the rotation irrep. 

Operators corresponding to the examples considered in \cref{eqn:tasteSingletVec,eqn:tasteVecVec,eqn:tasteVecPion} are given by
\begin{itemize}
    \item taste singlet, spin vector, $(0,0,0) \rtimes[(0, 0,0,0),\pi] \rtimes T_0^{-}$
    \begin{align}
        \mathcal{O}^{\gamma_i \otimes \mathrm{1}}(\vec{p}=0)  \;:\;  \varphi(n)=\eta_i(n),\delta_j=\delta_{j,i}. \label{eqn:onelinkOperator}
    \end{align}
    \item taste vector, spin vector, $(0,0,0) \rtimes [(\pi, 0, \pi, \pi), 0] \rtimes A_0^{+}$
    \begin{align}
        \mathcal{O}^{\gamma_i \otimes \gamma_i}(\vec{p}=0)  \;:\;  \varphi(n)=( - 1 ) ^ { \sum _ { \rho \neq i } n _ { \rho } },\delta=(0,0,0,0).
    \end{align}
    \item taste vector, spin pseudo-scalar, $(0,0,\ell) \rtimes [( \pi, 0, \pi, \pi),\pi] \rtimes A_2$
    \begin{align}
        \mathcal{O}^{\gamma_5 \gamma_0 \otimes \gamma_i}(\vec{p}=[0,0,l])  \;:\;  \varphi(n),\ \delta = \begin{cases}
        (-1)^{n_0 + n_3 }, \hspace*{1.625em} (0,0,1,1) & \text{if } i=1\\
        (-1)^{n_0 + n_2 + n_3 }, \; (0,1,0,1)  & \text{if } i=2
    \end{cases}. \label{eqn:pionTimeLinkTreat}
    \end{align}
\end{itemize}
As mentioned, the first operator corresponds to taste-singlet vector meson; it contains a one component shift. To preserve gauge invariance, these operators have gauge links connecting fields on different sites. Hence, this operator is referred to as a `1-link' operator for the single link connecting $\bar \chi$ and $\chi$.
The second operator is local and is named so.
For the last irrep, the taste vector pseudo-scalar, the operator $\mathcal{O}^{\gamma_5 \otimes \gamma_i}$ also excites the same states but contains a shift in the time direction, so we do not use it.

For the non-zero momentum example irreps considered above, \cref{eqn:zeroMomIrrepEx,eqn:nonzeroMomIrrepEx}, we have
\begin{itemize}
    \item $(0,0,0) \rtimes[(\pi,  \pi,0,0), \pi] \rtimes A_1^{+}$
    \begin{align}
        \mathcal{O}^{\gamma_5 \otimes \gamma_i}(\vec{p}=0) \;:\;  \varphi(n),\ \delta = \begin{cases}
        (-1)^{n_0 + n_1 + n_2}, \ (0,0,1,1) & \text{if } i=1\\
        (-1)^{n_0 + n_1 },     (0,1,0,1) & \text{if } i=2\\
        (-1)^{n_0 + n_1 + n_3 }, \ (0,1,1,0)  & \text{if } i=3
    \end{cases}.
    \end{align}
    \item $(0,0,\ell) \rtimes[(\pi, \pi, 0,0),\pi] \rtimes A_3$
    \begin{align}
        \mathcal{O}^{\gamma_5 \otimes \gamma_i}(\vec{p}=[0,0,l]) \;:\;  \varphi(n),\ \delta = \begin{cases}
        (-1)^{n_0 + n_1 + n_2}, (0,0,1,1) & \text{if } i=1\\
        (-1)^{n_0 + n_1 },     (0,1,0,1) & \text{if } i=2
        \end{cases}.
    \end{align}
    \item $(0,0,\ell) \rtimes[(\pi, 0,0, \pi),\pi] \rtimes A_1$
    \begin{align}
        \mathcal{O}^{\gamma_5 \otimes \gamma_3}(\vec{p}=[0,0,l]) \;:\;  \varphi(n)=(-1)^{n_0 + n_1 + n_3},\delta = (0,1,1,0).
    \end{align}
\end{itemize}
Here, the first operator excites the rows of an irrep which then splits into two non-zero momentum irreps which are excited by the second and third operators.

\subsection{Connecting staggered observables to the continuum}\label{sec:continuumStag}

There are two considerations when connecting an observable computed with staggered quarks to a continuum observable. The first is the subduction from the states in the continuum to the states of the staggered lattice group.
As mentioned in \cref{sec:groupStruct}, for staggered quarks a $\text{SU}(4)_T$ symmetry emerges in the continuum, meaning all states with the same quantum numbers, but different tastes, are degenerate and have the same properties as the same physical state. This degeneracy is lifted at finite lattice spacing, hence there is a non-trivial spectrum of states for each physical state. A central part of this work is understanding this taste-split spectrum as it pertains to two-pion states. The second consideration is the contribution of the four quark-tastes to the staggered fermion determinant. This is resolved by so-called (fourth-) rooting, the effect of this on the observables computed in this work is discussed in \cref{sec:rooting}.

\subsubsection{Continuum decomposition}\label{sec:contDecomp}

The decomposition from the continuum symmetry group irreps to the lattice irreps discussed above is laid out in Ref.~\cite{sharpe.1}. However, there are some errors in that work, so we reproduce the full discussion here with corrections. Ignoring flavor, subduction from the continuum group to the lattice group is given by the following map:
\begin{center}
\begin{tikzcd}
\begin{array} {c}
\textrm{Continuum Group}\\
\text{SU}(4)_T \times \text{SU}(2)_S \times P \times C
\end{array} \arrow{d} \\
\left[ \text{SU}(2)_L  \times \text{SU}(2)_S \right] \times \left[\text{SU}(2)_R \times P \times C\right] \arrow{d}\\
\begin{array}{c}
\textrm{Staggered Rest Frame Group}\\
 \displaystyle\frac{(\text{SW}_4 \times \Gamma_{2,2})}{(-E \times -E)} \\
\{R_{ij}, R_{j k} \Xi_{k j}\} \times \{C_0, \Xi_0, \Xi_{123}, C_0\Xi_0I_S\}
\end{array} \arrow{r} &
\begin{array}{c}
\textrm{Staggered Group w/o Translations}\\
\Gamma_{4,1} \rtimes \text{O}_\text{h}  \\
\{\Xi_\mu, C_0 \} \rtimes \{R_{ij}, I_S\}
\end{array} \arrow{d}\\
& \begin{array}{c}
\textrm{Full Staggered Group: $G_{T_0}$}\\
(Z_{N_s/2}^3 \times \Gamma_{4,1}) \rtimes \text{O}_\text{h} \\
\end{array} ,
\end{tikzcd}
\end{center}
where the meaning and role of $\text{SW}_4\times\Gamma_{2,2}$ is explained below.
The symmetry group in the continuum for a flavorless state is $\text{SU}(4)_T \times \text{SU}(2)_S \times P \times C$. \footnote{Apart from the inconsequential $U_V(1)$ that corresponds to baryon number conservation.}
Here, $P$ and $C$ are continuum parity and charge conjugation, which are distinct but related to spatial inversion $I_S$ and staggered charge conjugation $C_0$; $\text{SU}(2)_S$ is the standard continuum spin group with integer and half-integer spin representations; $\text{SU}(4)_T$ is the continuum symmetry group of four degenerate tastes.We have two bosonic representations of $\text{SU}(4)_T$ labelled $\mathbf{0}$ and $\mathbf{15}$.
The $\text{SU}(4)$ singlet $\mathbf{0}$ is one dimensional and decomposes to the taste-singlet, $\pi_\xi = (0,0,0,0)$, while the $\text{SU}(4)$ fundamental irrep $\mathbf{15}$ is fifteen dimensional and decomposes to all other tastes.

The 15 taste transformations $\Xi_\mu$, $\Xi_5$, $\Xi_\mu \Xi_5$, $\Xi_{\mu \nu}$ are generators of the continuum $\text{SU}(4)_T$ but also exist as a subgroup inside it as is the case for the doubling symmetry, \cref{eqn:doublegens}, which has the equivalent group structure. By examining the action of these transformations in momentum space \cite{Golterman:1984cy}, one finds that  $\Xi_{i j}$ lie in a $\text{SU}(2)_L$ subgroup while $\Xi_0$ and $\Xi_{123}\equiv\Xi_1\Xi_2\Xi_3$ lie in a commuting $\text{SU}(2)_R$ subgroup, i.e. $\text{SU}(2)_L \times \text{SU}(2)_R \subset \text{SU}(4)_T$. The bosonic irrep decomposition for this step is given by,
\begin{align}
    \text{SU}(4) &\to \text{SU}(2) \times \text{SU}(2),\\
    \mathbf{0} &\to (0, 0),\\
    \mathbf{15} &\to  (0,1) \oplus (1,0)  \oplus  (1,1) ,
\end{align}
where $0$ and $1$ on the RHS are the  familiar one-dimensional `spin 0' and three-dimensional `spin 1' irreps of $\text{SU}(2)$.

For the second step of the decomposition, parity and charge conjugation correspond to their continuum counterparts, with an additional taste transformation.
The relationship between spatial inversion and parity is straightforward to read off from \cref{eqn:stagParity}
\begin{align}
I_S = P \Xi_0. \label{eqn:paritySubDuce}
\end{align}
Charge conjugation is more complicated, but the process of extracting it is described Ref.~\cite{sharpe.1}. It amounts to the following procedure, first count the zeros, $\#0\text{s}$, in the taste irrep orbit representative vector $\pi_\xi$. Then the relationship between the continuum charge conjugation $C$ and lattice charge conjugation $C_0$ is
\begin{align}
        C_0 =  \begin{cases}
            C,  & \quad \text{if } \#0\text{s} = 0,3,4  \\
            -C, & \quad \text{if } \#0\text{s} = 1,2
            \end{cases}. \label{eqn:chargeSubDuce}
\end{align}

Lattice rotations correspond to simultaneous rotations of staggered taste and spin, \cref{eqn:rotOP}, and sit inside the diagonal subgroup of $\text{SU}(2)_L \times \text{SU}(2)_S$, which is subduced into the group generated by $\{\Xi_{k j}, R_{ij} \}$. Rewriting this generating set as
\begin{align}
    &\{\tilde{R}_{4 i}, R_{ij}\} ,\label{eqn:SW4Gens}\\
    &\tilde{R}_{4 i} \equiv R_{j k} \Xi_{k j},
\end{align} 
gives a group isomorphic to $\text{SW}_4$, the symmetry group of the hypercube, which appears in the map shown above. Below, $\text{SW}_4$ is a useful tool for decomposing continuum states into staggered irreps.

The group $\text{SU}(2)_{R} \times P \times C$ is subduced into the group generated by the remaining staggered symmetries $\{\Xi_0, \Xi_{123}, I_S, C_0\}$.
Using \cref{eqn:paritySubDuce,eqn:chargeSubDuce},  where again, rewriting generators
\begin{align}
 &\{\Xi_0, \Xi_{123}, C_0\Xi_0I_S, C_0 \}, \label{eqn:gamma22Gens}
 \end{align}
gives the defining set of mutually anti-commuting generators of $\Gamma_{2,2}$,
\begin{align}
    \Xi_0^2=(C_0\Xi_0I_S)^2=-C_0^2=-\Xi_{123}^2=1.
\end{align}
The group,
\begin{align}
     (\text{SW}_4 \times \Gamma_{2,2})/(-E \times -E),
\end{align}
is the staggered rest frame group and is isomorphic to the group $\Gamma_{4,1}\rtimes \text{O}_\text{h}$ in \cref{eqn:stagGroupStruc}.
The quotient factor, $(-E \times -E)$,\footnote{$E$ denotes the identity element of the respective group.} ensures only bosonic-type and fermionic-type irreps exist in the direct product, i.e., Abelian irreps of $\Gamma_{2,2}$ are combined with non-projective irreps of $\text{SW}_4$, while the faithful four dimensional irrep of $\Gamma_{2,2}$ is combined with the projective irreps of $\text{SW}_4$.
The irreps and characters of $\text{SW}_4$ are given in Refs.~\cite{Mandula:1983ut,sharpe.1}, however, there are some errors in Table~6 of Ref.~\cite{sharpe.1}. In Ref.~\cite{Mandula:1983ut}, Table~3\,2a for $\text{SW}_4$ is correct, even though it is subduced from Table~3\,3b, which interchanges the characters for $(1,0)$ and $(0,1)$. The bosonic irrep part of the character table is reproduced in \cref{tab:charTableSWBos} with the classes labelled by class representatives corresponding to \cref{eqn:SW4Gens}.
In the case of bosonic (Abelian) irreps of $\Gamma_{2,2}$, the homomorphism $\Gamma_{2,2} \to \text{Z}_2^4$ furnishes 16 one-dimensional irreps.
These irreps are labeled by $\pi_\Gamma=(\xi_0, \xi_{123}, \xi_{I_{S}}, \xi_C)$, taking values $0$ or $\pi$. The characters are straightforward and are given in \cref{eqn:gamma22Chars}.

The subduction from $\text{SU}(2)_L \times \text{SU}(2)_S \cong O(4) \to \text{SW}_4$ is described in Ref.~\cite{Mandula:1983ut}.\footnote{The ordering of $\text{SU}(2)_L \times \text{SU}(2)_S$ is the correct one given the definitions of the $(0,1)$ and $(1,0)$ irreps in Refs.~\cite{sharpe.1,Mandula:1983ut}, but Ref.~\cite{sharpe.1} subsequently flips the order when carrying out its subduction analysis.}
One restricts $\text{SU}(2)_L \times \text{SU}(2)_S$ to $\text{SW}_4$ using the natural mapping with $O(4)$. It is then straightforward to decompose the representations of the restricted group using the standard character vector algebra (the same process as in \cref{sec:nonzeromom}). Explicit results for spin $0, 1, 2$ are given here,
\begin{align}
    (0,0) &\to \text{\tiny\yng(4)\normalsize}, \\
    (1,0) &\to (1,0), \\
    (0,1) &\to (0,1), \\
    (1,1) &\to \text{\tiny\yng(3,1)\normalsize}\oplus\mathbf{6}, \\
    (0,2) &\to \text{\tiny\yng(2,2)\normalsize} \oplus\overline{(0,1)},\\
    (1,2) &\to \text{\tiny\tiny\yng(2,1,1)\normalsize} \oplus\mathbf{6} \oplus(1,0) \oplus\overline{(1,0)}.
\end{align}
The irrep {\tiny\yng(1,1,1,1)} in \cref{tab:charTableSWBos} appears first in the spin 3 subduction.

For $\text{SU}(2)_R \times P \times C \to \Gamma_{2,2}$, one just needs to subduce $\text{SU}(2)_R \to \{\Xi_0, \Xi_{123}\} \cong\text{D}_4$, which  is straightforward for the bosonic case via the homomorphism $\text{D}_4\to \text{Z}_2\times \text{Z}_2$.
$C_0$ and $P$ follow from \cref{eqn:paritySubDuce,eqn:chargeSubDuce}. The mapping is given in Ref.~\cite{sharpe.1},
\begin{align}
    \text{SU}(2)_R &\to \text{Z}_2 \times \text{Z}_2 , \\
    0 &\to (0,0) , \\
    1 &\to (\pi,0) \oplus (0,\pi) \oplus (\pi,\pi) ,
\end{align}
where the irreps and characters for the group $\text{Z}_2 \times \text{Z}_2$ are given in \cref{tab:charGammaZ2}. This completes the second step of the continuum decomposition map.
The isomorphism between the rest frame group and \cref{eqn:stagGroupStruc} without translations is straightforward, as they contain the same generating elements, just rearranged. One makes the identification between the classes and matches the character vectors of the irreps. There are $17$ irreps/classes in $\Gamma_{2,2}$ and $13$ irreps/classes in $\text{SW}_4$ giving a total of $221$\footnote{It is a coincidence that the number of irreps/classes coincides with the recurrence of periodical cicadas.} in the direct product, however 58 of them are removed through the $\text{Z}_2 \times \text{Z}_2$ quotient giving 163 classes corresponding to the 160 zero momentum bosonic irreps in \cref{table:tasteorbit} and the 3 fermionic irreps which are not considered here. The similarity between the irreps are given in \cref{tab:restFrameIso}.

\begin{table}[t]
\centering
\caption[Bosonic irreps for the isomorphic groups $(\text{SW}_4 \times \Gamma_{2,2})/(-E \times -E)$ and $\Gamma_{4,1} \rtimes \text{O}_\text{h}$]{Irreps for the isomorphic groups $(\text{SW}_4 \times \Gamma_{2,2})/(-E \times -E)$ and $\Gamma_{4,1} \rtimes \text{O}_\text{h}$. The first five rows are the irreps corresponding to row one in \cref{table:tasteorbit}, the next five are the irreps in row two of that table.} \label{tab:restFrameIso}
\begin{tabular}{ccc}
\hline \hline
    $(\text{SW}_4 \times \Gamma_{2,2})/(-E \times -E) $  &                    $\Gamma_{4,1} \rtimes \text{O}_\text{h}  $ &  Dimension \\
\hline
    \multirow{2}{*}{\tiny\yng(4)\normalsize$\otimes (\xi_0, \xi_{123}, \xi_{I_{S}}, \xi_C)$} &  \multirow{2}{*}{$[(\xi_0, \xi_1, \xi_2, \xi_3), \xi_C] \rtimes A_0^{e^{\xi_{I_{S}}}}$} &           \multirow{2}{*}{1} \\
      & & \\
    \multirow{2}{*}{\tiny\yng(1,1,1,1)\normalsize$ \otimes (\xi_0, \xi_{123}, \xi_{I_{S}}, \xi_C)$} &  \multirow{2}{*}{$[(\xi_0, \xi_1, \xi_2, \xi_3), \xi_C] \rtimes A_1^{e^{\xi_{I_{S}}}}$} &           \multirow{2}{*}{1} \\
& & \\
     \multirow{2}{*}{\tiny\yng(2,2)\normalsize$ \otimes (\xi_0, \xi_{123}, \xi_{I_{S}}, \xi_C)$} &  \multirow{2}{*}{$[(\xi_0, \xi_1, \xi_2, \xi_3), \xi_C] \rtimes E_0^{e^{\xi_{I_{S}}}}$} &           \multirow{2}{*}{2} \\
    & & \\
              \multirow{2}{*}{$(0,1) \otimes (\xi_0, \xi_{123}, \xi_{I_{S}}, \xi_C)$} &  \multirow{2}{*}{$[(\xi_0, \xi_1, \xi_2, \xi_3), \xi_C] \rtimes T_0^{e^{\xi_{I_{S}}}}$} &           \multirow{2}{*}{3} \\
             & & \\
   \multirow{2}{*}{$\overline{(0,1)} \otimes (\xi_0, \xi_{123}, \xi_{I_{S}}, \xi_C)$} &  \multirow{2}{*}{$[(\xi_0, \xi_1, \xi_2, \xi_3), \xi_C] \rtimes T_1^{e^{\xi_{I_{S}}}}$ } &        \multirow{2}{*}{3} \\
  & & \\
     \multirow{2}{*}{\tiny\yng(3,1)\normalsize$ \otimes (\xi_0, \xi_{123}, \xi_{I_{S}}, \xi_C)$} &  \multirow{2}{*}{$[(\xi_0, \xi_1, \xi_2, \xi_3), \xi_C] \rtimes A_0^{e^{\xi_{I_{S}}}}$} &          \multirow{2}{*}{3} \\
    & & \\
   \multirow{2}{*}{\tiny\yng(2,1,1) \normalsize$ \otimes (\xi_0, \xi_{123}, \xi_{I_{S}}, \xi_C)$} &  \multirow{2}{*}{$[(\xi_0, \xi_1, \xi_2, \xi_3), \xi_C] \rtimes A_1^{e^{\xi_{I_{S}}}}$} &          \multirow{2}{*}{ 3} \\
  & & \\
              \multirow{2}{*}{$(1,0) \otimes (\xi_0, \xi_{123}, \xi_{I_{S}}, \xi_C)$} &  \multirow{2}{*}{$[(\xi_0, \xi_1, \xi_2, \xi_3), \xi_C] \rtimes A_2^{e^{\xi_{I_{S}}}}$} &           \multirow{2}{*}{3} \\
             & & \\
   \multirow{2}{*}{$\overline{(1,0)} \otimes (\xi_0, \xi_{123}, \xi_{I_{S}}, \xi_C)$} &  \multirow{2}{*}{$[(\xi_0, \xi_1, \xi_2, \xi_3), \xi_C] \rtimes A_3^{e^{\xi_{I_{S}}}}$} &           \multirow{2}{*}{3} \\
  & & \\
         \multirow{2}{*}{$\mathbf{6} \otimes (\xi_0, \xi_{123}, \xi_{I_{S}, }\xi_C)$} &  \multirow{2}{*}{$[(\xi_0, \xi_1, \xi_2, \xi_3), \xi_C] \rtimes E_0^{e^{\xi_{I_{S}}}}$} &          \multirow{2}{*}{6} \\
        & & \\
\hline \hline
\end{tabular}
\end{table}

With this similarity, the first three steps of the decomposition are completed. The final step just follows what is described in \cref{sec:nonzeromom}. To illustrate the full procedure, the decomposition of the spin-zero meson with $P=-1$ and $C=1$\footnote{A pion if the correct isospin is chosen.} and momentum in the $z$ direction is performed. The decomposition for other momenta is given in \cref{sec:pionIrreps}. Also given in \cref{sec:rhoIrreps} is the decomposition of the $\rho$ meson, including the example from Ref.~\cite{sharpe.1} which is repeated but with a corrected decomposition. The continuum spin-zero state can be in the taste-singlet irrep $\mathbf{0}$ or the taste-fifteen irrep $\mathbf{15}$, hence we have
\begin{align}
(\mathbf{0}, 0) &\to (0, 0) \otimes  0 \to \text{\tiny\yng(4)\normalsize} \otimes(0, 0),\\
(\mathbf{15}, 0) &\to (1, 0) \otimes  1 \oplus (1, 0) \otimes  0\ \oplus\ (0, 0) \otimes 1 \to \nonumber \\
&\hphantom{IM} (1,0) \otimes (\pi, 0) \oplus (1,0) \otimes (0, \pi) \oplus (1,0) \otimes (\pi, \pi) \oplus (1,0) \otimes (0, 0)\ \oplus \nn\\
&\hphantom{IM} \mbox{\text{\tiny\yng(4)\normalsize}} \otimes (\pi, 0) \oplus \text{\tiny\yng(4)\normalsize} \otimes (0, \pi) \oplus \text{\tiny\yng(4)\normalsize} \otimes (\pi, \pi).
\end{align}
Proceeding with the mapping from \cref{tab:restFrameIso}, using \cref{eqn:paritySubDuce,eqn:chargeSubDuce} with $P=-1$ and $C=1$ and the characters from \cref{tab:charGammaZ2}
\begin{align}
\text{\tiny\yng(4)\normalsize} \otimes (0, 0) &\sim (0, 0, 0) \rtimes [(0, 0, 0, 0), 0]  \rtimes   A_0^{-} \;:\; \mathcal{O}^{ \gamma_5 \otimes 1}(0, 0, 0),\\
\text{\tiny\yng(4)\normalsize} \otimes (\pi, 0) &\sim (0, 0, 0) \rtimes  [(\pi, 0, 0, 0), 0]   \rtimes A_0^{+}  \;:\; \mathcal{O}^{ \gamma_5 \otimes  \gamma_5\gamma_0}(0, 0, 0),\\
\text{\tiny\yng(4)\normalsize} \otimes (0, \pi) &\sim (0, 0, 0) \rtimes  [(0, \pi, \pi, \pi), \pi]  \rtimes   A_0^{-}  \;:\; \mathcal{O}^{ \gamma_5  \otimes  \gamma_0}(0, 0, 0),\\
\text{\tiny\yng(4)\normalsize} \otimes (\pi, \pi) &\sim (0, 0, 0)  \rtimes  [(\pi, \pi, \pi, \pi), 0]    \rtimes    A_0^{+}    \;:\; \mathcal{O}^{ \gamma_5  \otimes  \gamma_5}(0, 0, 0),\\
(1,0) \otimes (\pi, 0) &\sim (0, 0, 0)  \rtimes  [(\pi, \pi, \pi, 0), \pi]    \rtimes    A_2^{+}    \;:\; \mathcal{O}^{ \gamma_5  \otimes  \gamma_i}(0, 0, 0),\\
(1,0) \otimes (0, \pi) &\sim (0, 0, 0)  \rtimes  [(0, 0, 0, \pi), 0]    \rtimes    A_2^{-}    \;:\; \mathcal{O}^{ \gamma_5  \otimes  \gamma_5\gamma_i}(0, 0, 0),\\
(1,0) \otimes (\pi, \pi) &\sim (0, 0, 0)  \rtimes  [(\pi, 0, 0, \pi), \pi]    \rtimes    A_2^{+}    \;:\; \mathcal{O}^{ \gamma_5  \otimes  \gamma_i\gamma_0}(0, 0, 0),\\
(1,0) \otimes (0, 0) &\sim (0, 0, 0)  \rtimes  [(0, \pi, \pi, 0), \pi]    \rtimes    A_2^{-}    \;:\; \mathcal{O}^{ \gamma_5  \otimes  \gamma_i\gamma_j}(0, 0, 0),
\end{align}
where corresponding operators, using \cref{sec:stagOPS}, are also given. The first four irreps are one-dimensional, the last four are three-dimensional, giving $4+12=16$ `pion' states, as expected. For non-zero momentum in the continuum, one decomposes the zero momentum lattice irreps. For momentum  $(0, 0, p_z)$ one has,
\begin{align}
(0, 0, 0) \rtimes [(0, 0, 0, 0), 0]   \rtimes   A_0^{-}  &\to \begin{cases}(0, 0, 1) \rtimes [(0, 0, 0, 0), 0]   \rtimes   A_1 :  \mathcal{O}^{ \gamma_5 \otimes 1}(0, 0, 1)\end{cases}\\
(0, 0, 0) \rtimes [(\pi, 0, 0, 0), 0]   \rtimes   A_0^{+}  &\to \begin{cases}(0, 0, 1) \rtimes [(\pi, 0, 0, 0), 0]   \rtimes   A_0 :  \mathcal{O}^{ \gamma_5 \otimes \gamma_5\gamma_0}(0, 0, 1)\end{cases}\\
(0, 0, 0) \rtimes [(0, \pi, \pi, \pi), \pi]   \rtimes   A_0^{-}  &\to \begin{cases}(0, 0, 1) \rtimes [(0, \pi, \pi, \pi), \pi]   \rtimes   A_1 :  \mathcal{O}^{ \gamma_5 \otimes \gamma_0}(0, 0, 1)\end{cases}\\
(0, 0, 0) \rtimes [(\pi, \pi, \pi, \pi), 0]   \rtimes   A_0^{+}  &\to \begin{cases}(0, 0, 1) \rtimes [(\pi, \pi, \pi, \pi), 0]   \rtimes   A_0 :  \mathcal{O}^{ \gamma_5 \otimes \gamma_5}(0, 0, 1)\end{cases}\\
(0, 0, 0) \rtimes [(\pi, \pi, \pi, 0), \pi]   \rtimes   A_2^{+}  &\to \begin{cases}(0, 0, 1) \rtimes [(\pi, 0, \pi, \pi), \pi]   \rtimes   A_2 :  \mathcal{O}^{ \gamma_5 \otimes \gamma_{i \neq 3}}(0, 0, 1) \\ (0, 0, 1) \rtimes [(\pi, \pi, \pi, 0), \pi]   \rtimes   A_1 :  \mathcal{O}^{ \gamma_5 \otimes \gamma_3}(0, 0, 1)\end{cases}\\
(0, 0, 0) \rtimes [(0, 0, 0, \pi), 0]   \rtimes   A_2^{-}  &\to \begin{cases}(0, 0, 1) \rtimes [(0, 0, 0, \pi), 0]   \rtimes   A_0 :  \mathcal{O}^{ \gamma_5 \otimes \gamma_5\gamma_3}(0, 0, 1) \\ (0, 0, 1) \rtimes [(0, 0, \pi, 0), 0]   \rtimes   A_2 :  \mathcal{O}^{ \gamma_5 \otimes \gamma_5\gamma_{i \neq 3}}(0, 0, 1)\end{cases}\\
(0, 0, 0) \rtimes [(\pi, 0, 0, \pi), \pi]   \rtimes   A_2^{+}  &\to \begin{cases}(0, 0, 1) \rtimes [(\pi, 0, 0, \pi), \pi]   \rtimes   A_1 :  \mathcal{O}^{ \gamma_5 \otimes \gamma_3\gamma_0}(0, 0, 1) \\ (0, 0, 1) \rtimes [(\pi, 0, \pi, 0), \pi]   \rtimes   A_3 ,\  \mathcal{O}^{ \gamma_5 \otimes \gamma_{i \neq 3}\gamma_0}(0, 0, 1)\end{cases}\\
(0, 0, 0) \rtimes [(0, \pi, \pi, 0), \pi]   \rtimes   A_2^{-}  &\to \begin{cases}(0, 0, 1) \rtimes [(0, 0, \pi, \pi), \pi]   \rtimes   A_3 :  \mathcal{O}^{ \gamma_5 \otimes \gamma_{i \neq 3}\gamma_3}(0, 0, 1) \\ (0, 0, 1) \rtimes [(0, \pi, \pi, 0), \pi]   \rtimes   A_0 :  \mathcal{O}^{ \gamma_5 \otimes \gamma_{i \neq 3}\gamma_{j \neq 3}}(0, 0, 1)\end{cases}
\end{align}
Here, again the last four irreps undergo taste orbit splitting at non-zero momentum.
It is also important to note that the operators given will excite states in irreps of both parities for $\xi_0$ and $\xi_3$.

\section{Character tables}\label{sec:charTables}

This appendix contains the character tables used to construct the irreducible representations in \cref{sec:stagGroupPrimer}. These are, again, contained in Ref.~\cite{sharpe.1}. However, we repeat them here to address slight notational differences in irrep labelling and minor errors in some tables in that work.

\subsection{Character tables for little groups}\label{sec:charTablesTaste}

Here we given the character tables for the little groups of the taste-orbit under rotations defined in \cref{table:tasteorbit}. The two $(0,0,0)$ momentum taste little groups, $O_{\rm h}$ and $\text{D}_{4 \rm h}$, are given in \cref{tab:LG000T000,tab:LG000T00p}. For momentum $(0, 0, n)$, the two little groups, $\text{D}_{\rm 4}$ and $\text{C}_{2 \rm v}$, are given in \cref{tab:LG001T000,tab:LG001T00p}. The remaining unique group character tables, $\text{D}_{3}$ and $\text{Z}_{2}$, are given in \cref{tab:LG111T000,tab:LG110T00p}. Where the little group structure is repeated for different momentum and taste orbits, we indicate in the caption and include the unique rotation group elements for each case in the table.

\begin{table}[!htbp]
\centering
\caption{Character table for the $\vec{p}=2\pi(0, 0, 0)/L$, $\pi_\xi= (\xi_0, \pi,\pi,\pi)$ little group $\text{O}_\text{h}$.}
\label{tab:LG000T000}
\begin{tabular}{lrrrrrrrrrrr}
\hline \hline
Rep. element &  Class size &  $A_0^+$ &  $A_0^-$ &  $A_1^+$ &  $A_1^-$ &  $E_0^+$ &  $E_0^-$ &  $T_0^+$ &  $T_0^-$ &  $T_1^+$ &  $T_1^-$ \\
\hline
$E$                     &           1 &      1 &      1 &      1 &      1 &      2 &      2 &      3 &      3 &      3 &      3 \\
$I_S$                   &           1 &      1 &     $-1$ &      1 &     $-1$ &      2 &     $-2$ &      3 &     $-3$ &      3 &     $-3$ \\
$R_{12}R_{12}$          &           3 &      1 &      1 &      1 &      1 &      2 &      2 &     $-1$ &     $-1$ &     $-1$ &     $-1$ \\
$R_{12}R_{12}I_S$       &           3 &      1 &     $-1$ &      1 &     $-1$ &      2 &     $-2$ &     $-1$ &      1 &     $-1$ &      1 \\
$R_{12}$                &           6 &      1 &      1 &     $-1$ &     $-1$ &      0 &      0 &      1 &      1 &     $-1$ &     $-1$ \\
$R_{12}I_S$             &           6 &      1 &     $-1$ &     $-1$ &      1 &      0 &      0 &      1 &     $-1$ &     $-1$ &      1 \\
$R_{12}R_{12}R_{23}$    &           6 &      1 &      1 &    $ -1$ &     $-1$ &      0 &      0 &     $-1$ &     $-1$ &      1 &      1 \\
$R_{12}R_{12}R_{23}I_S$ &           6 &      1 &     $-1$ &     $-1$ &      1 &      0 &      0 &     $-1$ &      1 &      1 &    $ -1$ \\
$R_{12}R_{23}$          &           8 &      1 &      1 &      1 &      1 &     $-1$ &     $-1$ &      0 &      0 &      0 &      0 \\
$R_{12}R_{23}I_S$       &           8 &      1 &     $-1$ &      1 &     $-1$ &     $-1$ &      1 &      0 &      0 &      0 &      0 \\
\hline \hline
\end{tabular}
\end{table}

\begin{table}[!htbp]
\centering
\caption{Character table for the $\vec{p}=2\pi(0, 0, 0)/L$, $\pi_\xi= (\xi_0, 0,\pi,\pi)$ little group $\text{D}_\text{4h}$.}
\label{tab:LG000T00p}
\begin{tabular}{lccccccccccc}
\hline \hline
Rep. element &  Class size & $A_0^+$ &  $A_0^-$ &  $A_1^+$ &  $A_1^-$ & $A_2^+$ &  $A_2^-$ &  $A_3^+$ &  $A_3^-$ &  $E_0^+$ &  $E_0^-$ \\
\midrule
$E$                     &           1 &      1 &      1 &      1 &      1 &      1 &      1 &      1 &      1 &      2 &      2 \\
$I_S$                   &           1 &      1 &     $-1$ &      1 &     $-1$ &      1 &     $-1$ &      1 &     $-1$ &      2 &     $-2$ \\
$R_{23}R_{23}$          &           1 &      1 &      1 &      1 &      1 &      1 &      1 &      1 &      1 &     $-2$ &     $-2$ \\
$R_{23}R_{23}I_S$       &           1 &      1 &     $-1$ &      1 &     $-1$ &      1 &     $-1$ &      1 &     $-1$ &     $-2$ &      2 \\
$R_{12}R_{12}$          &           2 &      1 &      1 &      1 &      1 &     $-1$ &     $-1$ &     $-1$ &     $-1$ &      0 &      0 \\
$R_{12}R_{12}I_S$       &           2 &      1 &     $-1$ &      1 &     $-1$ &     $-1$ &      1 &     $-1$ &      1 &      0 &      0 \\
$R_{12}R_{12}R_{23}$    &           2 &      1 &      1 &     $-1$ &     $-1$ &     $-1$ &     $-1$ &      1 &      1 &      0 &      0 \\
$R_{12}R_{12}R_{23}I_S$ &           2 &      1 &     $-1$ &     $-1$ &      1 &     $-1$ &      1 &      1 &     $-1$ &      0 &      0 \\
$R_{23}$                &           2 &      1 &      1 &     $-1$ &     $-1$ &      1 &      1 &     $-1$ &     $-1$ &      0 &      0 \\
$R_{23}I_S$             &           2 &      1 &     $-1$ &     $-1$ &      1 &      1 &     $-1$ &     $-1$ &      1 &      0 &      0 \\
\hline \hline
\end{tabular}
\end{table}

\begin{table}[!htbp]
\centering
\caption{Character table for the $\vec{p}=2\pi(0, 0, \ell)/L$, $\pi_\xi= (\xi_0, \pi,\pi,\xi_3)$ little group $\text{D}_{4}$.}
\label{tab:LG001T000}
\begin{tabular}{lcccccc}
\hline \hline
Rep. element &  Class size &  $A_0$ &  $A_1$ &  $A_2$ &  $A_3$ &  $E_0$ \\
\midrule
$E$                     &           1 &      1 &      1 &      1 &      1 &      2 \\
$R_{12}R_{12}$          &           1 &      1 &      1 &      1 &      1 &     $-2$ \\
$R_{12}$                &           2 &      1 &      1 &     $-1$ &     $-1$ &      0 \\
$R_{12}R_{23}R_{23}I_S$ &           2 &      1 &     $-1$ &    $-1$ &      1 &      0 \\
$R_{23}R_{23}I_S$       &           2 &      1 &     $-1$ &      1 &     $-1$ &      0 \\
\hline \hline
\end{tabular}
\end{table}

\begin{table}[!htbp]
\centering
\caption{Character table for the $\vec{p}=2\pi(0, 0, \ell)/L$, $\pi_\xi= (\xi_0, 0,\pi,\xi_3)$ and the $\vec{p}=2\pi/(\ell, \ell, 0)/L$, $\pi_\xi= (\xi_0, \pi,\pi,\xi_3)$ little group $\text{C}_\text{2v}$.}
\label{tab:LG001T00p}
\begin{tabular}{lccccc}
\hline \hline
Rep. element &  Class size &  $A_0$ &  $A_1$ &  $A_2$ &  $A_3$ \\
\midrule
$E$               &           1 &      1 &      1 &      1 &      1 \\
$R_{12}R_{12}$ / $R_{12}R_{12}I_S$    &           1 &      1 &      1 &     $-1$ &     $-1$ \\
$R_{23}R_{23}I_S$ / $R_{12}R_{23}R_{23}I_S$ &           1 &      1 &     $-1$ &      1 &     $-1$ \\
$R_{31}R_{31}I_S$ / $R_{23}R_{23}R_{12}$ &           1 &      1 &     $-1$ &     $-1$ &      1 \\
\hline \hline
\end{tabular}
\end{table}

\begin{table}[!htbp]
\centering
\caption[Character table for the $\vec{p}=2\pi(\ell, \ell, \ell)/L$, $\pi_\xi= (\xi_0, \pi,\pi,\pi)$ little group $\text{D}_{3}$]{Character table for the $\vec{p}=2\pi(\ell, \ell, \ell)/L$, $\pi_\xi= (\xi_0, \pi,\pi,\pi)$ little group $\text{D}_{3}$.}
\label{tab:LG111T000}
\begin{tabular}{lcccc}
\hline \hline
Rep. element &  Class size &  $A_0$ &  $A_1$ &  $E_0$ \\
\hline
$E$                     &           1 &      1 &      1 &      2 \\
$R_{23}R_{12}$          &           2 &      1 &      1 &     $-1$ \\
$R_{12}R_{12}R_{23}I_S$ &           3 &      1 &     $-1$ &      0 \\
\hline \hline
\end{tabular}
\end{table}

\begin{table}[!htbp]
\centering
\caption{Character table for the $\vec{p}=2\pi/(\ell, \ell, 0)/L$, $\pi_\xi= (\xi_0, 0,\pi,\xi_3)$, $\vec{p}=2\pi(\ell, \ell, \ell)/L$, $\pi_\xi= (\xi_0, 0,\pi,\pi)$, $\vec{p}=2\pi(\ell, \ell, m)/L$, $\pi_\xi= (\xi_0, \pi,\pi,\pi)$ and $\vec{p}=2\pi(0, \ell, m)/L$, $\pi_\xi= (\xi_0, \pi,\pi,\pi)$ little group $\text{Z}_{2}$.}
\label{tab:LG110T00p}
\begin{tabular}{lccc}
\hline \hline
Rep. element &  Class size &  $A_0$ &  $A_1$ \\
\hline
$E$               &           1 &      1 &      1 \\
$R_{12}R_{12}I_S$ / $R_{12}R_{12}R_{23}I_S$ / $R_{12}R_{23}R_{23}I_S$ / $R_{23}R_{23}I_S$&           1 &      1 &     $-1$ \\
\hline \hline
\end{tabular}
\end{table}

\clearpage

\subsection{Character tables for staggered rest frame groups}\label{sec:charTablesRestFrame}

The characters for the bosonic irreps of the group $\text{SW}_4$ are given in \cref{tab:charTableSWBos}. A mapping to the class labelling in Refs.~\cite{sharpe.1,Mandula:1983ut} is given in the first column. The first four irreps are induced from the symmetric group $S_4$. The irreps, $(1,0)$ and  $(0,1)$, are subduced from the full four-dimensional rotation $O(4)\cong \text{SU}(2) \times \text{SU}(2)$ and remain irreducible.
The penultimate two, $\overline{(1,0)}$ and $\overline{(0,1)}$ are the product of the previous two with {\tiny\yng(1,1,1,1)\hspace*{1pt}}.
In the last column, $\mathbf{6}$ is a six-dimensional irrep obtained from $O(4)$ \cite{Mandula:1983ut}.

\begin{table}
\centering
\caption[Character table for the bosonic irreps of $\text{SW}_4$]{Character table for the bosonic irreps of $\text{SW}_4$, in agreement with Ref.~\cite{Mandula:1983ut} but correcting errors in {\tiny\yng(2,2)}, {\tiny\yng(3,1)}, and {\tiny\yng(2,1,1)} of Ref.~\cite{sharpe.1}.}\label{tab:charTableSWBos}
\begin{tabular}{llccccccccccc}
\hline \hline
Label & Rep. element &  Class size &  \tiny\yng(4) &  \tiny\yng(1,1,1,1) &  \tiny\yng(2,2) &  \tiny\yng(3,1) &  \tiny\yng(2,1,1) &  $(1,0)$ &  $(0,1)$ &  $\overline{(1,0)}$ &  $\overline{(0,1)}$ & $\mathbf{6}$  \\
\hline
I &$E$           &           1 &             1 &                   1 &               2 &               3 &                 3 &        3 &        3 &                   3 &                   3     &        6 \\
II & $R_{12}$$R_{12}$                               &           6 &             1 &                   1 &               2 &               3 &                 3 &       $-1$ &       $-1$ &                  $-1$ &                  $-1$ &        $-2$\\
III &$R_{12}$$R_{12}$$\tilde R_{43}$$\tilde R_{43}$ &           1 &             1 &                   1 &               2 &               3 &                 3 &        3 &        3 &                   3 &                   3&       6\\
IV &$R_{12}$      &          12 &             1 &                  $-1$ &               0 &               1 &                $-1$ &        1 &        1 &                  $-1$ &                  $-1$ & 0\\
V &$R_{12}$$R_{12}$$R_{23}$                       &          24 &             1 &                  $-1$ &               0 &               1 &                $-1$ &       $-1$ &       $-1$ &                   1 &                   1& 0\\
VI &$\tilde R_{43}$$R_{12}$$R_{12}$                &          12 &             1 &                  $-1$ &               0 &               1 &                $-1$ &        1 &        1 &                  $-1$ &                  $-1$& 0\\
VII & $R_{12}$$R_{23}$                               &          32 &             1 &                   1 &              $-1$ &               0 &                 0 &        0 &        0 &                   0 &                   0& 0\\
VIII & $\tilde R_{43}$$R_{12}$$R_{12}$$R_{23}$        &          32 &             1 &                   1 &              $-1$ &               0 &                 0 &        0 &        0 &                   0 &                   0& 0\\
IX & $R_{12}$$R_{23}$$\tilde R_{43}$                &          24 &             1 &                  $-1$ &               0 &              $-1$ &                 1 &        1 &       $-1$ &                  $-1$ &                   1& 0\\
X & $R_{12}$$R_{23}$$\tilde R_{42}$                &          24 &             1 &                  $-1$ &               0 &              $-1$ &                 1 &       $-1$ &        1 &                   1 &                  $-1$& 0\\
XI&$R_{12}$$\tilde R_{43}$$R_{23}$$R_{23}$        &          12 &             1 &                   1 &               2 &              $-1$ &                $-1$ &       $-1$ &       $-1$ &                  $-1$ &                  $-1$& 2\\
XII &$\tilde R_{43}$$R_{12}$$R_{12}$$R_{12}$        &           6 &             1 &                   1 &               2 &              $-1$ &                $-1$ &       $-1$ &        3 &                  $-1$ &                   3& $-2$\\

XIII &$R_{12}$$\tilde R_{43}$                        &           6 &             1 &                   1 &               2 &              $-1$ &                $-1$ &        3 &       $-1$ &                   3 &                  $-1$& $-2$\\
\hline \hline
\end{tabular}
\end{table}

For the bosonic case, any group element $g$ of $\Gamma_{2,2}$ can be represented as a four vector $\Gamma$ which takes values $0$ or $1$ depending on what generators $g$ contains. The characters for the bosonic irreps, labelled by $\pi_\Gamma$, of $\Gamma_{2,2}$ are then given by
\begin{align}
    \chi^{\pi_\Gamma}(g) &= e^{i \pi_\Gamma \cdot \Gamma} \label{eqn:gamma22Chars}
\end{align}
Leaving charge conjugation, $C_0$, and spatial inversion, $I_S$, out of this group gives the corresponding character table for the bosonic representations of $\text{Z}_2\times \text{Z}_2$, \cref{tab:charGammaZ2}.

\begin{table}
\centering
\caption[Character table for the $\text{Z}_2\times \text{Z}_2$ group corresponding to the bosonic representations of $\{\Xi_0, \Xi_1\Xi_2\Xi_3\}$ ]{Character table for the $\text{Z}_2\times \text{Z}_2$ group corresponding to the bosonic representations of $\{\Xi_0, \Xi_{123}\}$.}
\label{tab:charGammaZ2}
\begin{tabular}{lccccc}
\hline \hline
Rep. element &  Class size &  $(0,0)$ &  $(\pi,0)$ &  $(0,\pi)$ &  $(\pi,\pi)$ \\
\hline
$E$              & 1 & 1 & 1 & 1 & 1\\
$\Xi_0$          & 1 & 1 &$-1$ & 1 &$-1$\\
$\Xi_{123}$      & 1 & 1 & 1 &$-1$ &$-1$\\
$\Xi_0\Xi_{123}$ & 1 & 1 &$-1$ &$-1$ & 1\\
\hline \hline
\end{tabular}
\end{table}

\subsection{Rest frame groups isomorphism}

The staggered lattice group has two useful representations at zero momentum, the first representation,
\begin{align}
    (\text{SW}_4 \times \Gamma_{2,2})/(-E \times -E),
\end{align}
is useful for subducing from the continuum. It has the generating elements,
\begin{align}
    &\{R_{ij}, \tilde{R}_{4 i} \equiv R_{j k} \Xi_{k j}\} \times \{\Xi_0, \Xi_{123}, C_0\Xi_0I_S, C_0\} \\
    &\{\Xi_0, \Xi_{123}, C_0\Xi_0I_S, C_0\} \equiv \{\Gamma_1, \Gamma_2, \Gamma_3, \Gamma_4\}.
\end{align}
The second representation of the group,
\begin{align}
    \Gamma_{4,1} \rtimes \text{O}_\text{h},
\end{align}
is useful for considering states at non-zero momentum by the natural extension, $T_i \rtimes \Gamma_{4,1} \rtimes \text{O}_\text{h}$. It has generating elements,
\begin{align}
    \{\Xi_0,  \Xi_{1},  \Xi_{2}, \Xi_{3}, C_0\} \rtimes \{R_{ij}, I_s\}
\end{align}
The similarity between the irreps is given in \cref{tab:restFrameIso}.

\section{Staggered irreps}\label{sec:stagIrrepsLists}

\subsection{The staggered rho}\label{sec:rhoIrreps}

Using the tools described in \cref{sec:contDecomp}, the full the decomposition of the vector meson with negative parity and negative charge conjugation is given here for zero momentum.
\begin{align}
(0, 1) &\to (0, 1)\otimes 0 \to (0,1)\otimes(0, 0)\\
(15, 1) &\to (1, 1)\otimes 1\ \oplus\ (1, 1)\otimes 0\ \oplus\ (0, 1)\otimes 1 \to \nn \\
&\hphantom{IM} \text{\tiny\yng(3,1)\normalsize}\otimes(\pi, 0)\ \oplus\ \text{\tiny\yng(3,1)\normalsize}\otimes(0, \pi)\ \oplus\ \text{\tiny\yng(3,1)\normalsize}\otimes(\pi, \pi)\ \oplus\ \text{\tiny\yng(3,1)\normalsize}\otimes(0, 0)\ \oplus \nonumber \\
&\hphantom{IM} \mathbf{6}\otimes(\pi, 0)\ \oplus\ \mathbf{6}\otimes(0, \pi)\ \oplus\ \mathbf{6}\otimes(\pi, \pi)\ \oplus\ \mathbf{6}\otimes(0, 0)\ \oplus \nonumber \\
&\hphantom{IM} (0,1)\otimes(\pi, 0)\ \oplus\ (0,1)\otimes(0, \pi)\ \oplus\ (0,1)\otimes(\pi, \pi)
\end{align}
The term $(0,1) \otimes\left( (\pi,0) \oplus (0,\pi) \oplus (\pi,\pi)\right)$ was mistakenly written with as  $(1,0)  \otimes \ldots$ in Ref.~\cite{sharpe.1}. Proceeding with the mapping from \cref{tab:restFrameIso}, using \cref{eqn:paritySubDuce,eqn:chargeSubDuce} with $P=-1$ and $C=-1$ and the characters from \cref{tab:charGammaZ2}
\begin{align}
(0,1)\otimes(0, 0) &\sim (0, 0, 0)  \rtimes  [(0, 0, 0, 0), \pi]    \rtimes    T_0^{-}    \;:\; \mathcal{O}^{ \gamma_i  \otimes  1}(0, 0, 0) \label{eqn:1lrho}\\
(0,1)\otimes(\pi, 0) &\sim (0, 0, 0)  \rtimes  [(\pi, 0, 0, 0), \pi]    \rtimes    T_0^{+}    \;:\; \mathcal{O}^{ \gamma_i  \otimes  \gamma_5\gamma_0}(0, 0, 0)\\
(0,1)\otimes(0, \pi) &\sim (0, 0, 0)  \rtimes  [(0, \pi, \pi, \pi), 0]    \rtimes    T_0^{-}    \;:\; \mathcal{O}^{ \gamma_i  \otimes  \gamma_0}(0, 0, 0)\\
(0,1)\otimes(\pi, \pi) &\sim (0, 0, 0)  \rtimes  [(\pi, \pi, \pi, \pi), \pi]    \rtimes    T_0^{+}    \;:\; \mathcal{O}^{ \gamma_i  \otimes  \gamma_5}(0, 0, 0)\\
\text{\tiny\yng(3,1)\normalsize}\otimes(0, 0) &\sim (0, 0, 0)  \rtimes  [(0, \pi, \pi, 0), 0]    \rtimes    A_0^{-}    \;:\; \mathcal{O}^{ \gamma_i  \otimes  \gamma_j\gamma_k}(0, 0, 0)\\
\text{\tiny\yng(3,1)\normalsize}\otimes(\pi, 0) &\sim (0, 0, 0)  \rtimes  [(\pi, \pi, \pi, 0), 0]    \rtimes    A_0^{+}    \;:\; \mathcal{O}^{ \gamma_i  \otimes  \gamma_i}(0, 0, 0) \label{eqn:localRho}\\
\text{\tiny\yng(3,1)\normalsize}\otimes(0, \pi) &\sim (0, 0, 0)  \rtimes  [(0, 0, 0, \pi), \pi]    \rtimes    A_0^{-}    \;:\; \mathcal{O}^{ \gamma_i  \otimes  \gamma_5\gamma_i}(0, 0, 0)\\
\text{\tiny\yng(3,1)\normalsize}\otimes(\pi, \pi) &\sim (0, 0, 0)  \rtimes  [(\pi, 0, 0, \pi), 0]    \rtimes    A_0^{+}    \;:\; \mathcal{O}^{ \gamma_i  \otimes  \gamma_i\gamma_0}(0, 0, 0)\\
\mathbf{6}\otimes(\pi, 0) &\sim (0, 0, 0)  \rtimes  [(\pi, \pi, \pi, 0), 0]    \rtimes    E_0^{+}    \;:\; \mathcal{O}^{ \gamma_i  \otimes  \gamma_j}(0, 0, 0)\\
\mathbf{6}\otimes(0, \pi) &\sim (0, 0, 0)  \rtimes  [(0, 0, 0, \pi), \pi]    \rtimes    E_0^{-}    \;:\; \mathcal{O}^{ \gamma_i  \otimes  \gamma_5\gamma_j}(0, 0, 0)\\
\mathbf{6}\otimes(\pi, \pi) &\sim (0, 0, 0)  \rtimes  [(\pi, 0, 0, \pi), 0]    \rtimes    E_0^{+}    \;:\; \mathcal{O}^{ \gamma_i  \otimes  \gamma_j\gamma_0}(0, 0, 0)\\
\mathbf{6}\otimes(0, 0) &\sim (0, 0, 0)  \rtimes  [(0, \pi, \pi, 0), 0]    \rtimes    E_0^{-}    \;:\; \mathcal{O}^{ \gamma_i  \otimes  \gamma_i\gamma_j}(0, 0, 0)
\end{align}
In this work, we use the states and operators associated with \cref{eqn:1lrho}, called the `one-link' or taste-singlet $\rho$. Continuing with the example from Ref.~\cite{sharpe.1}, giving the $\rho$ momentum $(0,0,p_z)$ results in
\begin{align}
    (0, 0, 0)  \rtimes  [(\pi, 0, 0, 0), \pi]    \rtimes    T_0^{+}  &\to  \begin{cases}
        (0, 0, 1)  \rtimes  [(\pi, 0, 0, 0), \pi]    \rtimes    A_1 \;:\; \mathcal{O}^{\gamma_3 \otimes \gamma_5 \gamma_0} (0, 0, 1)\\
        (0, 0, 1)  \rtimes  [(\pi, 0, 0, 0), \pi]    \rtimes    E_0 \;:\; \mathcal{O}^{\gamma_{i \neq 3} \otimes \gamma_5 \gamma_0} (0, 0, 1)
    \end{cases}\\
    (0, 0, 0)  \rtimes  [(0, \pi, \pi, \pi), 0]    \rtimes    T_0^{-}  &\to  \begin{cases}
        (0, 0, 1)  \rtimes  [(0, \pi, \pi, \pi), 0]    \rtimes    A_0 \;:\;   \mathcal{O}^{\gamma_3 \otimes \gamma_0} (0, 0, 1)\\
        (0, 0, 1)  \rtimes  [(0, \pi, \pi, \pi), 0]    \rtimes    E_0 \;:\; \mathcal{O}^{\gamma_{i \neq 3} \otimes \gamma_0} (0, 0, 1)
    \end{cases}\\
    (0, 0, 0)  \rtimes  [(\pi, \pi, \pi, \pi), \pi]    \rtimes    T_0^{+}  &\to  \begin{cases}
        (0, 0, 1)  \rtimes  [(\pi, \pi, \pi, \pi), \pi]    \rtimes    A_1 \;:\;   \mathcal{O}^{\gamma_3 \otimes \gamma_5} (0, 0, 1)\\
        (0, 0, 1)  \rtimes  [(\pi, \pi, \pi, \pi), \pi]    \rtimes    E_0 \;:\;   \mathcal{O}^{\gamma_{i \neq 3} \otimes \gamma_5} (0, 0, 1)
    \end{cases}\\
    (0, 0, 0)  \rtimes  [(0, \pi, \pi, 0), 0]    \rtimes    A_0^{-} &\to  \begin{cases}
        (0, 0, 1)  \rtimes  [(0, \pi, \pi, 0), 0]    \rtimes    A_1 \;:\;    \mathcal{O}^{\gamma_3 \otimes \gamma_{j \neq i} \gamma_{k \neq i}}(0, 0, 1)\\
        (0, 0, 1)  \rtimes  [(0, 0, \pi, \pi), 0]    \rtimes    A_1 \;:\;   \mathcal{O}^{\gamma_{i \neq 3} \otimes \gamma_{j \neq i} \gamma_{k \neq i}}(0, 0, 1)
    \end{cases}\\
    (0, 0, 0)  \rtimes  [(\pi, \pi, \pi, 0), 0]    \rtimes    A_0^{+} &\to  \begin{cases}
        (0, 0, 1)  \rtimes  [(\pi, \pi, \pi, 0), 0]    \rtimes    A_0 \;:\;   \mathcal{O}^{\gamma_3 \otimes \gamma_{3}}(0, 0, 1)\\
        (0, 0, 1)  \rtimes  [(\pi, 0, \pi, \pi), 0]    \rtimes    A_0 \;:\;    \mathcal{O}^{\gamma_{i \neq 3} \otimes \gamma_{i} }(0, 0, 1)
    \end{cases}\\
    (0, 0, 0)  \rtimes  [(0, 0, 0, \pi), \pi]    \rtimes    A_0^{-} &\to  
    \begin{cases}
        (0, 0, 1)  \rtimes  [(0, 0, 0, \pi), \pi]    \rtimes    A_1 \;:\;    \mathcal{O}^{\gamma_3 \otimes \gamma_{5} \gamma_{3}}(0, 0, 1)\\
        (0, 0, 1)  \rtimes  [(0, 0, \pi, 0), \pi]    \rtimes    A_1 \;:\;    \mathcal{O}^{\gamma_{i \neq 3} \otimes\gamma_{5} \gamma_{i} }(0, 0, 1)
    \end{cases}\\
    (0, 0, 0)  \rtimes  [(\pi, 0, 0, \pi), 0]    \rtimes    A_0^{+} &\to 
    \begin{cases}
        (0, 0, 1)  \rtimes  [(\pi, 0, 0, \pi), 0]    \rtimes    A_0 \;:\;    \mathcal{O}^{\gamma_3 \otimes  \gamma_{3} \gamma_{0}}(0, 0, 1)\\
        (0, 0, 1)  \rtimes  [(\pi, 0, \pi, 0), 0]    \rtimes    A_0 \;:\;    \mathcal{O}^{\gamma_{i \neq 3} \otimes \gamma_{i} \gamma_{0}}(0, 0, 1)
    \end{cases}\\
    (0, 0, 0)  \rtimes  [(0, \pi, \pi, 0), 0]    \rtimes    E_0^{-} &\to  
    \begin{cases}
        (0, 0, 1)  \rtimes  [(0, \pi, \pi, 0), 0]    \rtimes    E_0 \;:\;    \mathcal{O}^{\gamma_{i \neq 3} \otimes \gamma_{i} \gamma_{j \neq i,3}}(0, 0, 1)\\
        (0, 0, 1)  \rtimes  [(0, 0, \pi, \pi), 0]    \rtimes    A_0 \;:\;    \mathcal{O}^{\gamma_{i \neq 3} \otimes \gamma_{i} \gamma_{3}}(0, 0, 1)\\
        (0, 0, 1)  \rtimes  [(0, 0, \pi, \pi), 0]    \rtimes    A_2 \;:\;
        \mathcal{O}^{\gamma_{3} \otimes \gamma_{i \neq 3} \gamma_{3}}(0, 0, 1)
    \end{cases}\\
    (0, 0, 0)  \rtimes  [(\pi, \pi, \pi, 0), 0]    \rtimes    E_0^{+} &\to  
    \begin{cases}
        (0, 0, 1)  \rtimes  [(\pi, \pi, \pi, 0), 0]    \rtimes    E_0 \;:\;    \mathcal{O}^{\gamma_{i \neq 3} \otimes \gamma_{3}} (0, 0, 1)\\
        (0, 0, 1)  \rtimes  [(\pi, 0, \pi, \pi), 0]    \rtimes    A_1 \;:\;    \mathcal{O}^{\gamma_{i \neq 3} \otimes \gamma_{j \neq 3} \gamma_{3}}(0, 0, 1)\\
        (0, 0, 1)  \rtimes  [(\pi, 0, \pi, \pi), 0]    \rtimes    A_3 \;:\;
        \mathcal{O}^{\gamma_{3} \otimes \gamma_{j \neq 3}}(0, 0, 1)
    \end{cases}\\
     (0, 0, 0)  \rtimes  [(0, 0, 0, \pi), \pi]    \rtimes    E_0^{-} &\to  
    \begin{cases}
        (0, 0, 1)  \rtimes  [(0, 0, 0, \pi), \pi]    \rtimes    E_0 \;:\;    \mathcal{O}^{\gamma_{i \neq 3} \otimes \gamma_{5} \gamma_{3}} (0, 0, 1)\\
        (0, 0, 1)  \rtimes  [(0, 0, \pi, 0), \pi]    \rtimes    A_0 \;:\;    \mathcal{O}^{\gamma_{i \neq 3} \otimes \gamma_{5} \gamma_{j \neq 3} }(0, 0, 1)\\
        (0, 0, 1)  \rtimes  [(0, 0, \pi, 0), \pi]    \rtimes    A_3 \;:\;
        \mathcal{O}^{\gamma_{3} \otimes \gamma_{5} \gamma_{j \neq 3}}(0, 0, 1)
    \end{cases}\\
     (0, 0, 0)  \rtimes  [(\pi, 0, 0, \pi), 0]    \rtimes    E_0^{+} &\to  
    \begin{cases}
        (0, 0, 1)  \rtimes  [(\pi, 0, 0, \pi), 0]    \rtimes    E_0 \;:\;    \mathcal{O}^{\gamma_{i \neq 3} \otimes  \gamma_{3} \gamma_{0}} (0, 0, 1)\\
        (0, 0, 1)  \rtimes  [(\pi, 0, \pi, 0), 0]    \rtimes    A_1 \;:\;   \mathcal{O}^{\gamma_{i \neq 3} \otimes  \gamma_{j \neq 3} \gamma_{0}}(0, 0, 1)\\
        (0, 0, 1)  \rtimes  [(\pi, 0, \pi, 0), 0]    \rtimes    A_2 \;:\;
        \mathcal{O}^{\gamma_{3} \otimes  \gamma_{j \neq 3} \gamma_{0}}(0, 0, 1)
    \end{cases}
\end{align}
Giving a different breakdown as to what is described in Ref.~\cite{sharpe.1}. The breakdown in that work is likely incorrect due to the aforementioned issue in footnote 8 of this appendix. Alongside this, there is also a further error in the decomposition to non-zero momentum which the above breakdown corrects.

\subsection{The staggered pion}\label{sec:pionIrreps}

The full the decomposition of the pseudo-scalar with $P=-1$, and $C=1$ is given here for the range of momentum considered in this work.
Using the results from \cref{sec:contDecomp}, the zero-momentum irreps and operators are
\begin{align}
(0, 0) &\to (0, 0)\otimes 0 \nonumber \to \text{\tiny\yng(4)\normalsize}\otimes(0, 0),\\
(15, 0) &\to (1, 0)\otimes 1\ \oplus\ (1, 0)\otimes 0\ \oplus\ (0, 0)\otimes 1 \to \nonumber \\
&\hphantom{IM} (1,0)\otimes(\pi, 0)\ \oplus\ (1,0)\otimes(0, \pi)\ \oplus\ (1,0)\otimes(\pi, \pi)\ \oplus\ (1,0)\otimes(0, 0)\ \oplus \nonumber \\
&\hphantom{IM} \text{\tiny\yng(4)\normalsize}\otimes(\pi, 0)\ \oplus\ \text{\tiny\yng(4)\normalsize}\otimes(0, \pi)\ \oplus\ \text{\tiny\yng(4)\normalsize}\otimes(\pi, \pi).
\end{align}
Relating these irreps to the irreps of \cref{eqn:stagGroupStruc},
\begin{align}
\text{\tiny\yng(4)\normalsize}\otimes(0, 0) &\sim (0, 0, 0)  \rtimes  [(0, 0, 0, 0), 0]    \rtimes    A_0^{-}    \;:\; \mathcal{O}^{ \gamma_5  \otimes  1}(0, 0, 0), \label{eqn:pion1}\\
\text{\tiny\yng(4)\normalsize}\otimes(\pi, 0) &\sim (0, 0, 0)  \rtimes  [(\pi, 0, 0, 0), 0]    \rtimes    A_0^{+}    \;:\; \mathcal{O}^{ \gamma_5  \otimes  \gamma_5\gamma_0}(0, 0, 0), \label{eqn:pion50}\\
\text{\tiny\yng(4)\normalsize}\otimes(0, \pi) &\sim (0, 0, 0)  \rtimes  [(0, \pi, \pi, \pi), \pi]    \rtimes    A_0^{-}    \;:\; \mathcal{O}^{ \gamma_5  \otimes  \gamma_0}(0, 0, 0), \label{eqn:pion0}\\
\text{\tiny\yng(4)\normalsize}\otimes(\pi, \pi) &\sim (0, 0, 0)  \rtimes  [(\pi, \pi, \pi, \pi), 0]    \rtimes    A_0^{+}    \;:\; \mathcal{O}^{ \gamma_5  \otimes  \gamma_5}(0, 0, 0), \label{eqn:pion5}\\
(1,0)\otimes(\pi, 0) &\sim (0, 0, 0)  \rtimes  [(\pi, \pi, \pi, 0), \pi]    \rtimes    A_2^{+}    \;:\; \mathcal{O}^{ \gamma_5  \otimes  \gamma_i}(0, 0, 0), \label{eqn:pioni}\\
(1,0)\otimes(0, \pi) &\sim (0, 0, 0)  \rtimes  [(0, 0, 0, \pi), 0]    \rtimes    A_2^{-}    \;:\; \mathcal{O}^{ \gamma_5  \otimes  \gamma_5\gamma_i}(0, 0, 0), \label{eqn:pion5i}\\
(1,0)\otimes(\pi, \pi) &\sim (0, 0, 0)  \rtimes  [(\pi, 0, 0, \pi), \pi]    \rtimes    A_2^{+}    \;:\; \mathcal{O}^{ \gamma_5  \otimes  \gamma_i\gamma_0}(0, 0, 0), \label{eqn:pioni0}\\
(1,0)\otimes(0, 0) &\sim (0, 0, 0)  \rtimes  [(0, \pi, \pi, 0), \pi]    \rtimes    A_2^{-}    \;:\; \mathcal{O}^{ \gamma_5  \otimes  \gamma_i\gamma_j}(0, 0, 0). \label{eqn:pionij}
\end{align}

The $(0,0,1)$ momentum subduction is given by
\begin{align}
(0, 0, 0)  \rtimes  [(0, 0, 0, 0), 0]    \rtimes    A_0^{-}  &\to  \begin{cases}(0, 0, 1)  \rtimes  [(0, 0, 0, 0), 0]    \rtimes    A_1 :  \mathcal{O}^{ \gamma_5  \otimes  1}(0, 0, 1)\end{cases}\\
(0, 0, 0)  \rtimes  [(\pi, 0, 0, 0), 0]    \rtimes    A_0^{+}  &\to  \begin{cases}(0, 0, 1)  \rtimes  [(\pi, 0, 0, 0), 0]    \rtimes    A_0 :  \mathcal{O}^{ \gamma_5  \otimes  \gamma_5\gamma_0}(0, 0, 1)\end{cases}\\
(0, 0, 0)  \rtimes  [(0, \pi, \pi, \pi), \pi]    \rtimes    A_0^{-}  &\to  \begin{cases}(0, 0, 1)  \rtimes  [(0, \pi, \pi, \pi), \pi]    \rtimes    A_1 :  \mathcal{O}^{ \gamma_5  \otimes  \gamma_0}(0, 0, 1)\end{cases}\\
(0, 0, 0)  \rtimes  [(\pi, \pi, \pi, \pi), 0]    \rtimes    A_0^{+}  &\to  \begin{cases}(0, 0, 1)  \rtimes  [(\pi, \pi, \pi, \pi), 0]    \rtimes    A_0 :  \mathcal{O}^{ \gamma_5  \otimes  \gamma_5}(0, 0, 1)\end{cases} \label{eqn:100pion5}\\
(0, 0, 0)  \rtimes  [(\pi, \pi, \pi, 0), \pi]    \rtimes    A_2^{+}  &\to  \begin{cases}(0, 0, 1)  \rtimes  [(\pi, 0, \pi, \pi), \pi]    \rtimes    A_2 :  \mathcal{O}^{ \gamma_5  \otimes  \gamma_{i \neq 3}}(0, 0, 1) \\ (0, 0, 1)  \rtimes  [(\pi, \pi, \pi, 0), \pi]    \rtimes    A_1 :  \mathcal{O}^{ \gamma_5  \otimes  \gamma_3}(0, 0, 1)\end{cases}\\
(0, 0, 0)  \rtimes  [(0, 0, 0, \pi), 0]    \rtimes    A_2^{-}  &\to  \begin{cases}(0, 0, 1)  \rtimes  [(0, 0, 0, \pi), 0]    \rtimes    A_0 :  \mathcal{O}^{ \gamma_5  \otimes  \gamma_5\gamma_3}(0, 0, 1) \\ (0, 0, 1)  \rtimes  [(0, 0, \pi, 0), 0]    \rtimes    A_2 :  \mathcal{O}^{ \gamma_5  \otimes  \gamma_5\gamma_{i \neq 3}}(0, 0, 1)\end{cases} \label{eqn:pion001G5I}\\
(0, 0, 0)  \rtimes  [(\pi, 0, 0, \pi), \pi]    \rtimes    A_2^{+}  &\to  \begin{cases}(0, 0, 1)  \rtimes  [(\pi, 0, 0, \pi), \pi]    \rtimes    A_1 :  \mathcal{O}^{ \gamma_5  \otimes  \gamma_3\gamma_0}(0, 0, 1) \\ (0, 0, 1)  \rtimes  [(\pi, 0, \pi, 0), \pi]    \rtimes    A_3 :  \mathcal{O}^{ \gamma_5  \otimes  \gamma_{i \neq 3}\gamma_0}(0, 0, 1)\end{cases}\\
(0, 0, 0)  \rtimes  [(0, \pi, \pi, 0), \pi]    \rtimes    A_2^{-}  &\to  \begin{cases}(0, 0, 1)  \rtimes  [(0, 0, \pi, \pi), \pi]    \rtimes    A_3 :  \mathcal{O}^{ \gamma_5  \otimes  \gamma_{i \neq 3}\gamma_3}(0, 0, 1) \label{eqn:pion001GIT}\\ (0, 0, 1)  \rtimes  [(0, \pi, \pi, 0), \pi]    \rtimes    A_0 :  \mathcal{O}^{ \gamma_5  \otimes  \gamma_{i \neq 3}\gamma_{j \neq 3}}(0, 0, 1)\end{cases}
\end{align}

The $(1,1,0)$ momentum momentum subduction is given by
\begin{align}
(0, 0, 0)  \rtimes  [(0, 0, 0, 0), 0]    \rtimes    A_0^{-}  &\to  \begin{cases}(1, 1, 0)  \rtimes  [(0, 0, 0, 0), 0]    \rtimes    A_1 :  \mathcal{O}^{ \gamma_5  \otimes  1}(1, 1, 0)\end{cases}\\
(0, 0, 0)  \rtimes  [(\pi, 0, 0, 0), 0]    \rtimes    A_0^{+}  &\to  \begin{cases}(1, 1, 0)  \rtimes  [(\pi, 0, 0, 0), 0]    \rtimes    A_0 :  \mathcal{O}^{ \gamma_5  \otimes  \gamma_5\gamma_0}(1, 1, 0)\end{cases}\\
(0, 0, 0)  \rtimes  [(0, \pi, \pi, \pi), \pi]    \rtimes    A_0^{-}  &\to  \begin{cases}(1, 1, 0)  \rtimes  [(0, \pi, \pi, \pi), \pi]    \rtimes    A_1 :  \mathcal{O}^{ \gamma_5  \otimes  \gamma_0}(1, 1, 0)\end{cases}\\
(0, 0, 0)  \rtimes  [(\pi, \pi, \pi, \pi), 0]    \rtimes    A_0^{+}  &\to  \begin{cases}(1, 1, 0)  \rtimes  [(\pi, \pi, \pi, \pi), 0]    \rtimes    A_0 :  \mathcal{O}^{ \gamma_5  \otimes  \gamma_5}(1, 1, 0)\end{cases}\\
(0, 0, 0)  \rtimes  [(\pi, \pi, \pi, 0), \pi]    \rtimes    A_2^{+}  &\to  \begin{cases}(1, 1, 0)  \rtimes  [(\pi, 0, \pi, \pi), \pi]    \rtimes    A_1 :  \mathcal{O}^{ \gamma_5  \otimes  \gamma_{i\neq 3}}(1, 1, 0) \\ (1, 1, 0)  \rtimes  [(\pi, \pi, \pi, 0), \pi]    \rtimes    A_2 :  \mathcal{O}^{ \gamma_5  \otimes  \gamma_3}(1, 1, 0)\end{cases}\\
(0, 0, 0)  \rtimes  [(0, 0, 0, \pi), 0]    \rtimes    A_2^{-}  &\to  \begin{cases}(1, 1, 0)  \rtimes  [(0, 0, 0, \pi), 0]    \rtimes    A_3 :  \mathcal{O}^{ \gamma_5  \otimes  \gamma_5\gamma_3}(1, 1, 0) \\ (1, 1, 0)  \rtimes  [(0, 0, \pi, 0), 0]    \rtimes    A_0 :  \mathcal{O}^{ \gamma_5  \otimes  \gamma_5\gamma_{i\neq 3}}(1, 1, 0)\end{cases}\\
(0, 0, 0)  \rtimes  [(\pi, 0, 0, \pi), \pi]    \rtimes    A_2^{+}  &\to  \begin{cases}(1, 1, 0)  \rtimes  [(\pi, 0, 0, \pi), \pi]    \rtimes    A_2 :  \mathcal{O}^{ \gamma_5  \otimes  \gamma_3\gamma_0}(1, 1, 0) \\ (1, 1, 0)  \rtimes  [(\pi, 0, \pi, 0), \pi]    \rtimes    A_1 :  \mathcal{O}^{ \gamma_5  \otimes  \gamma_{i\neq 3}\gamma_0}(1, 1, 0)\end{cases}\\
(0, 0, 0)  \rtimes  [(0, \pi, \pi, 0), \pi]    \rtimes    A_2^{-}  &\to  \begin{cases}(1, 1, 0)  \rtimes  [(0, 0, \pi, \pi), \pi]    \rtimes    A_0 :  \mathcal{O}^{ \gamma_5  \otimes  \gamma_{i\neq 3}\gamma_3}(1, 1, 0) \\ (1, 1, 0)  \rtimes  [(0, \pi, \pi, 0), \pi]    \rtimes    A_3 :  \mathcal{O}^{ \gamma_5  \otimes  \gamma_{i\neq 3}\gamma_{j\neq 3}}(1, 1, 0)\end{cases}
\end{align}

The $(1,1,1)$ momentum subduction is given by
\begin{align}
(0, 0, 0)  \rtimes  [(0, 0, 0, 0), 0]    \rtimes    A_0^{-}  &\to  \begin{cases}(1, 1, 1)  \rtimes  [(0, 0, 0, 0), 0]    \rtimes    A_1:  \mathcal{O}^{ \gamma_5  \otimes  1}(1, 1, 1)\end{cases}\\
(0, 0, 0)  \rtimes  [(\pi, 0, 0, 0), 0]    \rtimes    A_0^{+}  &\to  \begin{cases}(1, 1, 1)  \rtimes  [(\pi, 0, 0, 0), 0]    \rtimes    A_0 :  \mathcal{O}^{ \gamma_5  \otimes  \gamma_5\gamma_0}(1, 1, 1)\end{cases}\\
(0, 0, 0)  \rtimes  [(0, \pi, \pi, \pi), \pi]    \rtimes    A_0^{-}  &\to  \begin{cases}(1, 1, 1)  \rtimes  [(0, \pi, \pi, \pi), \pi]    \rtimes    A_1 :  \mathcal{O}^{ \gamma_5  \otimes  \gamma_0}(1, 1, 1)\end{cases}\\
(0, 0, 0)  \rtimes  [(\pi, \pi, \pi, \pi), 0]    \rtimes    A_0^{+}  &\to  \begin{cases}(1, 1, 1)  \rtimes  [(\pi, \pi, \pi, \pi), 0]    \rtimes    A_0 :  \mathcal{O}^{ \gamma_5  \otimes  \gamma_5}(1, 1, 1)\end{cases}\\
(0, 0, 0)  \rtimes  [(\pi, \pi, \pi, 0), \pi]    \rtimes    A_2^{+}  &\to  \begin{cases}(1, 1, 1)  \rtimes  [(\pi, 0, \pi, \pi), \pi]    \rtimes    A_1 :  \mathcal{O}^{ \gamma_5  \otimes  \gamma_i}(1, 1, 1)\end{cases}\\
(0, 0, 0)  \rtimes  [(0, 0, 0, \pi), 0]    \rtimes    A_2^{-}  &\to  \begin{cases}(1, 1, 1)  \rtimes  [(0, 0, 0, \pi), 0]    \rtimes    A_0 :  \mathcal{O}^{ \gamma_5  \otimes  \gamma_5\gamma_i}(1, 1, 1)\end{cases}\\
(0, 0, 0)  \rtimes  [(\pi, 0, 0, \pi), \pi]    \rtimes    A_2^{+}  &\to  \begin{cases}(1, 1, 1)  \rtimes  [(\pi, 0, 0, \pi), \pi]    \rtimes    A_1 :  \mathcal{O}^{ \gamma_5  \otimes  \gamma_i\gamma_0}(1, 1, 1)\end{cases}\\
(0, 0, 0)  \rtimes  [(0, \pi, \pi, 0), \pi]    \rtimes    A_2^{-}  &\to  \begin{cases}(1, 1, 1)  \rtimes  [(0, 0, \pi, \pi), \pi]    \rtimes    A_0 :  \mathcal{O}^{ \gamma_5  \otimes  \gamma_i\gamma_j}(1, 1, 1)\end{cases}
\end{align}

\section{Staggered two-pion Clebsch-Gordan coefficients}\label{sec:stag2piStatesApp}

In this Appendix, the Clebsch-Gordan coefficents (CGs) for the two cases of staggered two-pion states which couple to the taste-singlet vector current are given. These are two-pion states built out of single pion states, which are either one- or three-dimensional at zero momentum. We only consider the case for momentum $\vec{p}$, $p_i=0,1$ i.e the irreps and operators described in \cref{sec:pionIrreps} (see \cref{sec:OpBasis} for en explanation of this).

\subsection{One-dimensional pion irreps}

These irreps correspond to \cref{eqn:pion1,eqn:pion50,eqn:pion0,eqn:pion5}. All these cases are equivalent and have the same decomposition as appears with Wilson fermions. We use the pseudo-Goldstone boson pion, \cref{eqn:pion5}, for illustration. The case for one unit of momentum, $(0,0,1)$, is given in \cref{sec:opconstruction} but we give the results again here in \cref{tab:cgg5p001App}, along with the higher momentum CGs, $(1,1,0)$ and $(1,1,1)$ in \cref{tab:cgg5p110App,tab:cgg5p111App}, respectively.
\begin{table}[!htbp]
\centering
\caption{Clebsch-Gordan table for $(0, 0, 1)  \rtimes  [(\pi, \pi, \pi, \pi), 0]    \rtimes  A_0 \otimes(0, 0, 1)  \rtimes  [(\pi, \pi, \pi, \pi), \pi]    \rtimes    A_0 = (0, 0, 0)  \rtimes  [(0, 0, 0, 0), \pi]    \rtimes   T_0^{-} \oplus\cdots$. The irreps in the rows and columns are labeled by the corresponding operators.}
\label{tab:cgg5p001App}
\begin{tabular}{lccc}
\hline \hline
Tensor product row & $\mathcal{O}^{ \gamma_1  \otimes  1}(0, 0, 0)$ & $\mathcal{O}^{ \gamma_2  \otimes  1}(0, 0, 0)$ & $\mathcal{O}^{ \gamma_3  \otimes  1}(0, 0, 0)$ \\
\hline
$\mathcal{O}^{ \gamma_5  \otimes  \gamma_5}(1, 0, 0)\otimes\mathcal{O}^{\gamma_5  \otimes  \gamma_5}(-1, 0, 0)$ & $\frac{1}{\sqrt{2}}$ &  0 &  0 \\
$\mathcal{O}^{ \gamma_5  \otimes  \gamma_5}(-1, 0, 0)\otimes\mathcal{O}^{ \gamma_5  \otimes  \gamma_5}(1, 0, 0)$ &   $-\frac{1}{\sqrt{2}}$ &  0 &  0 \\
$\mathcal{O}^{ \gamma_5  \otimes  \gamma_5}(0, 1, 0)\otimes\mathcal{O}^{ \gamma_5  \otimes  \gamma_5}(0, -1, 0)$ &  0 & $\frac{1}{\sqrt{2}}$ &  0 \\
$\mathcal{O}^{ \gamma_5  \otimes  \gamma_5}(0, -1, 0)\otimes\mathcal{O}^{ \gamma_5  \otimes  \gamma_5}(0, 1, 0)$ &  0 &   $-\frac{1}{\sqrt{2}}$ &  0 \\
$\mathcal{O}^{ \gamma_5  \otimes  \gamma_5}(0, 0, 1)\otimes\mathcal{O}^{ \gamma_5  \otimes  \gamma_5}(0, 0, -1)$ &  0 &  0 & $\frac{1}{\sqrt{2}}$ \\
$\mathcal{O}^{ \gamma_5  \otimes  \gamma_5}(0, 0, -1)\otimes\mathcal{O}^{ \gamma_5  \otimes  \gamma_5}(0, 0, 1)$ &  0 &  0 &   $-\frac{1}{\sqrt{2}}$ \\
\hline \hline
\end{tabular}
\end{table}
\begin{table}[!htbp]
\centering
\caption{Clebsch-Gordan table for $(1, 1, 0)  \rtimes  [(\pi, \pi, \pi, \pi), 0]    \rtimes    A_0 \otimes(1, 1, 0)  \rtimes  [(\pi, \pi, \pi, \pi), \pi]    \rtimes    A_0 = (0, 0, 0)  \rtimes  [(0, 0, 0, 0), \pi]    \rtimes    T_0^{-}  \oplus\cdots$. The irreps in the rows and columns are labeled by the corresponding operators.}
\label{tab:cgg5p110App}
\begin{tabular}{lccc}
\hline \hline
Tensor product row & $\mathcal{O}^{ \gamma_1  \otimes  1}(0, 0, 0)$ & $\mathcal{O}^{ \gamma_2  \otimes  1}(0, 0, 0)$ & $\mathcal{O}^{ \gamma_3  \otimes  1}(0, 0, 0)$ \\
\hline
$\mathcal{O}^{ \gamma_5  \otimes  \gamma_5}(1, 1, 0)\otimes\mathcal{O}^{ \gamma_5  \otimes  \gamma_5}(1, 1, 0)$     & $\frac{1}{\sqrt{8}}$ & $\frac{1}{\sqrt{8}}$ &  0 \\
$\mathcal{O}^{ \gamma_5  \otimes  \gamma_5}(-1, -1, 0)\otimes\mathcal{O}^{ \gamma_5  \otimes  \gamma_5}(-1, -1, 0)$ &   $-\frac{1}{\sqrt{8}}$ &   $-\frac{1}{\sqrt{8}}$ &  0 \\
$\mathcal{O}^{ \gamma_5  \otimes  \gamma_5}(-1, 1, 0)\otimes\mathcal{O}^{ \gamma_5  \otimes  \gamma_5}(-1, 1, 0)$   &   $-\frac{1}{\sqrt{8}}$ & $\frac{1}{\sqrt{8}}$ &  0 \\
$\mathcal{O}^{ \gamma_5  \otimes  \gamma_5}(1, -1, 0)\otimes\mathcal{O}^{ \gamma_5  \otimes  \gamma_5}(1, -1, 0)$   & $\frac{1}{\sqrt{8}}$ &   $-\frac{1}{\sqrt{8}}$ &  0 \\
$\mathcal{O}^{ \gamma_5  \otimes  \gamma_5}(1, 0, 1)\otimes\mathcal{O}^{ \gamma_5  \otimes  \gamma_5}(1, 0, 1)$     & $\frac{1}{\sqrt{8}}$ &  0 & $\frac{1}{\sqrt{8}}$ \\
$\mathcal{O}^{ \gamma_5  \otimes  \gamma_5}(-1, 0, -1)\otimes\mathcal{O}^{ \gamma_5  \otimes  \gamma_5}(-1, 0, -1)$ &   $-\frac{1}{\sqrt{8}}$ &  0 &   $-\frac{1}{\sqrt{8}}$ \\
$\mathcal{O}^{ \gamma_5  \otimes  \gamma_5}(-1, 0, 1)\otimes\mathcal{O}^{ \gamma_5  \otimes  \gamma_5}(-1, 0, 1)$   &   $-\frac{1}{\sqrt{8}}$ &  0 & $\frac{1}{\sqrt{8}}$ \\
$\mathcal{O}^{ \gamma_5  \otimes  \gamma_5}(1, 0, -1)\otimes\mathcal{O}^{ \gamma_5  \otimes  \gamma_5}(1, 0, -1)$   & $\frac{1}{\sqrt{8}}$ &  0 &   $-\frac{1}{\sqrt{8}}$ \\
$\mathcal{O}^{ \gamma_5  \otimes  \gamma_5}(0, 1, -1)\otimes\mathcal{O}^{ \gamma_5  \otimes  \gamma_5}(0, 1, -1)$   &  0 & $\frac{1}{\sqrt{8}}$ &   $-\frac{1}{\sqrt{8}}$ \\
$\mathcal{O}^{ \gamma_5  \otimes  \gamma_5}(0, -1, 1)\otimes\mathcal{O}^{ \gamma_5  \otimes  \gamma_5}(0, -1, 1)$   &  0 &   $-\frac{1}{\sqrt{8}}$ & $\frac{1}{\sqrt{8}}$ \\
$\mathcal{O}^{ \gamma_5  \otimes  \gamma_5}(0, 1, 1)\otimes\mathcal{O}^{ \gamma_5  \otimes  \gamma_5}(0, 1, 1)$     &  0 & $\frac{1}{\sqrt{8}}$ & $\frac{1}{\sqrt{8}}$ \\
$\mathcal{O}^{ \gamma_5  \otimes  \gamma_5}(0, -1, -1)\otimes\mathcal{O}^{ \gamma_5  \otimes  \gamma_5}(0, -1, -1)$ &  0 &   $-\frac{1}{\sqrt{8}}$ &   $-\frac{1}{\sqrt{8}}$ \\
\hline \hline
\end{tabular}
\end{table}

\begin{table}[!htbp]
\centering
\caption{Clebsch-Gordan table for $(1, 1, 1)  \rtimes  [(\pi, \pi, \pi, \pi), 0]    \rtimes    A_0\ \otimes(1, 1, 1)  \rtimes  [(\pi, \pi, \pi, \pi), \pi]    \rtimes    A_0 = (0, 0, 0)  \rtimes  [(0, 0, 0, 0), \pi]    \rtimes    T_0^{-}  \oplus\cdots$. The irreps in the rows and columns are labeled by the corresponding operators.}
\label{tab:cgg5p111App}
\begin{tabular}{lccc}
\hline \hline
Tensor product row & $\mathcal{O}^{ \gamma_1  \otimes  1}(0, 0, 0)$ & $\mathcal{O}^{ \gamma_2  \otimes  1}(0, 0, 0)$ & $\mathcal{O}^{ \gamma_3  \otimes  1}(0, 0, 0)$  \\
\hline
$\mathcal{O}^{ \gamma_5  \otimes  \gamma_5}(1, 1, 1)\otimes\mathcal{O}^{ \gamma_5  \otimes  \gamma_5}(1, 1, 1)$       & $\frac{1}{\sqrt{8}}$ & $\frac{1}{\sqrt{8}}$ & $\frac{1}{\sqrt{8}}$ \\
$\mathcal{O}^{ \gamma_5  \otimes  \gamma_5}(-1, -1, -1)\otimes\mathcal{O}^{ \gamma_5  \otimes  \gamma_5}(-1, -1, -1)$ &   $-\frac{1}{\sqrt{8}}$ &   $-\frac{1}{\sqrt{8}}$ &   $-\frac{1}{\sqrt{8}}$ \\
$\mathcal{O}^{ \gamma_5  \otimes  \gamma_5}(-1, 1, 1)\otimes\mathcal{O}^{ \gamma_5  \otimes  \gamma_5}(-1, 1, 1)$     &   $-\frac{1}{\sqrt{8}}$ & $\frac{1}{\sqrt{8}}$ & $\frac{1}{\sqrt{8}}$ \\
$\mathcal{O}^{ \gamma_5  \otimes  \gamma_5}(1, -1, -1)\otimes\mathcal{O}^{ \gamma_5  \otimes  \gamma_5}(1, -1, -1)$   & $\frac{1}{\sqrt{8}}$ &   $-\frac{1}{\sqrt{8}}$ &   $-\frac{1}{\sqrt{8}}$ \\
$\mathcal{O}^{ \gamma_5  \otimes  \gamma_5}(1, -1, 1)\otimes\mathcal{O}^{ \gamma_5  \otimes  \gamma_5}(1, -1, 1)$     & $\frac{1}{\sqrt{8}}$ &   $-\frac{1}{\sqrt{8}}$ & $\frac{1}{\sqrt{8}}$ \\
$\mathcal{O}^{ \gamma_5  \otimes  \gamma_5}(-1, 1, -1)\otimes\mathcal{O}^{ \gamma_5  \otimes  \gamma_5}(-1, 1, -1)$   &   $-\frac{1}{\sqrt{8}}$ & $\frac{1}{\sqrt{8}}$ &   $-\frac{1}{\sqrt{8}}$ \\
$\mathcal{O}^{ \gamma_5  \otimes  \gamma_5}(1, 1, -1)\otimes\mathcal{O}^{ \gamma_5  \otimes  \gamma_5}(1, 1, -1)$     & $\frac{1}{\sqrt{8}}$ & $\frac{1}{\sqrt{8}}$ &   $-\frac{1}{\sqrt{8}}$ \\
$\mathcal{O}^{ \gamma_5  \otimes  \gamma_5}(-1, -1, 1)\otimes\mathcal{O}^{ \gamma_5  \otimes  \gamma_5}(-1, -1, 1)$   &   $-\frac{1}{\sqrt{8}}$ &   $-\frac{1}{\sqrt{8}}$ & $\frac{1}{\sqrt{8}}$ \\
\hline \hline
\end{tabular}
\end{table}

\clearpage

\subsection{Three-dimensional pion irreps}

For the case of the irreps which are three-dimensional at zero momentum, \cref{eqn:pioni,eqn:pion5i,eqn:pioni0,eqn:pionij}, we use the taste-pseudo vector `one-link' pion as the representative example. Again we start with repeating the results from \cref{sec:opconstruction} while including the one-dimensional split irrep here before the higher momentum cases. The $(0,0,1)$ momentum irreps and operators are given in \cref{tab:cgg5ip001App} for the one-dimensional case and \cref{tab:cgg5jp001App} for the two-dimensional case. For $(1,1,0)$ momentum, the one- and two-dimensional irrep CGs are given in \cref{tab:cgg5ip110App,tab:cgg5jp110App} respectively. Finally at $(1,1,1)$, we have a restoration of the three-dimensional symmetry and hence only set of CGs given in \cref{tab:cgg5ip111App}.
\begin{table}[htbp]
\centering
\caption{Clebsch-Gordan table for $(0, 0, 1)  \rtimes  [(0, 0, 0, \pi), 0]    \rtimes    A_0\ \otimes(0, 0, 1)  \rtimes  [(0, 0, 0, \pi), \pi]    \rtimes    A_0 = (0, 0, 0)  \rtimes  [(0, 0, 0, 0), \pi]    \rtimes    T_0^{-}  \oplus\cdots$. The irreps in the rows and columns are labeled by the corresponding operators.}
\label{tab:cgg5ip001App}
\begin{tabular}{lccc}
\hline \hline
{} & $\mathcal{O}^{ \gamma_1  \otimes  1}(0, 0, 0)$ & $\mathcal{O}^{ \gamma_2  \otimes  1}(0, 0, 0)$ & $\mathcal{O}^{ \gamma_3  \otimes  1}(0, 0, 0)$ \\
\hline
$\mathcal{O}^{ \gamma_5  \otimes  \gamma_5\gamma_1}(1, 0, 0)\otimes\mathcal{O}^{  \gamma_5  \otimes  \gamma_5\gamma_1}(-1, 0, 0)$ & $\frac{1}{\sqrt{4}}$ &  0 &  0 \\
$\mathcal{O}^{ \gamma_5  \otimes  \gamma_5\gamma_1}(-1, 0, 0)\otimes\mathcal{O}^{  \gamma_5  \otimes  \gamma_5\gamma_1}(1, 0, 0)$ &   $-\frac{1}{\sqrt{4}}$ &  0 &  0 \\
$\mathcal{O}^{ \gamma_5  \otimes  \gamma_5\gamma_2}(0, -1, 0)\otimes\mathcal{O}^{ \gamma_5  \otimes  \gamma_5\gamma_2}(0, 1, 0)$ &  0 &   $-\frac{1}{\sqrt{4}}$ &  0 \\
$\mathcal{O}^{ \gamma_5  \otimes  \gamma_5\gamma_2}(0, 1, 0)\otimes\mathcal{O}^{ \gamma_5  \otimes  \gamma_5\gamma_2}(0, -1, 0)$ &  0 & $\frac{1}{\sqrt{4}}$ &  0 \\
$\mathcal{O}^{ \gamma_5  \otimes  \gamma_5\gamma_3}(0, 0, 1)\otimes\mathcal{O}^{ \gamma_5  \otimes  \gamma_5\gamma_3}(0, 0, -1)$ &  0 &  0 & $\frac{1}{\sqrt{4}}$ \\
$\mathcal{O}^{ \gamma_5  \otimes  \gamma_5\gamma_3}(0, 0, -1)\otimes\mathcal{O}^{ \gamma_5  \otimes  \gamma_5\gamma_3}(0, 0, 1)$ &  0 &  0 &   $-\frac{1}{\sqrt{4}}$ \\
\hline \hline
\end{tabular}
\end{table}
\begin{table}[!htbp]
\centering
\caption{Clebsch-Gordan table for $(0, 0, 1)  \rtimes  [(0, 0, \pi, 0), 0]    \rtimes    A_2\ \otimes(0, 0, 1)  \rtimes  [(0, 0, \pi, 0), \pi]    \rtimes    A_2 = (0, 0, 0)  \rtimes  [(0, 0, 0, 0), \pi]    \rtimes    T_0^{-}  \oplus\cdots$. The irreps in the rows and columns are labeled by the corresponding operators.} 
\label{tab:cgg5jp001App}
\begin{tabular}{lccc}
\hline \hline
Tensor product row & $\mathcal{O}^{ \gamma_1  \otimes  1}(0, 0, 0)$ & $\mathcal{O}^{ \gamma_2  \otimes  1}(0, 0, 0)$ & $\mathcal{O}^{ \gamma_3  \otimes  1}(0, 0, 0)$ \\
\hline
$\mathcal{O}^{ \gamma_5  \otimes  \gamma_5\gamma_2}(1, 0, 0)\otimes\mathcal{O}^{ \gamma_5  \otimes  \gamma_5\gamma_2}(-1, 0, 0)$ & $\frac{1}{\sqrt{4}}$ &  0 &  0 \\
$\mathcal{O}^{ \gamma_5  \otimes  \gamma_5\gamma_3}(1, 0, 0)\otimes\mathcal{O}^{ \gamma_5  \otimes  \gamma_5\gamma_3}(-1, 0, 0)$ & $\frac{1}{\sqrt{4}}$ &  0 &  0 \\
$\mathcal{O}^{ \gamma_5  \otimes  \gamma_5\gamma_2}(-1, 0, 0)\otimes\mathcal{O}^{ \gamma_5  \otimes  \gamma_5\gamma_2}(1, 0, 0)$ &   $-\frac{1}{\sqrt{4}}$ &  0 &  0 \\
$\mathcal{O}^{ \gamma_5  \otimes  \gamma_5\gamma_3}(-1, 0, 0)\otimes\mathcal{O}^{ \gamma_5  \otimes  \gamma_5\gamma_3}(1, 0, 0)$ &   $-\frac{1}{\sqrt{4}}$ &  0 &  0 \\
$\mathcal{O}^{ \gamma_5  \otimes  \gamma_5\gamma_3}(0, 1, 0)\otimes\mathcal{O}^{ \gamma_5  \otimes  \gamma_5\gamma_3}(0, -1, 0)$ &  0 & $\frac{1}{\sqrt{4}}$ &  0 \\
$\mathcal{O}^{ \gamma_5  \otimes  \gamma_5\gamma_1}(0, 1, 0)\otimes\mathcal{O}^{ \gamma_5  \otimes  \gamma_5\gamma_1}(0, -1, 0)$ &  0 & $\frac{1}{\sqrt{4}}$ &  0 \\
$\mathcal{O}^{ \gamma_5  \otimes  \gamma_5\gamma_3}(0, -1, 0)\otimes\mathcal{O}^{ \gamma_5  \otimes  \gamma_5\gamma_3}(0, 1, 0)$ &  0 &   $-\frac{1}{\sqrt{4}}$ &  0 \\
$\mathcal{O}^{ \gamma_5  \otimes  \gamma_5\gamma_1}(0, -1, 0)\otimes\mathcal{O}^{ \gamma_5  \otimes  \gamma_5\gamma_1}(0, 1, 0)$ &  0 &  $ -\frac{1}{\sqrt{4}}$ &  0 \\
$\mathcal{O}^{ \gamma_5  \otimes  \gamma_5\gamma_1}(0, 0, 1)\otimes\mathcal{O}^{ \gamma_5  \otimes  \gamma_5\gamma_1}(0, 0, -1)$ &  0 &  0 & $\frac{1}{\sqrt{4}}$ \\
$\mathcal{O}^{ \gamma_5  \otimes  \gamma_5\gamma_2}(0, 0, 1)\otimes\mathcal{O}^{ \gamma_5  \otimes  \gamma_5\gamma_2}(0, 0, -1)$ &  0 &  0 & $\frac{1}{\sqrt{4}}$ \\
$\mathcal{O}^{ \gamma_5  \otimes  \gamma_5\gamma_1}(0, 0, -1)\otimes\mathcal{O}^{ \gamma_5  \otimes  \gamma_5\gamma_1}(0, 0, 1)$ &  0 &  0 &   $-\frac{1}{\sqrt{4}}$ \\
$\mathcal{O}^{\gamma_5  \otimes  \gamma_5\gamma_2}(0, 0, -1)\otimes\mathcal{O}^{ \gamma_5  \otimes  \gamma_5\gamma_2}(0, 0, 1)$ &  0 &  0 &   $-\frac{1}{\sqrt{4}}$ \\
\hline \hline
\end{tabular}
\end{table}
The $(1,1,0)$ momentum irreps and operators are given in \cref{tab:cgg5ip110App} for the one-dimensional case and \cref{tab:cgg5jp110App} for the two-dimensional case .
\begin{table}[htbp]
\centering
\caption{Clebsch-Gordan table for $(1, 1, 0)  \rtimes  [(0, 0, 0, \pi), 0]    \rtimes    A_3\ \otimes(1, 1, 0)  \rtimes  [(0, 0, 0, \pi), \pi]    \rtimes    A_3 =  (0, 0, 0)  \rtimes  [(0, 0, 0, 0), \pi]    \rtimes    T_0^{-}  \oplus\cdots$. The irreps in the rows and columns are labeled by the corresponding operators.}
\label{tab:cgg5ip110App}
\begin{tabular}{lccc}
\hline \hline
Tensor product row & $\mathcal{O}^{ \gamma_1  \otimes  1}(0, 0, 0)$ & $\mathcal{O}^{ \gamma_2  \otimes  1}(0, 0, 0)$ & $\mathcal{O}^{ \gamma_3  \otimes  1}(0, 0, 0)$ \\
\hline
$\mathcal{O}^{ \gamma_5  \otimes  \gamma_5\gamma_3}(1, 1, 0)\otimes\mathcal{O}^{ \gamma_5  \otimes  \gamma_5\gamma_3}(1, 1, 0)$     & $\frac{1}{\sqrt{8}}$ & $\frac{1}{\sqrt{8}}$ &  0 \\
$\mathcal{O}^{ \gamma_5  \otimes  \gamma_5\gamma_3}(-1, -1, 0)\otimes\mathcal{O}^{ \gamma_5  \otimes  \gamma_5\gamma_3}(-1, -1, 0)$ &   $-\frac{1}{\sqrt{8}}$ &   $-\frac{1}{\sqrt{8}}$ &  0 \\
$\mathcal{O}^{ \gamma_5  \otimes  \gamma_5\gamma_3}(-1, 1, 0)\otimes\mathcal{O}^{ \gamma_5  \otimes  \gamma_5\gamma_3}(-1, 1, 0)$   &   $-\frac{1}{\sqrt{8}}$ & $\frac{1}{\sqrt{8}}$ &  0 \\
$\mathcal{O}^{ \gamma_5  \otimes  \gamma_5\gamma_3}(1, -1, 0)\otimes\mathcal{O}^{ \gamma_5  \otimes  \gamma_5\gamma_3}(1, -1, 0)$   & $\frac{1}{\sqrt{8}}$ &   $-\frac{1}{\sqrt{8}}$ &  0 \\
$\mathcal{O}^{ \gamma_5  \otimes  \gamma_5\gamma_2}(1, 0, 1)\otimes\mathcal{O}^{ \gamma_5  \otimes  \gamma_5\gamma_2}(1, 0, 1)$     & $\frac{1}{\sqrt{8}}$ &  0 & $\frac{1}{\sqrt{8}}$ \\
$\mathcal{O}^{ \gamma_5  \otimes  \gamma_5\gamma_2}(-1, 0, -1)\otimes\mathcal{O}^{ \gamma_5  \otimes  \gamma_5\gamma_2}(-1, 0, -1)$ &   $-\frac{1}{\sqrt{8}}$ &  0 &   $-\frac{1}{\sqrt{8}}$ \\
$\mathcal{O}^{ \gamma_5  \otimes  \gamma_5\gamma_2}(-1, 0, 1)\otimes\mathcal{O}^{ \gamma_5  \otimes  \gamma_5\gamma_2}(-1, 0, 1)$   &   $-\frac{1}{\sqrt{8}}$ &  0 & $\frac{1}{\sqrt{8}}$ \\
$\mathcal{O}^{ \gamma_5  \otimes  \gamma_5\gamma_2}(1, 0, -1)\otimes\mathcal{O}^{ \gamma_5  \otimes  \gamma_5\gamma_2}(1, 0, -1)$   & $\frac{1}{\sqrt{8}}$ &  0 &   $-\frac{1}{\sqrt{8}}$ \\
$\mathcal{O}^{ \gamma_5  \otimes  \gamma_5\gamma_1}(0, 1, -1)\otimes\mathcal{O}^{ \gamma_5  \otimes  \gamma_5\gamma_1}(0, 1, -1)$   &  0 & $\frac{1}{\sqrt{8}}$ &   $-\frac{1}{\sqrt{8}}$ \\
$\mathcal{O}^{ \gamma_5  \otimes  \gamma_5\gamma_1}(0, -1, 1)\otimes\mathcal{O}^{ \gamma_5  \otimes  \gamma_5\gamma_1}(0, -1, 1)$   &  0 &   $-\frac{1}{\sqrt{8}}$ & $\frac{1}{\sqrt{8}}$ \\
$\mathcal{O}^{ \gamma_5  \otimes  \gamma_5\gamma_1}(0, 1, 1)\otimes\mathcal{O}^{ \gamma_5  \otimes  \gamma_5\gamma_1}(0, 1, 1)$     &  0 & $\frac{1}{\sqrt{8}}$ & $\frac{1}{\sqrt{8}}$ \\
$\mathcal{O}^{ \gamma_5  \otimes  \gamma_5\gamma_1}(0, -1, -1)\otimes\mathcal{O}^{ \gamma_5  \otimes  \gamma_5\gamma_1}(0, -1, -1)$ &  0 &   $-\frac{1}{\sqrt{8}}$ &   $-\frac{1}{\sqrt{8}}$ \\
\hline \hline
\end{tabular}
\end{table}
\begin{table}[htbp]
\centering
\caption[Clebsch-Gordan table for $\pi_{\otimes \gamma_5 \gamma_{i \neq 3}}\pi_{\otimes  \gamma_5  \gamma_{i \neq 3}}$ with momenta (1,1,0) to $\rho_{\otimes 1} (0,0,0)$.]{Clebsch-Gordan table for $(1, 1, 0)  \rtimes  [(0, 0, \pi, 0), 0]    \rtimes    A_0\ \otimes(1, 1, 0)  \rtimes  [(0, 0, \pi, 0), \pi]    \rtimes    A_0 = (0, 0, 0)  \rtimes  [(0, 0, 0, 0), \pi]    \rtimes    T_0^{-}  \oplus\cdots$. The irreps in the rows and columns are labeled by the corresponding operators.}
\label{tab:cgg5jp110App}
\begin{tabular}{lccc}
\hline \hline
Tensor product row & $\mathcal{O}^{ \gamma_1  \otimes  1}(0, 0, 0)$ & $\mathcal{O}^{ \gamma_2  \otimes  1}(0, 0, 0)$ & $\mathcal{O}^{ \gamma_3  \otimes  1}(0, 0, 0)$ \\
\hline
$\mathcal{O}^{ \gamma_5  \otimes  \gamma_5\gamma_1}(1, 1, 0)\otimes\mathcal{O}^{\gamma_5  \otimes  \gamma_5\gamma_1}(-1, -1, 0)$ & $\frac{1}{4}$ &  $\frac{1}{4}$ & 0 \\
$\mathcal{O}^{ \gamma_5  \otimes  \gamma_5\gamma_2}(1, 1, 0)\otimes\mathcal{O}^{ \gamma_5  \otimes  \gamma_5\gamma_2}(-1, -1, 0)$ & $\frac{1}{4}$ & $\frac{1}{4}$ & 0 \\
$\mathcal{O}^{ \gamma_5  \otimes  \gamma_5\gamma_1}(-1, -1, 0)\otimes\mathcal{O}^{ \gamma_5  \otimes  \gamma_5\gamma_1}(1, 1, 0)$ & $-\frac{1}{4}$ & $-\frac{1}{4}$ & 0 \\
$\mathcal{O}^{ \gamma_5  \otimes  \gamma_5\gamma_2}(-1, -1, 0)\otimes\mathcal{O}^{ \gamma_5  \otimes  \gamma_5\gamma_2}(1, 1, 0)$ & $-\frac{1}{4}$ & $-\frac{1}{4}$ & 0 \\
$\mathcal{O}^{ \gamma_5  \otimes  \gamma_5\gamma_1}(-1, 1, 0)\otimes\mathcal{O}^{ \gamma_5  \otimes  \gamma_5\gamma_1}(1, -1, 0)$ &$-\frac{1}{4}$ & $\frac{1}{4}$ & 0 \\
$\mathcal{O}^{ \gamma_5  \otimes  \gamma_5\gamma_2}(-1, 1, 0)\otimes\mathcal{O}^{ \gamma_5  \otimes  \gamma_5\gamma_2}(1, -1, 0)$ &$-\frac{1}{4}$ & $\frac{1}{4}$ &0 \\
$\mathcal{O}^{ \gamma_5  \otimes  \gamma_5\gamma_1}(1, -1, 0)\otimes\mathcal{O}^{ \gamma_5  \otimes  \gamma_5\gamma_1}(-1, 1, 0)$ &$\frac{1}{4}$ & $-\frac{1}{4}$ & 0 \\
$\mathcal{O}^{ \gamma_5  \otimes  \gamma_5\gamma_2}(1, -1, 0)\otimes\mathcal{O}^{ \gamma_5  \otimes  \gamma_5\gamma_2}(-1, 1, 0)$ &$\frac{1}{4}$ & $-\frac{1}{4}$ & 0 \\
$\mathcal{O}^{ \gamma_5  \otimes  \gamma_5\gamma_2}(1, 0, 1)\otimes\mathcal{O}^{ \gamma_5  \otimes  \gamma_5\gamma_2}(-1, 0, -1)$ &$\frac{1}{4}$ & 0 & $\frac{1}{4}$ \\
$\mathcal{O}^{ \gamma_5  \otimes  \gamma_5\gamma_3}(1, 0, 1)\otimes\mathcal{O}^{ \gamma_5  \otimes  \gamma_5\gamma_3}(-1, 0, -1)$ & $\frac{1}{4}$ &  0 & $\frac{1}{4}$ \\
$\mathcal{O}^{ \gamma_5  \otimes  \gamma_5\gamma_2}(-1, 0, -1)\otimes\mathcal{O}^{ \gamma_5  \otimes  \gamma_5\gamma_2}(1, 0, 1)$ & $-\frac{1}{4}$ & 0 & $-\frac{1}{4}$\\
$\mathcal{O}^{ \gamma_5  \otimes  \gamma_5\gamma_3}(-1, 0, -1)\otimes\mathcal{O}^{\gamma_5  \otimes  \gamma_5\gamma_3}(1, 0, 1)$ &$-\frac{1}{4}$ &  0 &$-\frac{1}{4}$ \\
$\mathcal{O}^{ \gamma_5  \otimes  \gamma_5\gamma_2}(-1, 0, 1)\otimes\mathcal{O}^{ \gamma_5  \otimes  \gamma_5\gamma_2}(1, 0, -1)$ &   $-\frac{1}{4}$ &  0 & $\frac{1}{4}$\\
$\mathcal{O}^{ \gamma_5  \otimes  \gamma_5\gamma_3}(-1, 0, 1)\otimes\mathcal{O}^{\gamma_5  \otimes  \gamma_5\gamma_3}(1, 0, -1)$ &   $-\frac{1}{4}$ &  0 & $\frac{1}{4}$\\
$\mathcal{O}^{ \gamma_5  \otimes  \gamma_5\gamma_2}(1, 0, -1)\otimes\mathcal{O}^{ \gamma_5  \otimes  \gamma_5\gamma_2}(-1, 0, 1)$ & $\frac{1}{4}$ &  0 &   $-\frac{1}{4}$\\
$\mathcal{O}^{ \gamma_5  \otimes  \gamma_5\gamma_3}(1, 0, -1)\otimes\mathcal{O}^{\gamma_5  \otimes  \gamma_5\gamma_3}(-1, 0, 1)$ & $\frac{1}{4}$ &  0 &   $-\frac{1}{4}$\\
$\mathcal{O}^{ \gamma_5  \otimes  \gamma_5\gamma_3}(0, 1, 1)\otimes\mathcal{O}^{ \gamma_5  \otimes  \gamma_5\gamma_3}(0, -1, -1)$ &  0 & $\frac{1}{4}$ & $\frac{1}{4}$ \\
$\mathcal{O}^{ \gamma_5  \otimes  \gamma_5\gamma_1}(0, 1, 1)\otimes\mathcal{O}^{ \gamma_5  \otimes  \gamma_5\gamma_1}(0, -1, -1)$ &  0 & $\frac{1}{4}$ & $\frac{1}{4}$ \\
$\mathcal{O}^{ \gamma_5  \otimes  \gamma_5\gamma_3}(0, -1, -1)\otimes\mathcal{O}^{\gamma_5  \otimes  \gamma_5\gamma_3}(0, 1, 1)$ &  0 &$-\frac{1}{4}$ & $-\frac{1}{4}$ \\
$\mathcal{O}^{ \gamma_5  \otimes  \gamma_5\gamma_1}(0, -1, -1)\otimes\mathcal{O}^{\gamma_5  \otimes  \gamma_5\gamma_1}(0, 1, 1)$ &  0 & $-\frac{1}{4}$ & $-\frac{1}{4}$ \\
$\mathcal{O}^{ \gamma_5  \otimes  \gamma_5\gamma_3}(0, 1, -1)\otimes\mathcal{O}^{ \gamma_5  \otimes  \gamma_5\gamma_3}(0, -1, 1)$ &  0 & $\frac{1}{4}$ & $-\frac{1}{4}$\\
$\mathcal{O}^{ \gamma_5  \otimes  \gamma_5\gamma_1}(0, 1, -1)\otimes\mathcal{O}^{ \gamma_5  \otimes  \gamma_5\gamma_1}(0, -1, 1)$ &  0 & $\frac{1}{4}$ & $-\frac{1}{4}$\\
$\mathcal{O}^{ \gamma_5  \otimes  \gamma_5\gamma_3}(0, -1, 1)\otimes\mathcal{O}^{ \gamma_5  \otimes  \gamma_5\gamma_3}(0, 1, -1)$ &  0 & $-\frac{1}{4}$ & $\frac{1}{4}$\\
$\mathcal{O}^{ \gamma_5  \otimes  \gamma_5\gamma_1}(0, -1, 1)\otimes\mathcal{O}^{ \gamma_5  \otimes  \gamma_5\gamma_1}(0, 1, -1)$ &  0 & $-\frac{1}{4}$ & $\frac{1}{4}$\\
\hline \hline
\end{tabular}
\end{table}
The $(1,1,1)$ momentum irreps and operators are given in \cref{tab:cgg5ip111App} for the three-dimensional case.
\begin{table}[hpt]
\centering
\caption{Clebsch-Gordan table for $(1, 1, 1)      \rtimes    [(0, 0, 0,   \pi), 0]        \rtimes      A_0   \otimes  (1, 1, 1)      \rtimes    [(0, 0, 0,   \pi),   \pi]        \rtimes      A_0 = (0, 0, 0)      \rtimes    [(0, 0, 0, 0),   \pi]        \rtimes      T_0^{-}  \oplus\cdots$. The irreps in the rows and columns are labeled by the corresponding operators.}
\label{tab:cgg5ip111App}
\begin{tabular}{lccc}
\hline \hline
Tensor product row & $  \mathcal{O}^{   \gamma_1      \otimes    1}  (0, 0, 0)$ & $  \mathcal{O}^{   \gamma_2      \otimes    1}  (0, 0, 0)$ & $  \mathcal{O}^{   \gamma_3      \otimes    1}  (0, 0, 0)$ \\
\hline
$  \mathcal{O}^{\gamma_5\otimes\gamma_5\gamma_1}(1, 1, 1)    \otimes    \mathcal{O}^{\gamma_5\otimes\gamma_5\gamma_1}(-1, -1, -1)$ & $\frac{1}{4}$ & $\frac{1}{4\sqrt{2}}$ & $\frac{1}{4\sqrt{2}}$ \\
$  \mathcal{O}^{\gamma_5\otimes\gamma_5\gamma_2}(1, 1, 1)    \otimes    \mathcal{O}^{\gamma_5\otimes\gamma_5\gamma_2}(-1, -1, -1)$ & $\frac{1}{4\sqrt{2}}$ & $\frac{1}{4}$ & $\frac{1}{4\sqrt{2}}$ \\
$  \mathcal{O}^{\gamma_5\otimes\gamma_5\gamma_3}(1, 1, 1)\ \otimes\ \mathcal{O}^{\gamma_5\otimes\gamma_5\gamma_3}(-1, -1, -1)$ & $\frac{1}{4\sqrt{2}}$ & $\frac{1}{4\sqrt{2}}$ & $\frac{1}{4}$ \\
$  \mathcal{O}^{\gamma_5\otimes\gamma_5\gamma1}(-1, -1, -1)    \otimes    \mathcal{O}^{\gamma_5\otimes\gamma_5\gamma_1}(1, 1, 1)$ &  $-\frac{1}{4}$ &  $-\frac{1}{4\sqrt{2}}$ &  $-\frac{1}{4\sqrt{2}}$ \\
$  \mathcal{O}^{\gamma_5\otimes\gamma_5\gamma_2}(-1, -1, -1)    \otimes    \mathcal{O}^{\gamma_5\otimes\gamma_5\gamma_2}(1, 1, 1)$ &  $-\frac{1}{4\sqrt{2}}$ &  $-\frac{1}{4}$ &  $-\frac{1}{4\sqrt{2}}$ \\
$  \mathcal{O}^{\gamma_5\otimes\gamma_5\gamma_3}(-1, -1, -1)    \otimes    \mathcal{O}^{\gamma_5\otimes\gamma_5\gamma_3}(1, 1, 1)$ &  $-\frac{1}{4\sqrt{2}}$ &  $-\frac{1}{4\sqrt{2}}$ &  $-\frac{1}{4}$ \\
$  \mathcal{O}^{\gamma_5\otimes\gamma_5\gamma_1}(-1, 1, 1)    \otimes    \mathcal{O}^{\gamma_5\otimes\gamma_5\gamma_1}(1, -1, -1)$ &  $-\frac{1}{4}$ & $\frac{1}{4\sqrt{2}}$ & $\frac{1}{4\sqrt{2}}$ \\
$  \mathcal{O}^{\gamma_5\otimes\gamma_5\gamma_2}(-1, 1, 1)    \otimes    \mathcal{O}^{\gamma_5\otimes\gamma_5\gamma_2}(1, -1, -1)$ &  $-\frac{1}{4\sqrt{2}}$ & $\frac{1}{4}$ & $\frac{1}{4\sqrt{2}}$ \\
$  \mathcal{O}^{\gamma_5\otimes\gamma_5\gamma_3}(-1, 1, 1)    \otimes    \mathcal{O}^{\gamma_5\otimes\gamma_5\gamma_3}(1, -1, -1)$ &  $-\frac{1}{4\sqrt{2}}$ & $\frac{1}{4\sqrt{2}}$ & $\frac{1}{4}$ \\
$  \mathcal{O}^{\gamma_5\otimes\gamma_5\gamma_1}(1, -1, -1)    \otimes    \mathcal{O}^{\gamma_5\otimes\gamma_5\gamma_1}(-1, 1, 1)$ & $\frac{1}{4}$ &  $-\frac{1}{4\sqrt{2}}$ &  $-\frac{1}{4\sqrt{2}}$ \\
$  \mathcal{O}^{\gamma_5\otimes\gamma_5\gamma_2}(1, -1, -1)    \otimes    \mathcal{O}^{\gamma_5\otimes\gamma_5\gamma_2}(-1, 1, 1)$ & $\frac{1}{4\sqrt{2}}$ &  $-\frac{1}{4}$ &  $-\frac{1}{4\sqrt{2}}$ \\
$  \mathcal{O}^{\gamma_5\otimes\gamma_5\gamma_3}(1, -1, -1)    \otimes    \mathcal{O}^{\gamma_5\otimes\gamma_5\gamma_3}(-1, 1, 1)$ & $\frac{1}{4\sqrt{2}}$ &  $-\frac{1}{4\sqrt{2}}$ &  $-\frac{1}{4}$ \\
$  \mathcal{O}^{\gamma_5\otimes\gamma_5\gamma_1}(1, -1, 1)    \otimes    \mathcal{O}^{\gamma_5\otimes\gamma_5\gamma_1}(-1, 1, -1)$ & $\frac{1}{4}$ &  $-\frac{1}{4\sqrt{2}}$ & $\frac{1}{4\sqrt{2}}$ \\
$  \mathcal{O}^{\gamma_5\otimes\gamma_5\gamma_2}(1, -1, 1)    \otimes    \mathcal{O}^{\gamma_5\otimes\gamma_5\gamma_2}(-1, 1, -1)$ & $\frac{1}{4\sqrt{2}}$ &  $-\frac{1}{4}$ & $\frac{1}{4\sqrt{2}}$ \\
$  \mathcal{O}^{\gamma_5\otimes\gamma_5\gamma_3}(1, -1, 1)    \otimes    \mathcal{O}^{\gamma_5\otimes\gamma_5\gamma_3}(-1, 1, -1)$ & $\frac{1}{4\sqrt{2}}$ &  $-\frac{1}{4\sqrt{2}}$ & $\frac{1}{4}$ \\
$  \mathcal{O}^{\gamma_5\otimes\gamma_5\gamma_1}(-1, 1, -1)    \otimes    \mathcal{O}^{\gamma_5\otimes\gamma_5\gamma_1}(1, -1, 1)$ &  $-\frac{1}{4}$ & $\frac{1}{4\sqrt{2}}$ &  $-\frac{1}{4\sqrt{2}}$ \\
$  \mathcal{O}^{\gamma_5\otimes\gamma_5\gamma_2}(-1, 1, -1)    \otimes    \mathcal{O}^{\gamma_5\otimes\gamma_5\gamma_2}(1, -1, 1)$ &  $-\frac{1}{4\sqrt{2}}$ & $\frac{1}{4}$ &  $-\frac{1}{4\sqrt{2}}$ \\
$  \mathcal{O}^{\gamma_5\otimes\gamma_5\gamma_3}(-1, 1, -1)    \otimes    \mathcal{O}^{\gamma_5\otimes\gamma_5\gamma_3}(1, -1, 1)$ &  $-\frac{1}{4\sqrt{2}}$ & $\frac{1}{4\sqrt{2}}$ &  $-\frac{1}{4}$ \\
$  \mathcal{O}^{\gamma_5\otimes\gamma_5\gamma_1}(1, 1, -1)    \otimes    \mathcal{O}^{\gamma_5\otimes\gamma_5\gamma_1}(-1, -1, 1)$ & $\frac{1}{4}$ & $\frac{1}{4\sqrt{2}}$ &  $-\frac{1}{4\sqrt{2}}$ \\
$  \mathcal{O}^{\gamma_5\otimes\gamma_5\gamma_2}(1, 1, -1)    \otimes    \mathcal{O}^{\gamma_5\otimes\gamma_5\gamma_2}(-1, -1, 1)$ & $\frac{1}{4\sqrt{2}}$ & $\frac{1}{4}$ &  $-\frac{1}{4\sqrt{2}}$ \\
$  \mathcal{O}^{\gamma_5\otimes\gamma_5\gamma_3}(1, 1, -1)    \otimes    \mathcal{O}^{\gamma_5\otimes\gamma_5\gamma_3}(-1, -1, 1)$ & $\frac{1}{4\sqrt{2}}$ & $\frac{1}{4\sqrt{2}}$ &  $-\frac{1}{4}$ \\
$  \mathcal{O}^{\gamma_5\otimes\gamma_5\gamma_1}(-1, -1, 1)    \otimes    \mathcal{O}^{\gamma_5\otimes\gamma_5\gamma_1}(1, 1, -1)$ &  $-\frac{1}{4}$ &  $-\frac{1}{4\sqrt{2}}$ & $\frac{1}{4\sqrt{2}}$ \\
$  \mathcal{O}^{\gamma_5\otimes\gamma_5\gamma_2}(-1, -1, 1)    \otimes    \mathcal{O}^{\gamma_5\otimes\gamma_5\gamma_2}(1, 1, -1)$ &  $-\frac{1}{4\sqrt{2}}$ &  $-\frac{1}{4}$ & $\frac{1}{4\sqrt{2}}$ \\
$  \mathcal{O}^{\gamma_5\otimes\gamma_5\gamma_3}(-1, -1, 1)    \otimes    \mathcal{O}^{\gamma_5\otimes\gamma_5\gamma_3}(1, 1, -1)$ &  $-\frac{1}{4\sqrt{2}}$ &  $-\frac{1}{4\sqrt{2}}$ & $\frac{1}{4}$ \\
\hline \hline
\end{tabular}
\end{table}

\clearpage

\section{Rooting and staggered observables}\label{sec:rooting}

In quantum field theory, observables can be obtained through taking functional derivatives of a generating functional. For the case of staggered fermions, the fermion determinant, $\det D_f[U]$, describes four tastes of staggered fermions for each flavor. This is addressed by taking the fourth root of the fermion determinant. The rooted-staggered path integral is then
\begin{align}
    Z= \int \mathcal{D}[U] \prod_f \left[\det D_f[U]\right]^{\frac{1}{4}} \,\mathrm{e}^{-S_G}. \label{eqn:pathIntRooted}
\end{align}
Rooting results in additional factors of $\frac{1}{4}$ that need to be included when computing physical observables. As illustration, we obtain the vector current two-point correlation function using the staggered action, \cref{eqn:stagFermionActionReduced}. Coupling to a background photon field $C^\mu(n)$, we have,
\begin{align}
S_{F}\left[\chi_f, \bar{\chi}_f, U, C\right]&=a^{4} \sum_f \sum_{n, m \in \Lambda} \bar\chi_f(n) D_f[U, C](n|m)\chi_f(m), \label{eqn:stagFermionActionReducedBackgroundC}\\
D_f[U, C](n|m) &=  \sum_\mu \eta_{\mu}(n) \left[\frac{U _ { \mu }(n)e^{iaQ_fC_\mu(n)}\delta_{m, n+\hat{\mu}}}{2 a} -\text{H.c.}\right]+m\,\delta_{n, m},
\end{align}
where H.c.is the Hermitian conjugate. The generating functional is
\begin{align}
    Z[C]= \int \mathcal{D}[U] \prod_f \det D_f[U,C] \,\mathrm{e}^{-S_G}.
\end{align}
The vector current two-point function is obtained by taking the second derivative with respect to the source field, giving
\begin{align}
    \left.\frac{\delta^{2} \log  Z[C]}{\delta C_{\mu}(x) \delta C_{\mu}(x^{\prime})}\right|_{C=0} &= \left. \frac{1}{Z[C(x^{\prime})]} \frac{\delta^{2} Z[C]}{\delta C_{\mu}(x) \delta C_{\mu}(x^{\prime})}\right|_{C=0} -\left. \frac{1}{Z^2[C(x^{\prime})]}\frac{\delta Z[C]}{\delta C_{\mu}(x) }\right|_{C=0}
\end{align}
The second term on the right-hand side is zero by the lattice rotation-reflection symmetries. Then using 
\begin{align}
    \det M = \exp \textrm{tr} \log M,
\end{align}
first without rooting, gives
\begin{align}
    \frac{1}{Z[C(x^{\prime})]} \frac{\delta^{2} Z[C]}{\delta C_{\mu}(x) \delta C_{\mu}(x^{\prime})}&= \frac{1}{Z[C(x^{\prime})]} \frac{\delta }{\delta C_{\mu}(x) }\int \mathcal{D}[U] \mathrm{e}^{-S_G}   \prod_f \det D_f[U,C] \nn\\
    &\times \textrm{tr} \left[\sum_fD^{-1}_f[U,C](x^{\prime})\, i Q_f \eta_{\mu}(x^{\prime})\left[ \frac{U _ { \mu }(x^{\prime})e^{iaQ_fC_\mu(x^{\prime})}\delta_{m, x^{\prime}+\hat{\mu}}}{2} -\text{h.c.}\right]\right]  \label{eqn:nonRootDeriv}
\end{align}
Taking the second derivative and setting $C=0$ gives
\begin{align}
    =\int \mathcal{D}[U] \mathrm{e}^{-S_G}   \prod_f &\det D_f[U,C] \Bigg\{  \textrm{tr} \Big[\sum_f D^{-1}_f[U,C](x)\,  Q_f j^{\mu}({x}) D^{-1}_f[U,C](x^{\prime})\,  Q_f j^{\mu}(x^{\prime}) \Big] \nn\\
    &-\textrm{tr}  \Big[ \sum_{f^\prime} D^{-1}_{f^\prime}[U,C](x)\, Q_f^\prime j^{\mu}({x})\Big] \times \textrm{tr}\Big[\sum_fD^{-1}_f[U,C](x^{\prime})\,  Q_f j^{\mu}({x^{\prime}})\Big] \Bigg\} \label{eqn:twoPointWickNonRooted}
\end{align}
where the lattice current operator, $j^{\mu}(x)$, was introduced 
\begin{align}
    J^{\mu}(x) &\equiv \frac{1}{2} \left[\eta_{\mu}(x) U _ { \mu }(x)\delta_{m, x+\hat{\mu}} -\text{h.c.}\right]
\end{align}
The trace on the first line \cref{eqn:twoPointWickNonRooted} is the `connected' Wick contraction while the product of the two traces on the second line is the `disconnected' contraction, corresponding to the following two diagrams,
\begin{align}
\sum_f 
\quad 
\vcenter{\hbox{\begin{tikzpicture}
  \begin{feynman}
    \vertex (a)  at (-1.5,0);
    \vertex (m1) at (-0.7,0);
    \vertex (m2)  at (0.7, 0);
    \vertex (b) at (1.5, 0);
    \diagram* {
      (a) -- [photon, edge label=$C_\mu$, insertion =1] (m1) -- [fermion, half left, edge label=$q_f$] (m2) -- [photon, edge label=$C_\nu$, insertion=0] (b),
      (m2) -- [fermion, half left, edge label=$\bar q_f$] (m1)};
  \end{feynman}
\end{tikzpicture}}} 
\quad
- \sum_{f, f^\prime}
\quad
\vcenter{\hbox{\begin{tikzpicture}
  \begin{feynman}
    \vertex (a)  at (-1.5,0);
    \vertex (m1) at (-0.6,0);
    \vertex (m2)  at (0.6, 0);
    \vertex (m3) at (0.9,0);
    \vertex (m4)  at (2.1, 0);
    \vertex (b) at (3.0, 0);
    \diagram* {
      (a) -- [photon, edge label=$C_\mu$, insertion =1] (m1) -- [fermion, half left, edge label=$q_f$] (m2),
      (m2) -- [fermion, half left, edge label=$\bar q_f$] (m1),
      (m4) -- [photon, edge label=$C_\mu$, insertion =0] (b),
      (m3) -- [fermion, half left, edge label=$q_{f^\prime}$] (m4) -- [fermion, half left, edge label=$\bar q_{f^\prime}$] (m3)};
  \end{feynman}
\end{tikzpicture}}}, \label{eqn:vecWickDiagrams}
\end{align} 
where the $\times$ implies the background fields are set to zero as above. The effect of the $4$ tastes is to add three additional quark loops for each flavor.

Re-computing  \cref{eqn:twoPointWickNonRooted} for the case of a rooted determinant results in 
\begin{align}
    =\int \mathcal{D}[U] \mathrm{e}^{-S_G}   \prod_f &\left[\det D_f[U,C]\right]^{\frac{1}{4}} \Bigg\{  \frac{1}{4}\textrm{tr} \Big[\sum_f D^{-1}_f[U,C](x)\,  Q_f J^{\mu}({x}) D^{-1}_f[U,C](x^{\prime})\,  Q_f J^{\mu}(x^{\prime}) \Big] \nn\\
    &-\frac{1}{16}\textrm{tr}  \Big[ \sum_{f^\prime} D^{-1}_{f^\prime}[U,C](x)\, Q_f^\prime J^{\mu}({x})\Big] \times \textrm{tr}\Big[\sum_fD^{-1}_f[U,C](x^{\prime})\,  Q_f J^{\mu}({x^{\prime}})\Big] \Bigg\}. \label{eqn:twoPointWickRooted}
\end{align}
There is now a factor of $\frac{1}{4}$ in the connected component and a factor of $\frac{1}{16}$ in the disconnected component. From this and \cref{eqn:vecWickDiagrams}, one can infer the diagrammatic rule that for every fermion loop in a staggered Wick contraction, a factor of $\frac{1}{4}$ is added to obtain the corresponding physical observable.

\section{Time-split two-pion operators}\label{sec:timeSplit}

\subsection{Operator definitions}
As mentioned in \cref{sec:corrFuncs}, the correlation functions used in this work are generated  with time-split two-pion operators.
This modification, introduced in Ref.~\cite{Fu:2016itp} to address possible Fierz rearrangement of pions on the same time slice, 
is not actually necessary with the random-wall sources used here.\footnote{This was not realized at the time of data generation.}
Here, for completeness, we describe the additional considerations these operators require.

The time-split operators, which are non-Hermitian to start with, are defined as 
\begin{align}
    \mathcal{O}^{\textrm{TS}}_{\pi \pi}(\vec{0}, t) &\equiv \pi_{\otimes \gamma_\xi}^+(p, t) \pi_{\otimes \gamma_\xi}^-(-p, t+1) - \pi_{\otimes \gamma_\xi}^+(-p, t) \pi_{\otimes \gamma_\xi}^-(p, t+1) , \\
    \mathcal{O}^{\textrm{TS}\,  \dagger}_{\pi \pi}(\vec{0}, t) &\equiv \pi_{\otimes \gamma_\xi}^+(p, t+1) \pi_{\otimes \gamma_\xi}^-(-p, t) - \pi_{\otimes \gamma_\xi}^+(-p, t+1) \pi^-(p, t) ,
\end{align}
using the notation of \cref{sec:opconstruction,sec:corrFuncs} on the right-hand side. To make apparent the effects of these operators on the correlation functions described in \cref{sec:corrFuncs}, it is useful to pull out the time dependence,
\begin{align}
\mathcal{O}^{\textrm{TS}}_{\pi \pi}(\vec{0}, t) 
&\equiv e^{Ht} \mathcal{O}_{\pi \pi}(\vec{0}) e^{-H(t+1)},\\
\mathcal{O}^{\textrm{TS}\, \dagger}_{\pi \pi}(\vec{0}, t) 
&\equiv e^{H(t+1)} \mathcal{O}_{\pi \pi}(\vec{0}) e^{-Ht},
\end{align}
with 
\begin{align}
    \mathcal{O}_{\pi \pi}(\vec{0}_{\vec{p}}) = \pi^+(\vec{p}) \pi^-(-\vec{p}) - \pi^+(-\vec{p}) \pi^-(\vec{p}),
\end{align}
now Hermitian.

\subsection{Correlation functions}

The two-point function, \cref{eq:corrFuncSpecRep}, is unchanged. The $C(t)_{\pi \pi \to \rho}$ three-point function, \cref{eqn:threePoint}, is modified as
\begin{align}
&C(t)_{\pi \pi \to \rho} =  \frac{1}{3}\sum_{i}\left\langle \rho^0_i(\vec{0}, t)  \mathcal{O}^{\textrm{TS}\, \dagger}_{\pi \pi}(\vec{0}_{\vec{p}}, 0) \right\rangle \nn = \sum_n \langle 0 | \rho^0 | n \rangle  \langle n |  \mathcal{O}^{\otimes \gamma_\xi}_{\pi \pi} | 0 \rangle  e^{-E_n (t-1)} \\
&=   4\cdot  \frac{1}{2} \cdot \frac{1}{4} \cdot \frac{1}{N_S^{9/2}} \cdot  \frac{1}{3} \times \sum_{i} \vcenter{\hbox{\begin{tikzpicture}
  \begin{feynman}
    \vertex[empty dot, label=below:{($\vec{p},0$)}]  (l) at (0.0,0) {$\gamma_5\otimes \gamma_\xi$};
    \vertex[empty dot, label=below:{($-\vec{p},1$)}] (r) at (3,0.2) {$\gamma_5\otimes \gamma_\xi$};
    \vertex[empty dot, label=above:{($\vec{0},t$)}]  (t) at (1.5,2.25) {$\gamma_i\otimes 1$};
    \diagram* {
      (l) -- [fermion, edge label'=\(D^{-1}_{l}\)] (r) -- [fermion, edge label'=\(D^{-1}_{l}\)] (t) -- [fermion, edge label'=\(D^{-1}_{l}\)] (l)};
  \end{feynman}
\end{tikzpicture}}},\\
&=\frac{1}{6N_S^{9/2}}\sum_{i,\vec{n}_0,\vec{n}_1,\vec{n}_2, \{\pm \delta_j \}} \varphi^{\gamma_5 \otimes \gamma_\xi}(n_0) \varphi^{\gamma_5 \otimes \gamma_\xi}(n_1) \varphi^{\gamma_i \otimes 1}(n_2) e^{i \vec{p} \cdot \left(\vec{n}_0-\vec{n}_1\right)} \nn \\
&\times \textrm{tr} \left[ D^{-1}_l(\vec{n}_0 + \delta^{\gamma_5 \otimes \gamma_\xi}, 0| \vec{n}_1, 1) D^{-1}_l(\vec{n}_1 + \delta^{\gamma_5 \otimes \gamma_\xi}, 1| \vec{n}_2, t) D^{-1}_l(\vec{n}_2 + \delta^{\gamma_i \otimes 1}, t| \vec{n}_0, 0) \right]. \label{eqn:threePointTS}
\end{align}
with the same normalizations defined in \cref{subsubsec:3pts}. The $\pi \pi \to \pi \pi$ four point function becomes
\begin{align}
&C(t)_{\pi \pi \to \pi \pi} =\left\langle \mathcal{O}^{\textrm{TS}}_{\pi \pi}(\vec{0}_{\vec{p}},t)  \mathcal{O}^{\textrm{TS}\, \dagger}_{\pi \pi}(\vec{0}_{\vec{p}},0)\right\rangle =  \sum_n  \langle 0 | \mathcal{O}^{\textrm{TS},\, \otimes \gamma_{\xi_1}}_{\pi \pi} | n\rangle  \langle n | \mathcal{O}^{\textrm{TS},\, \otimes \gamma_{\xi_2}}_{\pi \pi} | 0\rangle  e^{-E_n t}\nn \\
&=  -4\cdot \frac{1}{4} \cdot \frac{1}{N_S^6} \times \vcenter{\hbox{\begin{tikzpicture}
  \begin{feynman}
    \vertex[empty dot, label=below:{($\vec{p},0)$ }] (bl) at (0.0,0) {$\gamma_5 \otimes \gamma_{\xi_1}$};
    \vertex[empty dot, label=below:{($-\vec{p},1)$ }] (br) at (2.5,0.2) {$\gamma_5 \otimes \gamma_{\xi_1}$};
    \vertex[empty dot, label=above:{($-\vec{p},t)$ }] (tl) at (0,2.5) {$\gamma_5 \otimes \gamma_{\xi_2}$};
    \vertex[empty dot, label=above:{($\vec{p},t+1)$ }] (tr) at (2.5,2.7) {$\gamma_5 \otimes \gamma_{\xi_2}$};
    \diagram* {
      (bl) -- [fermion] (br) -- [fermion] (tr) -- [fermion] (tl) -- [fermion] (bl)};
  \end{feynman}
\end{tikzpicture}}}
\quad \quad +4\cdot \frac{1}{4}  \cdot \frac{1}{N_S^6} \times \vcenter{\hbox{\begin{tikzpicture}
  \begin{feynman}
    \vertex[empty dot, label=below:{($\vec{p},0)$ }] (bl) at (0.0,0) {$\gamma_5 \otimes \gamma_{\xi_1}$};
    \vertex[empty dot, label=below:{($-\vec{p},1)$ }] (br) at (2.5,0.2) {$\gamma_5 \otimes \gamma_{\xi_1}$};
    \vertex[empty dot, label=above:{($-\vec{p},t)$ }] (tl) at (0,2.5) {$\gamma_5 \otimes \gamma_{\xi_2}$};
    \vertex[empty dot, label=above:{($\vec{p},t+1)$ }] (tr) at (2.5,2.7) {$\gamma_5 \otimes \gamma_{\xi_2}$};
    \diagram* {
      (bl) -- [fermion] (br) -- [fermion] (tl) -- [fermion] (tr) -- [fermion] (bl)};
  \end{feynman}
\end{tikzpicture}}} \nn \\
 & \ \ +2 \cdot \frac{1}{16} \cdot \frac{1}{N_S^6} \times \vcenter{\hbox{\begin{tikzpicture}
  \begin{feynman}
    \vertex[empty dot, label=below:{($\vec{p},0)$ }] (bl) at (0.0,0) {$\gamma_5 \otimes \gamma_{\xi_1}$};
    \vertex[empty dot, label=below:{($-\vec{p}, 1)$ }] (br) at (2.5,0.2) {$\gamma_5 \otimes \gamma_{\xi_1}$};
    \vertex[empty dot, label=above:{($-\vec{p}, t)$ }] (tl) at (0,2.5) {$\gamma_5 \otimes \gamma_{\xi_2}$};
    \vertex[empty dot, label=above:{($\vec{p}, t+1)$ }] (tr) at (2.5,2.7) {$\gamma_5 \otimes \gamma_{\xi_2}$};
    \diagram* {
      (bl) -- [fermion, bend left=20] (tl) -- [fermion, bend left=20] (bl),
      (br) -- [fermion, bend left=20] (tr) -- [fermion, bend left=20] (br)};
  \end{feynman}
\end{tikzpicture}}}
\quad \quad  -2 \cdot \frac{1}{16} \cdot \frac{1}{N_S^6} \times \vcenter{\hbox{\begin{tikzpicture}
  \begin{feynman}
    \vertex[empty dot, label=below:{($\vec{p},0)$ }] (bl) at (0.0,0) {$\gamma_5 \otimes \gamma_{\xi_1}$};
    \vertex[empty dot, label=below:{($-\vec{p},1)$ }] (br) at (2.5,0.2) {$\gamma_5 \otimes \gamma_{\xi_1}$};
    \vertex[empty dot, label=above:{($-\vec{p},t)$ }] (tl) at (0,2.5) {$\gamma_5 \otimes \gamma_{\xi_2}$};
    \vertex[empty dot, label=above:{($\vec{p},t+1)$ }] (tr) at (2.5,2.7) {$\gamma_5 \otimes \gamma_{\xi_2}$};
    \diagram* {
      (bl) -- [fermion, bend left=20] (tr) -- [fermion, bend left=20] (bl),
      (br) -- [fermion, bend left=20] (tl) -- [fermion, bend left=20] (br)};
  \end{feynman}
\end{tikzpicture}}}, \\
&= \frac{1}{N_S^6}\sum_{\vec{n}_0,\vec{n}_1,\vec{n}_2, \vec{n}_3, \{\pm \delta_j \}} \varphi^{\gamma_5 \otimes \gamma_{\xi_1}}(n_0) \varphi^{\gamma_5 \otimes \gamma_{\xi_1}}(n_1) \varphi^{\gamma_5 \otimes \gamma_{\xi_2}}(n_2) \varphi^{\gamma_5 \otimes \gamma_{\xi_2}}(n_3) e^{i \vec{p} \cdot \left(\vec{n}_0-\vec{n}_1+\vec{n}_2-\vec{n}_3\right)} \Big[ \nn\\
&\quad \quad \quad \quad \quad \quad \quad \quad -\textrm{tr} \left[D^{-1}_l(\vec{n}_0 + \delta^{\gamma_5 \otimes \gamma_{\xi_1}}, 0| \vec{n}_1, 1) D^{-1}_l(\vec{n}_1 + \delta^{\gamma_5 \otimes \gamma_{\xi_1}}, 1| \vec{n}_2, t) \right.\nn \\
& \quad \quad \quad \quad \quad \quad \quad \quad \quad \quad \left. \times D^{-1}_l(\vec{n}_2 + \delta^{\gamma_5 \otimes \gamma_{\xi_2}}, t| \vec{n}_3, t+1) D^{-1}_l(\vec{n}_3 + \delta^{\gamma_5 \otimes \gamma_{\xi_2}}, t+1| \vec{n}_0, 0)\right] \nn\\
&\quad \quad \quad \quad \quad \quad \quad \quad +\textrm{tr} \left[D^{-1}_l(\vec{n}_0 + \delta^{\gamma_5 \otimes \gamma_{\xi_1}}, 0| \vec{n}_1, 1) D^{-1}_l(\vec{n}_1 + \delta^{\gamma_5 \otimes \gamma_{\xi_1}}, 1| \vec{n}_3, t+1) \right.\nn \\
& \quad \quad \quad \quad \quad \quad \quad \quad \quad \quad \left. \times D^{-1}_l(\vec{n}_3 + \delta^{\gamma_5 \otimes \gamma_{\xi_2}}, t+1| \vec{n}_2, t) D^{-1}_l(\vec{n}_2 + \delta^{\gamma_5 \otimes \gamma_{\xi_2}}, t| \vec{n}_0, 0)\right] \nn\\
&\quad \quad \quad \quad \quad \quad \quad \quad +\frac{1}{8}\textrm{tr} \left[D^{-1}_l(\vec{n}_0 + \delta^{\gamma_5 \otimes \gamma_{\xi_1}}, 0| \vec{n}_2, t) D^{-1}_l(\vec{n}_2 + \delta^{\gamma_5 \otimes \gamma_{\xi_2}}, t| \vec{n}_0, 0) \right]\nn \\
& \quad \quad \quad \quad \quad \quad \quad \quad \quad \quad \times \textrm{tr} \left[  D^{-1}_l(\vec{n}_1 + \delta^{\gamma_5 \otimes \gamma_{\xi_1}}, 1| \vec{n}_3, t+1) D^{-1}_l(\vec{n}_3 + \delta^{\gamma_5 \otimes \gamma_{\xi_2}}, t+1| \vec{n}_1, 1)\right] \nn\\
&\quad \quad \quad \quad \quad \quad \quad \quad -\frac{1}{8}\textrm{tr} \left[D^{-1}_l(\vec{n}_0 + \delta^{\gamma_5 \otimes \gamma_{\xi_1}}, 0| \vec{n}_3, t+1) D^{-1}_l(\vec{n}_3 + \delta^{\gamma_5 \otimes \gamma_{\xi_2}}, t+1| \vec{n}_0, 0) \right]\nn \\
& \quad \quad \quad \quad \quad \quad \quad \quad \quad \quad \times \textrm{tr} \left[  D^{-1}_l(\vec{n}_1 + \delta^{\gamma_5 \otimes \gamma_{\xi_1}}, 1| \vec{n}_2, t) D^{-1}_l(\vec{n}_2 + \delta^{\gamma_5 \otimes \gamma_{\xi_2}}, t| \vec{n}_1, 1)\right] \Big]. \label{eqn:fourPointsTS}
\end{align}

\section{Multi-exponential matrix fit}\label{sec:matrixFit}

An alternative approach to the GEVP method discussed in \cref{sec:gevp}, for extracting spectrum from the correlation function matrix in \cref{eqn:ctmat}, is to perform a multi-exponential fit directly to the matrix (MEM fit). Fitting to the following fit ansatz,
\begin{align}
C_{ij}(t)
=
\ \textrm{c}_{ij} & + \sum_n^{N_\textrm{states}} Z^i_{n} Z^j_{n} \left(e^{-E_n t} + e^{-E_n (T-t)}\right)
\nn \\
&+ (-1)^t 
\sum_m^{M_\textrm{states}} Z_{m,\,\mathrm{osc}}^{i} Z_{m,\,\mathrm{osc}}^{j}\left(e^{-E_{n,\,\mathrm{osc}}t}+ e^{-E_{n,\,\mathrm{osc}}(T-t)} \right)\, , \label{eq:matrixFit}
\end{align}
enables a complete determination of the spectral information obtained by the GEVP, as well as a direct determination of the constant-in-time finite-time contribution, $\textrm{c}_{ij}$, that was estimated in \cref{sec:finiteT}. In this appendix we provide results for this approach, which serves as a cross check on both the results obtained in \cref{sec:gevp} and \cref{sec:finiteT}. As this is to serve solely as a cross check, we do not perform a full systematic analysis of this approach, leaving it to a later work.

In line with the Bayesian fits performed in \cref{sec:fits}, we use as priors for the first seven energies $E_n$ and `diagonal' amplitudes, $ Z^i_{n},\ i=n$, the effective energy and amplitudes estimated from \cref{eqn:effectiveMass,eqn:effectiveAmp} applied to the diagonal entries of the correlation matrix. For the energies, we take the estimated energy as the central value and apply a 20\% prior width as in \cref{sec:fits}. For the amplitudes, as we cannot enforce $Z^i_{n} > 0$ for each amplitude due to relative sign differences between the operators in \cref{eqn:ctmat}, we use a prior centered at zero with the width given by twice the effective amplitude. For the off-diagonal amplitudes $ Z^i_{n},\ i\neq n$, we use a loose prior of $0(0.5)$\footnote{The width here is larger than the largest diagonal amplitude prior.}. For the states above these, we use the off-diagonal amplitude prior for all amplitudes, and we use a loose prior for the energy splitting of $\widetilde{\Delta E_n} =0.05(15)$ where the central value is characteristic of the splitting of the taste-broken spectrum in lattice units. For the diagonal constant terms, $\textrm{c}_{ij},\ i=j$, we take the estimates of \cref{sec:finiteT} for the central value with a $100\%$ prior width, allowing for a large sub-leading contribution. The priors for the off-diagonal constants are given by $\widetilde{\textrm{const}}_{ij} = 0(A_{ij})$, where the width is obtained using a conservative Cauchy-Schwartz bound from the diagonal constants,
\begin{align}
A_{ij}  =  \sqrt{\overline{\widetilde{\textrm{c}}_{ii}}\,\overline{\widetilde{\textrm{c}}_{jj}}}\,.
\end{align}
Here the overline indicates that the central value is taken from the diagonal priors.

\begin{figure}
  \centering
    \includegraphics[width=0.9\textwidth]{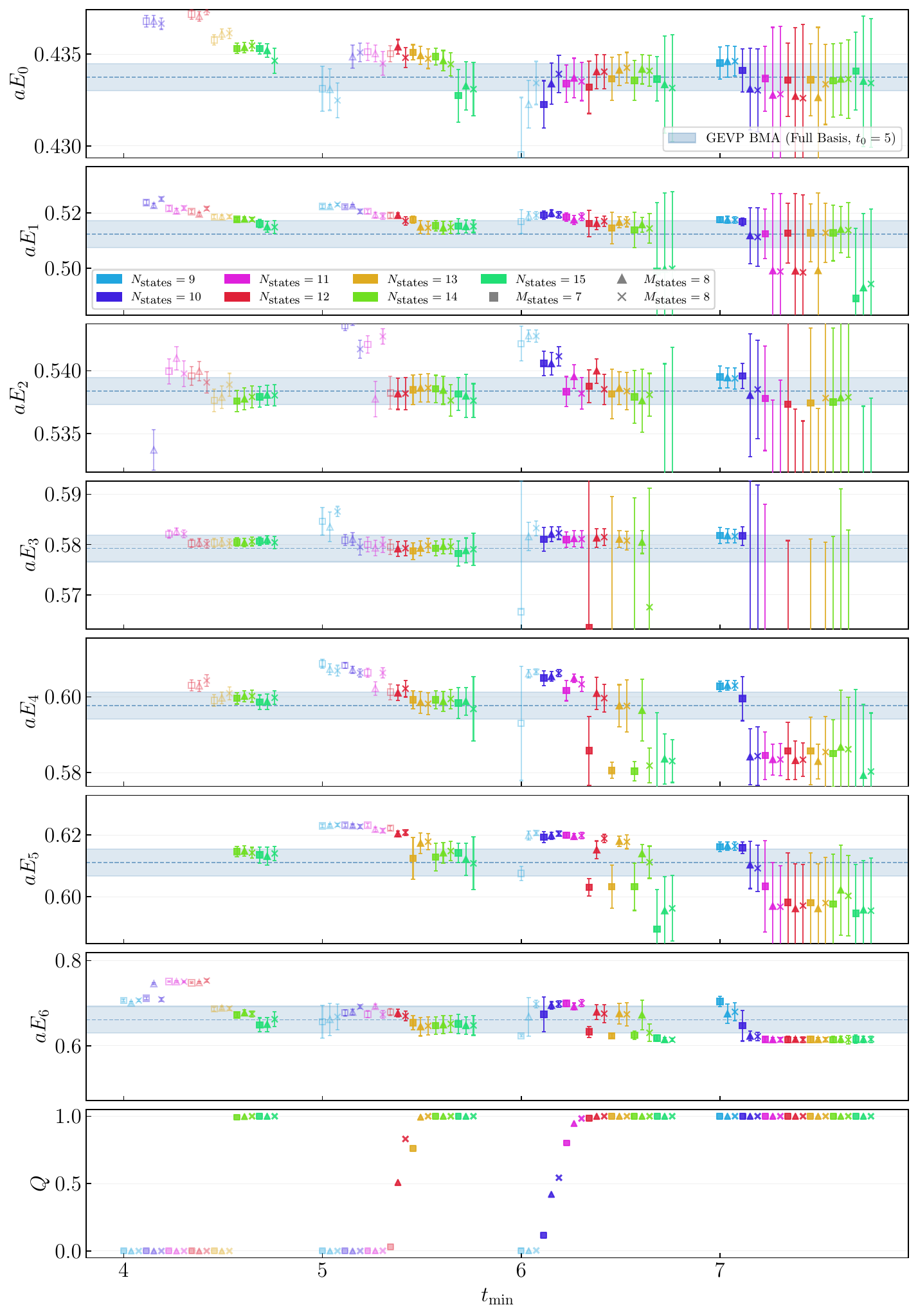}
    \caption{The stability of the extracted energies with respect to $t_\textrm{min}$, $N_\textrm{states}$, and $M_\textrm{states}$ of the MEM-fit ansatz of \cref{eq:matrixFit}. The GeVP energies determined in \cref{sec:gevp} are indicated with blue bands for each of the states.}\label{fig:MatrixFitStability}
\end{figure}

We fix the fit time ranges through the following procedure: The maximum times for diagonal correlators $C_{ii}(t)$, $t_{\textrm{max},ii}$, are obtained by choosing the largest $t_{\textrm{max},ii}$ that yields an acceptable fit quality for fits to just the diagonal $C_{ii}(t)$. The off-diagonal $t_{\textrm{max},ij}$ are then obtained by taking
\begin{equation}
t_{\textrm{max},ij}
=
\min\!\left(t_{\textrm{max},ii},\,t_{\textrm{max},jj}\right).
\end{equation}
For $t_{\textrm{min},ij}$ we enforce a common value, $t_{\textrm{min},ij}=t_{\textrm{min}}$, for all correlators and perform a stability analysis over a range of $t_{\textrm{min}}$. In addition to $t_{\textrm{min}}$, we also scan over the numbers of regular states, $N_{\textrm{states}}$, and oscillating states, $M_{\textrm{states}}$. 

The results of this scan are shown in \cref{fig:MatrixFitStability}, where we compare with the energies obtained from the GEVP procedure in \cref{sec:gevp}. We find the expected behavior, namely that the fits at early $t_{\textrm{min}}$ require a large number of excited states $N_{\textrm{states}}\geq 14$ for good fit quality, as indicated by the $Q$ value in the bottom panel. As we increase $t_{\textrm{min}}$, we find that the number of states needed drops to $N_{\textrm{states}} =  9$ at $t_{\textrm{min}}=7$. The results are very stable, especially above the ground state. Our preferred fit corresponds to the following hyper parameters $t_{\textrm{min}}=5$, $t_{\textrm{max}, ii} = \{15, 15, 17, 14, 17, 17, 17, 17\}$, $N_\textrm{states}=14$, $M_\textrm{states}=8$, indicated by the light-green triangles. The fit quality for this fit is exceptionally good at $\chi^2 / \text{d.o.f} [\text{d.o.f}] = 0.74 [418]$.

\subsection{Comparison with the GEVP spectrum}

\begin{figure}
  \centering
    \includegraphics[width=0.9\textwidth]{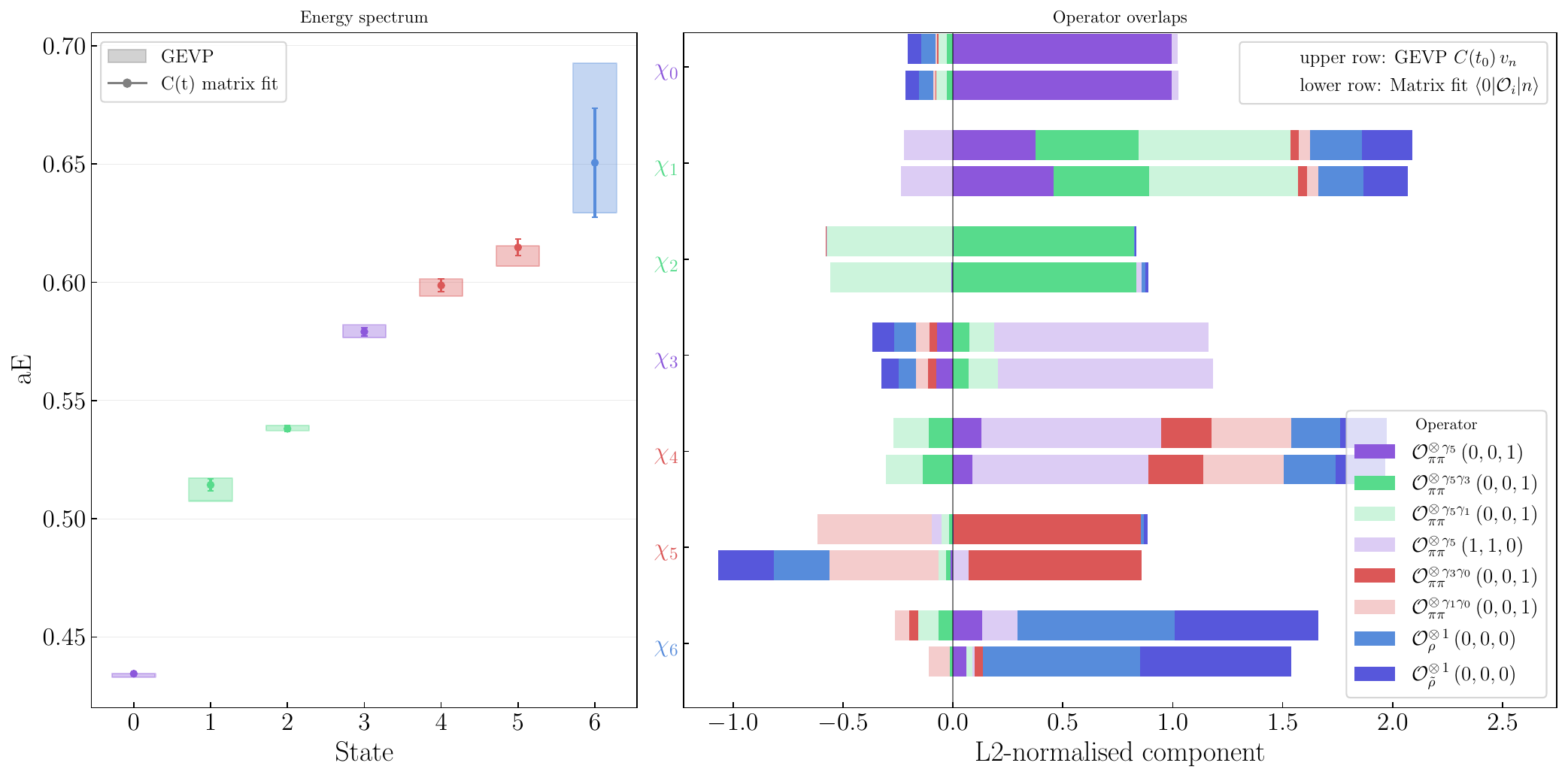}
    \caption{Comparison of the spectrum obtained from the GEVP performed \cref{sec:gevp} with that of the MEM fit described in this appendix. Left: Comparison of the energies obtained from the GEVP (bands) with the energies obtained from the fits (symbols). The colors correspond to the largest operator in the optimal operator of that state. Right: Comparison of the renormalized eigenvectors of the GEVP $C(t_0) v_n$ with the overlap amplitudes of the MEM fit.}\label{fig:MatrixFitGEVPCompareEnergy}
\end{figure}

In \cref{fig:MatrixFitGEVPCompareEnergy} (left), we compare the first seven energies obtained from the GEVP approach discussed in \cref{sec:gevp} with the spectrum obtained from the MEM approach discussed above. The results are all consistent within a standard deviation. We find that for some states the MEM approach gives more precise results, but we stress we have not performed a comprehensive estimate of the systematics in this approach, as it serves only as a cross check in this work. In \cref{fig:MatrixFitGEVPCompareEnergy} (right), we compare the renormalized eigenvectors of the GEVP $C(t_0)v_n$\footnote{The factor $C(t_0)$ relates eigenvectors of the GEVP equation with eigenvectors of $C(t)$.} (top row) and the overlap amplitudes $\langle 0 | \mathcal{O}_i | n \rangle$ obtained from the MEM approach (bottom row) for each of the first seven states. We observe exceptionally good agreement for all states except for the last two, where there are some differences, demonstrating that both approaches predict essentially the same operator-state mixing. The consistency highlights the validity of both approaches for extracting the desired spectrum.

\subsection{Finite-time estimate}

\begin{figure}
  \centering
    \includegraphics[width=0.7\textwidth]{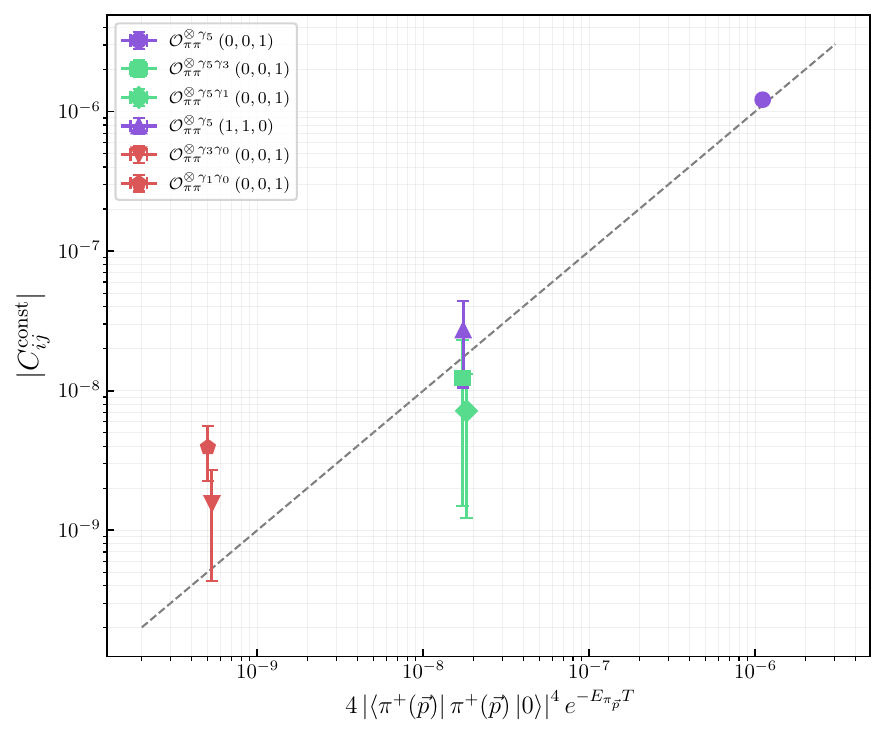}
    \caption{Comparison of the finite-time contribution estimated from procedure described in \cref{sec:finiteT} ($x$-value) to the constant term in the MEM fit ($y$-value). The dashed line indicates equivalence.}\label{fig:MatrixFitGEVPCompareFT}
\end{figure}

In addition to providing an alternative approach to extracting the spectrum, this method enables a more robust study of the finite-time effects estimated in \cref{sec:finiteT}. By including a constant term $c_{ij}$ in the fit for each correlator $C_{ij}(t)$ in \cref{eqn:ctmat}, we can estimate the size of the sub-leading finite-time effects not included in the estimates of \cref{sec:finiteT}. In \cref{fig:MatrixFitGEVPCompareFT}, we compare the estimates of the leading-order finite-time effects of the diagonal two-pion four-point functions ($x$-axis) with the values of the constants obtained from the MEM fit for these correlators ($y$-axis). With the dashed line indicating exact agreement, we find that for the largest contributions (the $\gamma_5 \otimes \gamma_5$ Goldstone pion) the estimate is in good agreement with the value of the constant obtained from the fit. As we get to the heavier states, the estimate is more of an underestimate, possibly due to mixing with lighter states. However, these contributions are almost two orders of magnitude smaller than the dominant well-determined estimate. Computing the difference between the constant and the estimate of \cref{sec:finiteT} for this dominant ground-state contribution, we find the sub-leading contribution to be roughly 9(5)\% of the total. 

\begin{figure}
  \centering
    \includegraphics[width=0.85\textwidth]{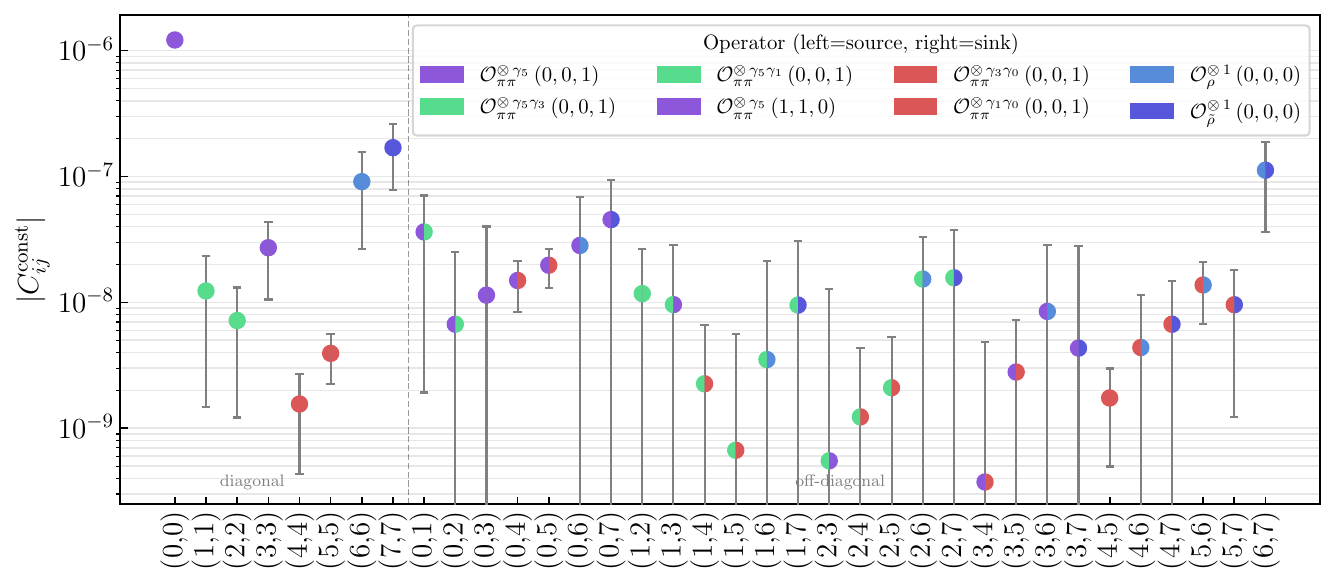}
    \caption{All constants obtained from the MEM fit \cref{eq:matrixFit}. On the left-side of the figure we show the diagonal constants including the $\rho$ two-point functions. On the right side we show the off-diagonal constants.}\label{fig:MatrixFitGEVPCompareFTOffDiag}
\end{figure}

In addition to the diagonal two-pion four-point functions of \cref{eqn:ctmat}, the MEM fit enables determinations of the finite-time contributions of the other correlators that appear in the matrix, namely the diagonal $\rho$ two-point functions $C_{\rho \to \rho}(t)$, which were also estimated in Ref.~\cite{FermilabLatticeHPQCD:2024ppc}, and the off-diagonal three- and four-point correlators. The results for these, along with results discussed above, are shown in \cref{fig:MatrixFitGEVPCompareFTOffDiag}. We find that the next largest finite-time contribution after the ground-state two-pion correlator comes from the $\rho$ two-point correlator. The size of this contribution is of the same order as the sub-leading contribution not included in the leading order estimate of \cref{sec:finiteT}. All other contributions are at least two orders of magnitude smaller. Finally, we note that the Euclidean time extent ($T$) of the HISQ ensemble used in this work is approximately 0.8~fm smaller than that of all the other physical-mass HISQ ensembles that will be used in future work; hence, these effects will be significantly reduced at finer lattice spacings.

\bibliographystyle{apsrev4-2}
\bibliography{refs}

\end{document}